\documentclass[final]{jpp}
\usepackage{graphicx}
\usepackage[T1]{fontenc}
\usepackage{amsmath}
\usepackage{bm}
\usepackage{bbold}
\usepackage{natbib}
\usepackage{upgreek}
\usepackage{float}
\usepackage{mathtools}
\usepackage[section]{placeins}
\usepackage{physics}
\usepackage{siunitx}
\usepackage{verbatim}
\usepackage{comment}
\usepackage[colorlinks=true,
  linkcolor=black, citecolor=blue, urlcolor=blue]{hyperref}   
\usepackage[dvipsnames,svgnames,x11names,hyperref]{xcolor}  
\usepackage[capitalize]{cleveref}
\newcommand\added[1]{{\color{black}#1}}

\renewcommand{\vec}[1]{\bm{#1}}

\renewcommand{\r}{\vec{r}}
\newcommand{\x}{\bm{x}}
\newcommand{\vi}{\bm{v}}

\newcommand{\e}{\bm{e}}
\newcommand{\R}{\bm{R}}

\renewcommand{\e}{\bm{e}}

\renewcommand{\b}{\vec{b}}

\renewcommand{\u}{\vec{u}}

\newcommand{\g}{\bm{g}}
\newcommand{\Z}{\bm{Z}} 

\renewcommand{\L}{\mathcal{L}}

\newcommand{\kernel}[1]{\mathcal{K}_{#1}}
\newcommand{\gyaver}[1]{\left<  #1 \right>}

\newcommand{\sparallel}{s_{\parallel a}}

\newcommand{\vparallel}{v_{\parallel}}

\newcommand{\T}{\mathcal{T}}         

\newcommand{\gyN}{N_a}

\newcommand{\gradv}{\grad_{\vi}}
\newcommand{\partialv}{\frac{\partial}{\partial v}}

\newcommand{\vTa}{v_{Ta}}
\newcommand{\vTb}{v_{Tb}}

\renewcommand{\sb}{s_b}
\newcommand{\sa}{s_a}

\newcommand{\M}{\mathsfbi{M}}
\newcommand{\m}{\mathsfbi{m}}
\newcommand{\Y}{\mathsfbi{Y}}

\newcommand{\corr}[1]{\textbf{\color[red]{}}}

\crefname{subsection}{sec.}{subsections}

\newcommand{\C}{\mathcal{C}}
\renewcommand{\g}{\textsl{g}}

\crefformat{subsection}{\S#1}

\shorttitle{Gyrokinetic Collision Operators Using a Moment Approach }
\shortauthor{B. J. Frei et al.}

\title{Development of Advanced Linearized Gyrokinetic Collision Operators Using a Moment Approach}

\author{B. J. Frei\aff{1}
  \corresp{\email{baptiste.frei@epfl.ch}},
  J. Ball \aff{1},
   A. C. D. Hoffmann \aff{1},
   R. Jorge\aff{2},
    P. Ricci\aff{1}
 \and  L. Stenger \aff{1}}
 
\affiliation{\aff{1} \'Ecole Polytechnique F\'ed\'erale de Lausanne (EPFL),
  Swiss Plasma Center (SPC),
CH-1015 Lausanne, Switzerland
\aff{2} Institute for Research in Electronics and Applied Physics, University of Maryland, College Park MD
20742, United States of America}

\begin{document}
\maketitle
\begin{abstract}
The derivation and numerical implementation of a linearized version of the gyrokinetic (GK) Coulomb collision operator (Jorge R. \textit{et al.}, J. Plasma Phys. \textbf{85}, 905850604 (2019)) and of the widely-used linearized GK Sugama collision operator (Sugama H. \textit{et al.}, Phys. Plasmas \textbf{16}, 112503 (2009)) is reported.
An approach based on a Hermite-Laguerre moment expansion of the perturbed gyrocenter distribution function is used, referred to as gyro-moment expansion. This approach allows considering arbitrary perpendicular wavenumber and expressing the two linearized GK operators as a linear combination of gyro-moments where the expansion coefficients are given by closed analytical expressions that depend on the perpendicular wavenumber and on the temperature and mass ratios of the colliding species. The drift-kinetic (DK) limits of the GK linearized Coulomb and Sugama operators are also obtained. \added{Comparisons between the gyro-moment approach and the DK Coulomb and GK Sugama operators in the continuum GK code GENE are reported, focusing on the ion-temperature-gradient instability and zonal flow damping, finding an excellent agreement. It is confirmed that stronger collisional damping of the zonal flow residual by the Sugama GK model compared to the GK linearized Coulomb (Pan Q. \textit{et al.}, Phys. Plasmas \textbf{27}, 042307 (2020)) persists at higher collisionality.} Finally, we show that the numerical efficiency of the gyro-moment approach increases with collisionality, a desired property for boundary plasma applications.
 \end{abstract}

\section{Introduction}

Understanding the plasma dynamics of the tokamak boundary is necessary to address some of the most crucial problems fusion is facing today and, for instance, to ensure the success of future devices, such as ITER \citep{Shimada2007,Holtkamp2009}. In fact, the boundary region, which extends from the external part of the closed flux surface region (typically referred to as the edge) to the scrape-off layer (SOL) where the magnetic field lines intercept the machine vessel walls, sets the particle and heat exhaust, helium ash removal and the overall confinement of the device. Most often, drift-reduced fluid models (see, e.g., \citep{Zeiler1997}) implemented in a number of simulation codes \citep{dudson2009,tamain2016,Halpern2016,Paruta2018,zhu2018,Stegmeir2019,Giacomin2020} are used to simulate SOL turbulence. However, the applicability of Braginskii-like fluid models is questionable to describe the whole boundary region, since they assume $k_\parallel \lambda_{mpf} \ll 1$ and $k_\perp \rho_i \ll 1$, where $k_\parallel$ and $k_\perp$ are the parallel and perpendicular components of the wavevector $\bm k$ to the equilibrium magnetic field lines, $\lambda_{mfp}$ is the electron mean free path, and $\rho_i$ is the ion Larmor radius. In particular, the $k_\parallel \lambda_{mpf} \ll 1$ assumption is expected to no longer be satisfied in larger and hotter machines near and inside of the edge, like ITER \citep{Holtkamp2009}. For this reason, an enormous effort is being carried out by the fusion community to extend core gyrokinetic (GK) codes, which are based on PIC or grid-based methods \citep{Dimits1996,Idomura2003,Gorler2011,Villard2013,Chang2017}, to simulate the boundary region. However, despite large recent progress, the application of core GK models and codes in the boundary region is challenged, for instance, by the nature of collisions. Collisions are expected to play an important role in the boundary region as they can substantially affect the properties of trapped electron (TEM) and ion temperature gradient (ITG) modes \citep{Belli2017}, the suppression of short wavelength structure \citep{Barnes2009}, TEM turbulent fluxes, the damping of zonal flows \citep{Pan2020,pan2021}\added{, and the growth rate of pedestal microtearing modes \citep{pan2021}.}

An accurate description of binary Coulomb collisions dominated by small angle deflection is provided by the nonlinear Fokker-Planck operator, that we refer to as the Coulomb collision operator \citep{Rosenbluth1957}. The nonlinear Coulomb collision operator is an integro-differential operator that acts on the full particle distribution function. Because of the complexity of the nonlinear Coulomb collision operator, its version linearized around a Maxwellian distribution function is usually considered \citep{Rosenbluth1957,Helander2002,Hazeltine2003}. However, despite being simpler than the nonlinear collision operator, the numerical implementation of the linearized Coulomb collision operator is still challenging. This inherent difficulty primarily stems in the evaluation of the component of the linearized collision operator that contains velocity space integrals of the perturbed distribution function. As a matter of fact, it is only recently that the implementation of a linearized GK Coulomb collision operator has been reported in the GK GENE code \citep{Pan2020,pan2021}. 

Because of the complexity associated with dealing with the Coulomb collision operator, simplified \added{model} collision operators have been developed \citep{Dougherty1964,Hirshman1976,Abel2008,Sugama2009,Sugama2019}. In particular, \citet{Sugama2009} derived an approximated linearized collision operator, that we refer to as the Sugama operator, \added{extending the work of \citet{Abel2008} to multiple ion species plasmas}. This operator is designed such that it conserves particle, momentum and energy and, in addition, it fulfils the self-adjoint relations for unlike-species, a useful property when solving collisional transport problems based on the drift-kinetic equation \citep{rosenbluth1972}. While the Sugama operator have been tested in neoclassical and turbulent studies at low collisionalities \citep{nakata2015improved,nunami2015development}, its deviation with respect to the Coulomb operator \citep{Crandall2020,Pan2020,pan2021} is expected to increase when applied at higher collisionalities, such as the ones in the plasma boundary. \added{We remark that the first GK Coulomb operators with exact field terms rather than simplified models, including finite Larmor radius effects, were formulated in \citet{Li2011,Madsen2013,Pan2019,Jorge2019}.}
 
  Advanced numerical algorithms have been developed for the implementation of these collision operators \citep{Xu1991,Dimits1994,Barnes2009,landreman2012,landreman2013,nakata2015improved,Hager2016,Belli2017,Crandall2020,Pan2020}, since numerical errors can potentially yield spurious noise and break the conservation laws producing artificial instabilities. Among the different numerical schemes, Monte-Carlo techniques \added{are used in some PIC codes} \citep{Xu1991,Dimits1994,Kolesnikov2010}, while GK continuum codes \added{and some PIC codes adopt} velocity-space grid-based discretization method \citep{Dorf2014,Hager2016,Pan2020,Crandall2020}. We remark that, recently, a novel conservative discontinuous Galerkin method has been developed and applied to a nonlinear Dougherty collision operator in \citet{Francisquez2020}, and present the potential of being generalized to more advanced collision operators. \added{Novel spectral methods have yielded significant improvements in accuracy and speed in neoclassical and gyrokinetic codes \citep{landreman2012,landreman2013,held2015,Belli2017}.} Also, orthogonal polynomials expansion techniques have been used in the numerical treatment of velocity-space derivatives \citep{Donnel2019}. 
  
  Recently, to provide a theoretical and numerical framework able to predict and study efficiently the dynamics of the boundary region, a GK model valid at an arbitrary level of collisionality has been developed by \citet{Frei2020}. This model is based on expanding the full distribution function on a Hermite-Laguerre polynomial basis. By projecting the GK equation onto the same basis, the evolution of the distribution function is reduced to the one of the coefficients of its expansion, referred to as \textit{gyro-moments}. To describe collisional effects, while \citet{Frei2020} presents a nonlinear full-F GK Dougherty collision operator for like-species, a nonlinear full-F Coulomb collision operator has been derived in \citet{Jorge2019} using the same Hermite-Laguerre gyro-moment approach. This operator is valid at arbitrary wavelengths perpendicular to the magnetic field, and arbitrary mass and temperature ratios. Coupled with the full-F nonlinear GK operator in \citet{Jorge2019}, the gyro-moment hierarchy in \citet{Frei2020} reduces, in the DK limit, to an improved set of Braginskii-like fluid equations where only the lowest order gyro-moments are considered \citep{Jorge2017}, while a progressively more detailed kinetic representation is obtained by increasing the number of gyro-moments.

The goal of the present paper is twofold. First, we derive the linearized GK Coulomb collision and the GK Sugama operators within the Hermite-Laguerre gyro-moment expansion. For treating the case of the GK Coulomb collision operator, we leverage the techniques introduced by \citet{Jorge2019}. Using a spherical harmonic moment expansion technique of the particle distribution function, we perform the gyro-average of the linearized Coulomb operator analytically, removing the fast particle gyro-motion, and project the result onto the Hermite-Laguerre basis. One of the advantages of the Hermite-Laguerre decomposition is that it allows one to reduce velocity integrals in the linearized collision operators, which are treated numerically in, e.g., \citep{Pan2020,Crandall2020}, to closed algebraic expressions that depend only on the perpendicular wavenumber, on the temperature and mass ratios of the colliding species and on the gyro-moments. The DK limits of the GK Coulomb and Sugama operators are also derived. Second, expressed in a form suitable for a numerical treatment, we implement these operators in a numerical simulation code. We test and compare the GK Sugama and DK Coulomb collision operators against the widely-used gyrokinetic continuum code GENE \citep{Jenko2000}, focusing on the ITG linear growth rate and collisional zonal flow damping. We report a good agreement with GENE, and show, in particular, that the new linearized GK Sugama produces a larger collisional damping of the ZF residual compared to the GK Coulomb. Finally, we illustrate the numerical efficiency of the gyro-moment approach against GENE as the collisionality increases, a desired property for boundary plasma applications. We remark that, while only valid for small deviations from equilibrium, the linearized GK operator presented in this work opens the way to a future numerical implementation of the full-F nonlinear GK Coulomb operator \citep{Jorge2019}.
 
The structure of the paper is the following. We first introduce the gyro-moment approach to the linearized gyrokinetic model in \cref{sec:GyrokineticModelEquation}. In \cref{sec:LinCollOp}, we derive the Hermite-Laguerre expansion the linearized GK Coulomb and  GK Sugama operators that we express in terms of gyro-moments. We describe the numerical implementations of the collision operators in \cref{sec:numericalimplementations}. Then, we test the linearized GK collision operators considering the ion-temperature-gradient (ITG) instability and collisional damping of the zonal flow (ZF) residual in \cref{sec:numericaltests}. By performing direct results comparisons with GENE, \cref{sec:convergence} illustrates the numerical efficiency of the gyro-moment approach as the collisionality increases. We conclude in \cref{sec:conclusion}.

\section{Gyrokinetic Model Equation}
\label{sec:GyrokineticModelEquation}

We consider the electrostatic gyrokinetic Boltzmann equation in the presence of magnetic field, density and temperature gradients. We use the gyrocenter phase-space coordinates  $\Z  = (\R, \mu, v_\parallel,\theta)$, where $\R = \r - \bm \rho_a$ is the gyrocenter position (with $a$ the species index), with $\r$ the particle position and $\bm \rho_a(\R, \mu, \theta) = \b \times \bm v/\Omega_a$ the gyroradius of species $a$ ($\b = \bm B / B$ and $\Omega_a = q_a B / m_a$), $\mu = m_a v_\perp^2/[2 B(\R)]$ is the magnetic moment, $v_\parallel = \bm b \cdot \bm v$ is the component of the velocity parallel to the magnetic field, and $\theta$ is the gyroangle. We linearize the GK Boltzmann equation by assuming that the \textit{gyrocenter} distribution function of species $a$, $F_a = F_a (\R, \mu, v_\parallel)$, is a perturbed Maxwellian, that is $F_{a} = F_{Ma} + \textsl{g}_a$, with $\g_a = \g_a(\R ,\mu, v_\parallel, t)$, the perturbation with respect to the local Maxwellian distribution function $F_{Ma} = N   e^{- s_{\parallel a}^2 - x_a} / (\pi^{3/2} v_{Ta}^3)$ such that $\g_{a}/ F_{Ma} \ll1$,, with $N = N_i(\R) = N_e(\R) $ the background gyrocenter density, $s_{ \parallel a} = v_\parallel / v_{Ta}(\R)$, $x_a= \mu B(\R) / T_{a}(\R)$, and $v_{T_a}^2(\R) = 2 T_{a}(\R)/m_a$ the thermal velocity based on the equilibrium temperature $T_a(\R)$. The linearized electrostatic GK Boltzmann equation that describes the time evolution of $\g_a(\bm k,t) = \int d \R \g_a(\R , t) e ^{- i \bm R \cdot \bm k}$ can be written as \citep{Hazeltine2003}

\begin{align} \label{eq:LinGK}        \frac{\partial}{\partial t } \g_a & + i \omega_{Ba} h_a + v_\parallel  \grad_\parallel h_a  - \frac{\mu}{m_s} (\b \cdot \grad B)\frac{\partial }{\partial v_\parallel}  h_a  -      i \omega_{Ta}^* J_0(b_a \sqrt{x_a})\phi F_{Ma}  = \sum_{b} \C_{ab},
\end{align}
\\
where $\phi = \phi(\bm k, t) $ is the Fourier component of the electrostatic potential and we introduce the nonadiabatic part, $h_a = h_a (\bm k, \mu, v_\parallel,t)$, of the perturbed gyrocenter distribution function $\g_a$, that is

\begin{align} \label{eq:hs}
    h_a = \g_a  + \frac{q_a}{T_{a}}     F_{Ma}  J_0(b_a \sqrt{x_a}) \phi.
\end{align}
\\
In \cref{eq:LinGK}, we define $\grad_\parallel = \b \cdot \grad$, $\omega_{Ba}    = v_{Ta}^2 ( x_a+ 2 s_{\parallel a}^2)  \omega_B/ (2 \Omega_a)$, with $\omega_B = \left(  \b \times \grad \ln B \right) \cdot \bm k$, and $\omega_{Ta}^*  =  \left[\omega_N + \omega_{T_a}\left( x_a + s_{\parallel a}^2 -3/2\right) \right]   $, with $\omega_N = \b \times \grad \ln N \cdot \bm k/ B$ and $\omega_{T_a} = \b \times \grad \ln T_a \cdot \bm k/B$ .  Finite Larmor radius (FLR) effects are taken into account through the zeroth order Bessel function, $J_0(b_a \sqrt{x_a})$, where $b_a = k_\perp v_{T_a} /\Omega_a$ is the normalized perpendicular wavevector, with $ k_\perp  = |\bm k - (\b \cdot \bm k) \b|$. On the right hand-side of \cref{eq:LinGK}, we define the linearized \textit{gyrokinetic} collision operator between species $a$ and $b$, $\C_{ab} = \C_{ab}(\R, \mu, v_\parallel)$, 

\begin{align} \label{eq:gyaverC}
    \C_{ab}= \gyaver{C_{ab}}_{\R}  = \frac{1}{2 \pi}\int_0^{2 \pi} d \theta C_{ab},
\end{align}
\\
that describes the effects of small angle Coulomb collisions between particle species $a$ and $b$. While the linearized collision operator $C_{ab}$ is the focus of \cref{sec:LinCollOp}, we remark that $\gyaver{\dots}_{\R} =\int_0^{2 \pi} d \theta \dots /(2 \pi) $ denotes the gyro-average operator evaluated at the gyrocenter position (the subscript $\R$ indicates that the integral over the gyroangle should be computed holding $\R$ constant). We remark that the linearized GK collision operator is defined as a function of the gyrocenter phase-space, i.e. $ \C_{ab} =  \C_{ab}(\R, \mu, v_\parallel)$, and it is therefore gyrophase independent.

The linearized Boltzmann equation is closed by the gyrokinetic quasi-neutrality condition that determines self-consistently the electrostatic potential, 

\begin{align} \label{eq:Poisson}
    \sum_a \frac{q_a^2}{T_a} \left[ 1 - \Gamma_0(a_a) \right] \phi = \sum_a q_a  \int d \mu d v_\parallel d \theta \frac{B}{m_a} J_0(b_a \sqrt{x_a}) \g_a
\end{align}
\\
where $a_a= b_a^2 / 2 $, and $\Gamma_0(x) = I_0(x) e^{- x}$ with $I_0$ the modified Bessel function. 

In order to approach the solution of the GK system, \cref{eq:LinGK,eq:Poisson}, we use an Hermite-Laguerre moment expansion of the gyrocenter perturbed distribution function $\g_a$ \citep{Mandell2018,Frei2020}. Then, by projecting \cref{eq:LinGK} onto the complete Hermite-Laguerre basis polynomials, we reduce its dimensionality removing the $(\mu, v_\parallel)$ dependence. More precisely, we decompose the perturbed \textit{gyrocenter} distribution function, $\g_a$, onto a set of Hermite-Laguerre polynomials \citep{Jorge2017,Jorge2019,Frei2020},

\begin{align} \label{eq:fHL}
  \g_a = \sum_{p = 0}^\infty \sum_{j = 0}^\infty N_a^{pj} \frac{H_p(s_{\parallel a}) L_j(x_a)}{\sqrt{2^p p!}} F_{Ma},
 \end{align}
 \\
where the Hermite-Laguerre velocity moments of $\g_a$ are defined as 

\begin{equation} \label{eq:Npjdef}
    N_a^{pj} = \frac{1}{N_a} \int d \mu  d \vparallel  d \theta\frac{B}{m_a} \g_a \frac{H_p(s_{\parallel a}) L_j(x_a)}{\sqrt{2^p p!}},
\end{equation}
\\
with $N_a =  \int d \mu  d \vparallel  d \theta B \g_a / m_a$ the gyrocenter density. In \cref{eq:fHL}, we introduce the Hermite and Laguerre polynomials, $H_p$ and $L_j$, via their Rodrigues' formulas  \citep{gradshteyn}

\begin{align}
H_p(x) & = (-1)^p e^{x^2} \frac{d^p}{d x^p} \left( e^{- x^2} \right), \label{eq:Hermite} \\
L_j(x)  &= \frac{e^{x}}{j!} \frac{d^j }{d x^j} \left ( e^{- x } x^j\right),
\end{align}
\\
 and we note their orthogonality relations over the intervals, $] -\infty, \infty[$ weighted by $e^{-x^2}$, and $[0,+ \infty[$ weighted by $e^{-x}$, respectively,
 
 \begin{align}
 \int_{- \infty}^\infty d x H_p(x) H_{p'}(x) e^{- x^2} & = 2^p p! \sqrt{\pi} \delta_{p}^{p'}, \\
 \quad  \int_0^\infty d x L_j(x) L_{j'}(x) e^{-x} & = \delta_j^{j'}.
 \end{align}
 \\
We denote the Hermite-Laguerre coefficients,  $N_a^{pj}$ in \cref{eq:fHL}, as the gyro-moments. We remark that any function, $f = f (\mu, v_\parallel)$, that satisfies 

 \begin{equation} \label{eq:L2cond}
 \int d \mu d v_\parallel  d \theta \frac{B}{m_a} |f|^2 e^{-\sparallel^2 -  x_a} < + \infty,
 \end{equation} 
\\
can be decomposed onto the orthogonal basis defined by the Hermite-Laguerre polynomials \citep{Wong1998}. This is, in particular, fulfilled by the perturbed gyrocenter distribution function $g_a$, and also by its nonadiabatic part, $h_a$.

 We now project \cref{eq:LinGK,eq:Poisson} onto the Hermite-Laguerre polynomial basis. For this purpose, we note that the Bessel function $J_0$ (appearing in both \cref{eq:LinGK,eq:Poisson} and arising from finite Larmor radius (FLR) effects) and, more generally $J_m$, with $m > 0$, can be conveniently expanded onto associated Laguerre polynomials, $L^m_n(x) = (-1)^m d^m L_{n + m}(x) / d x^m$, such as \citep{gradshteyn}

\begin{align} \label{eq:J02Laguerre}
J_m(b_a \sqrt{x_a}) = \left(\frac{b_a \sqrt{x_a}}{2}\right)^{m}\sum_{n=0}^\infty \frac{n!\kernel{n}(b_a )}{(n + m)!} L^m_n(x_a),
\end{align}
\\
where we introduce the velocity-independent expansion coefficients

\begin{align} \label{eq:kernel}
\kernel{n}(b_a)  = \frac{1}{n!}\left( \frac{b_a}{2}\right)^{2n } e^{- b_a^2 /4}.
\end{align}
\\
We refer to $\kernel{n}(b_a)$ as the $nth$-order kernel function, with velocity-independent argument $b_a$ \citep{Frei2020}. In fact, \cref{eq:J02Laguerre} allows us to isolate the velocity dependence, $\sqrt{x_a}$, appearing in the argument of the Bessel function $J_m$ from fluid quantities such as $b_a$.

Multiplying the gyrokinetic Boltzmann equation, \cref{eq:LinGK}, by the basis element $B H_p(s_{\parallel a}) L_j(x_a) / \sqrt{2^p p!}$, and integrating over the velocity space yields the gyro-moment hierarchy equation,

\begin{align} \label{eq:momenthierachyEquationNormalized}
     & \frac{\partial }{\partial t} N_a^{pj}   +  v_{Ta} \grad_\parallel \left( \sqrt{\frac{p+1}{2}}   n_a^{p+1j}+ \sqrt{\frac{p}{2}} n_a^{p-1j}  \right) \nonumber \\
      &   -  v_{Ta}\grad_\parallel \ln B \left[    (j+1)\sqrt{p+1} n_a^{p+1j}+ (j+1)\sqrt{p} n_a^{p-1j} - j \sqrt{p+1} n_a^{p+1j-1}- j \sqrt{p} n_a^{p-1j-1 }  \right] \nonumber \\
    & +   \frac{ T_a i \omega_B}{q_a B} \left[ \sqrt{(p+1)(p+2)} n_a^{p+2j}  
+(2p+1)  n_a^{pj}+ \sqrt{p(p-1)} n_a^{p-2j}
    \right. \nonumber \\
&    \left. +  (2 j +1)   n_a^{pj}  - j n_a^{pj-1} - (j+1)  n_a^{pj+1}    \right] \nonumber \\
& +   v_{Ta}  \grad_\parallel \ln B   \sqrt{ \frac{ p}{2}} \left[ (2 j +1) n_a^{p-1j} - (j +1) n_a^{p-1j+1} - j n_a^{p-1 j-1} \right] \nonumber \\
& - i \phi \left\{ \kernel{j} \delta_p^0 \omega_N +  \omega_{T_a}\left[\kernel{j} \left(\frac{\delta_p^2}{\sqrt{2}}  - \delta_p^0\right) + \delta_p^0 \left((2 j+1) \kernel{j} - j \kernel{j-1 } - (j+1)\kernel{j+1}\right) \right]  \right\} \nonumber \\
&  = \sum_{b} \C_{ab}^{pj},
\end{align}
\\
where we define $\C_{ab}^{pj} =\C_{ab}^{pj}(\R,t)$ as the Hermite-Laguerre gyro-moment expansion of the linearized GK collision operator $\C_{ab}$,

\begin{align} \label{eq:Cpj}
\C_{ab}^{pj} = \int  d \theta d \mu d v_\parallel \frac{B}{m_a} \frac{H_p(s_{\parallel a}) L_j(x_a)}{ \sqrt{2^p p!}} \C_{ab}.
\end{align}
\\
In \cref{eq:momenthierachyEquationNormalized}, the nonadiabatic gyro-moment $n_a^{pj}$, are defined by 

\begin{align} \label{eq:npjdef}
n_a^{pj} = \frac{1}{N_a} \int d \mu  d \vparallel  d \theta\frac{B}{m_a} h_a \frac{H_p(s_{\parallel a}) L_j(x_a)}{\sqrt{2^p p!}},
\end{align}
\\
and can be expressed in terms of the gyro-moments $N_a^{pj}$ by projecting \cref{eq:hs} onto the Hermite-Laguerre basis, yielding

\begin{equation} \label{eq:defnspj}
n_a^{pj} = N^{pj}_a + \frac{q_a}{ T_a} \kernel{j} \phi \delta_p^0.
\end{equation}
\\
Finally, the GK quasineutrality condition projected onto the Hermite-Laguerre basis is expressed as

\begin{equation} \label{eq:GKPoissonNpj}
\sum_{a} \frac{q_a^2}{T_a}\left( 1 - \sum_{n=0}^{\infty} \kernel{n}^2(b_a) \right) \phi = \sum_a q_a N_a \sum_{n=0}^{\infty}  \kernel{n}(b_a) N_a^{0n}.
\end{equation}
\\
The linearized gyro-moment hierarchy, \cref{eq:momenthierachyEquationNormalized}, with the GK quasineutrality condition \cref{eq:GKPoissonNpj}, describes the evolution of the gyro-moments as it results from the interplay of Landau damping, magnetic gradient drifts, magnetic trapping and, finally,  the driving density and temperature gradients, $\omega_N$ and $\omega_{Ta}$. Collisional effects are expressed by the collisional gyro-moment of the GK collision operator $\C_{ab}^{pj}$ defined in \cref{eq:Cpj}. Contrary to the GK Boltzmann equation, given in \cref{eq:LinGK}, that depends on the phase space coordinates, the gyro-moment hierarchy, \cref{eq:momenthierachyEquationNormalized}, constitutes an infinite system of fluid-like equations for the variables $N_a^{pj}$, which are only functions of the Fourier mode wavevector $\bm k$ and time $t$.

We remark that \cref{eq:momenthierachyEquationNormalized} is equivalent to the previous Hermite-Laguerre moment hierarchy derived by \citet{Mandell2018}, where the probabilist's Hermite polynomials, $He_{p}(x)$, are used instead, defined by $He_{p}(x)  = 2^{-p/2} H_p(x/2)$ (with $H_p(x)$ in \cref{eq:Hermite}), albeit with collisional effects modelled by a linearized GK Dougherty operator \citep{Dougherty1964}, which is characterized by a sparse Hermite-Laguerre representation. 

\section{Linearized Collision Operator Models}
\label{sec:LinCollOp}

In general, a linearized collision operator between particles of species $a$ and $b$, $C_{ab}$, is obtained by linearizing a nonlinear collision operator, $C_{ab}^{NL}$. Nonlinear collision operators are usually defined on the full \textit{particle} distribution function expressed in the particle phase-space $(\r, \bm v)$. This is because collisions occur at the particle position $\r$ (rather than at the gyrocenter position $\R$). Given $f_{Ma} = f_{Ma}(\bm r, \bm v)$, the particle Maxwellian distribution function of species $a$, and $f_a = f _a(\r,\bm v)$ its small amplitude perturbation, $f_a / f_{Ma} \ll 1$, the linearized collision operator, $C_{ab} = C_{ab}(\r, \vi) = C_{ab} (f_a,f_b)$, can be expressed as

\begin{align}
    C_{ab}  = C_{ab}^T + C_{ab}^F,    
\end{align}
\\
where we introduce the \textit{test} component of the linearized collision operator, $C_{ab}^T  = C_{ab}^T(\r, \bm \vi) =  C_{ab}^T(f_a(\r, \bm \vi)) =      C^{NL}(f_a,f_{Mb}) $, and the \textit{field} component, $ C_{ab}^F =  C_{ab}^F(\r, \bm \vi) = C_{ab}^F (f_b(\r, \bm \vi)) =   C^{NL}(f_{Ma},f_b) $. Local conservation properties of the collision operator constraint the \textit{test} and \textit{field} components. In fact, while particle conservation is satisfied by both test and field components \citep{Xu1991}, such that

 \begin{align} \label{eq:particleconservation}
      \int d \vi C_{ab}^T  = \int d \vi  C_{ab}^F  = 0,
 \end{align}
 \\
 the momentum and energy conservations require that 
 
 \begin{align} \label{eq:momentumconservation}
 \int d \vi m_a \bm v C_{ab}^T +  \int d \vi m_b \bm v C_{ba}^F = 0, \\
  \int d \vi m_a v^2  C_{ab}^T +  \int d \vi m_b v^2 C_{ba}^F = 0,\label{eq:energyconservation}
 \end{align}
 \\
 respectively. We remark that the velocity integrals in \cref{eq:particleconservation,eq:momentumconservation,eq:energyconservation} are evaluated at constant $\r$, and that local conservation laws do not hold in the gyrocenter coordinates $(\R, \mu, v_\parallel, \theta)$, because of the difference between the gyrocenter and particle positions, being $\r - \R = \bm \rho_a(\R,\mu,\theta)$. As an example of the implication of the difference between $\r$ and $\R$, we consider the zeroth-order gyro-moment of the GK linearized collision operator. In Fourier-space and performing the velocity at constant $\R$, one has

\begin{align} \label{eq:zeromomentC}
 \int  d \vi \gyaver{C^T_{ab}(\r,\vi)}_{\R} & = 2 \pi \int d \bm k  \int  d v_\parallel d \mu \frac{B}{m_a} \gyaver{e^{ i \bm k \cdot \bm \rho_a}C^T_{ab}(\bm k,\mu, v_\parallel, \theta)}_{\R} e^{i \bm k \cdot \R} \nonumber \\
&  = 2 \pi \int d \bm k  \sum_{j=0}^\infty \int  d v_\parallel d \mu \frac{B}{m_a} \gyaver{ \frac{(i\bm k \cdot \bm \rho_a)^j}{j!} C^T_{ab}(\bm k,\mu, v_\parallel, \theta)}_{\R} e^{i \bm k \cdot \R}.
\end{align}
\\
Since $C_{ab}^T $ conserves particles, i.e. $\int   d \vi C_{ab}^T = 0$, we expect only the $j = 0$ term in the $j$ sum to vanish when the gyro-averaged is performed, and the $j \neq 0$ terms to give finite contributions as the particle gyro-radius $\bm \rho_a =\bm \rho_a (\R, \mu,\theta) $ depends on the the gyro-angle $\theta$. In fact, \cref{eq:zeromomentC} does not vanish, and the linearized GK collision operator yields classical gyro-diffusion in gyrocenter phase-space. On the other hand, in the DK limit, where finite $\bm k \cdot \bm \rho_a $ effects in the collision operator are neglected, i.e. $\gyaver{C_{ab}(\r,\vi)}_{\R} \simeq \gyaver{C_{ab}(\R,\vi)}_{\R}$, the local conservation laws, in \cref{eq:particleconservation,eq:momentumconservation,eq:energyconservation}, for the gyrocenter positions are satisfied.



In the following of the present section, we derive the Hermite-Laguerre expansion of two advanced linearized GK collision operators, that are the linearized GK Coulomb collision operator \citep{Jorge2019} and the linearized GK Sugama collision operator \citep{Sugama2009}. First, following the procedure outlined in \citet{Jorge2019} in \Cref{sec:linearizedCoulomb}, we derive the linearized GK Coulomb collision operator using a spherical harmonic moment expansion. This allows us to perform the gyro-average and the expansion of the Coulomb collision operator in terms of gyro-moments analytically at arbitrary order in the perpendicular wavenumber. We also obtain the DK limit by explicitly taking the zeroth gyroradius limit, i.e. $\bm k \cdot \bm \rho_a \to 0$. Similarly, we project the GK Sugama collision operator \citep{Sugama2009} onto the Hermite-Laguerre basis in \Cref{sec:GKSugama}, and derive its DK limit.

\subsection{Linearized Gyrokinetic Coulomb Collision Operator}
\label{sec:linearizedCoulomb}

Starting from the nonlinear Coulomb collision operator, the \textit{test} component of the linearized Coulomb collision operator around a local \textit{particle} Maxwellian distribution function, $f_{Mb}$, is given in the particle pitch-angle coordinates $(\r, v,\xi, \theta)$ (with $v$ the modulus of $\bm v$ and $\xi = v_\parallel/v$ the pitch angle) by \citep{Rosenbluth1957,Helander2002,Hazeltine2003}

\begin{align} \label{eq:CCT}
C^T_{ab}(f_a) & =   \frac{m_a \nu_{ab} \vTa}{n_b} \left[ \left( \frac{2  \partial_v G(f_{Mb})}{ v^2} + \left( 1 -  \frac{m_a}{m_b} \right) \partial_v H(f_{Mb}) \right)  \partial_v f_{a} \right. \nonumber \\
& \left.  - \frac{1}{v^3} \partial_v G(f_{Mb}) \L^2 f_a +  \partial_{v}^2 G(f_{Mb})  \partial^2_v f_{a}  + \frac{m_a}{m_b} 8 \pi f_{Mb}  f_a\right].
\end{align}
\\
In \cref{eq:CCT}, we introduce the Rosenbluth potentials $H(f) = 2 \int d \bm v' f(\bm v') /u$ and $G(f)  = \int d \bm v' f(\bm v')u$ (with $u = |\vi - \vi'|$), the collision frequency between species $a$ and $b$, $\nu_{ab}  = 4 \sqrt{\pi} N_b q_a^2 q_b^2 \ln \Lambda/(3 m_a^{1/2} T_a^{3/2})] $, and the spherical angular operator $\mathcal{L}^2 f  = \partial_{v}(v^2 \partial_v f) - v^2 \grad_{\bm v}^2 f$. Finally, we note that the thermal velocity is defined as $v_{Ta}^2 = 2 T_a / m_a$. When evaluated with a Maxwellian distribution function, analytical expressions of the Rosenbluth potentials can be obtained, and are given by $H(f_{Mb}) = 2 n_b \erf(s_b)/v$, and $G(f_{Mb}) = n_b \vTb \left[ \left( 1 + 2 s_b^2 \right)\erf(s_b)/s_b + \erf'(\sb)\right]/2$, with $s_b = v / \vTb$, the error function, $\erf(x) = 2 \int_0^x d s e^{-s^2}/\sqrt{\pi}$, and its derivative, $\erf'(x) = 2 e^{-x^2}/\sqrt{\pi}$. 

The \textit{field} component of the linearized Coulomb collision operator, expressed in the particle coordinates $(\r, v,\xi, \theta)$, is 

\begin{align} \label{eq:CCF}
C^F_{ab}(f_b) = \frac{2 \nu_{ab} \vTa f_{Ma}}{n_b} \left[ 2 s_a^2 \partial_v^2 G(f_b) - H(f_b) - \left( 1 -  \frac{m_a}{m_b} \right) v \partial_v H(f_b) + \frac{m_a}{m_b} 4 \pi \vTa^2 f_b \right].
\end{align}
\\
We refer to \cref{eq:CCT,eq:CCF}, respectively, as the \textit{exact} test and field components of the Coulomb collision operator between species $a$ and $b$ \citep{Rosenbluth1957}. We remark that, while the test component, \cref{eq:CCT}, contains velocity derivatives evaluated at constant $\r$, the field component, \cref{eq:CCF}, has additional velocity-space integrals contained in the Rosenbluth potential to evaluate holding $\r$ constant. 

Since the gyrocenter coordinate transformation mixes spatial and velocity coordinates, such as $\r = \R + \bm \rho_a(\R,\mu, \theta)$), and challenges the analytical and numerical treatments of the \textit{exact} field component, \textit{ad hoc} field components have been developed to implement collisional effects in GK codes \citep{Abel2008,Sugama2009,Sugama2019}. Following closely the method adopted in \citet{Jorge2019}, we overcome the difficulty associated with the evaluation of the exact Coulomb collision operator by using a spherical harmonic moment expansion of the particle distribution function. Ultimately, this allows us to evaluate the exact test and field components of the linearized gyro-averaged Coulomb collision operator, $\C_{ab} = \gyaver{C_{ab}}_{\R}$, at arbitrary order in the perpendicular wavenumber.
 
 \subsubsection{Spherical Harmonics Moment Expansion and Gyro-average}

In order to derive an expression of the linearized Coulomb collision operator ready for an expansion in the Hermite-Laguerre basis, following \citet{Ji2006} and \citet{Jorge2019}, the perturbed \textit{particle} distribution function of species $a$, $f_a = f_a(\x, \vi)$, is expanded in spherical harmonics basis as 

\begin{align} \label{eq:fmpj}
    f_a = \sum_{p=0}^\infty \sum_{j=0}^\infty \frac{f_{Ma}}{\sigma_j^p} \M_a^{pj} \cdot \Y^{p}(\bm s_a) L_j^{p+1/2}(s_a^2),
    \end{align}
\\
with $\sigma_j^p = p!(p + j +1/2)!/[2^p (p +1/2)! j!]$ and $\bm s_a = \bm v / \vTa$. \added{We remark that the quantities $\M_a^{pj}$ are defined below in \cref{eq:Mpjdef}.} In \cref{eq:fmpj}, the velocity space basis, $ \Y^{p}(\bm s_a) L_j^{p+1/2}(s_a^2)$, is the product between associated Laguerre polynomials, $ L_j^{p+1/2}(s_s^2)$, defined as \citep{gradshteyn}

\begin{equation} \label{eq:Laguerre}
L_j^{p+1/2}(x) = \sum_{l=0}^j L_{jl}^{p} x^l,
\end{equation}
\\
with 

\begin{align} \label{eq:Ljlp}
L_{jl}^{p}  = \frac{(-1)^l (p + j + 1/2)!}{(j - l)!(l +p + 1/2 )! l!}    ,
\end{align}
\\
and the $p$th-order traceless symmetric spherical harmonic tensors, $\Y^p(\bm s_a)$. The spherical harmonic tensors have the property that  $\gradv^2 \Y^p =0$, such that $\mathcal{L}^2 \Y^p = p(p+1)\Y^p$ and $\Y^p(\vi) = v^p \Y^p(\hat{\vi})$ ($\hat{\vi} = \vi / v$) \citep{Jorge2019}, and can be explicitly defined by introducing the spherical harmonic basis, $\bm e^{pm}$, that satisfies the orthogonality relation $\e^{pm} \cdot\e^{pm'} = (-1)^m \delta_{-m}^{m'}$ \citep{Snider2017}. The expression of the spherical harmonic tensors, $\Y^p(\bm s_a)$, on that basis is 

\begin{align} \label{eq:Yp}
 \Y^p (\bm s_a)= s_a^p  \sqrt{\frac{2 \pi^{3/2} p!}{2^p (p+1/2)!}}\sum_{m=-p}^{p }  Y_p^m(\xi,\theta) \e^{pm},
 \end{align}
 \\
where

  \begin{align}\label{eq:Ypm2Ppm}
 Y_p^m(\xi,\theta) = \sqrt{\frac{(2p+1)(p-|m|)!}{4 \pi (p+|m|)!}} P_p^m(\xi) e^{im\theta},
\end{align}
\\
are scalar harmonic functions, with  $P_p^m(\xi) = (-1)^m (1 - \xi^2)^{{m/2}} d^{m} P_p(\xi) / d \xi^m$ the associated Legendre polynomials, being with $P_p(\xi) = d^p [(\xi^2 -1)^p]  / d \xi^p / [2^p p!] $ the Legendre polynomial \citep{gradshteyn}. We remark that \cref{eq:Yp} is particularly useful to perform analytically the gyro-average of the Coulomb collision operator since the $\theta$ dependence of $\Y^{p}$ is isolated in $Y_p^m(\xi, \theta)$. The spherical harmonic basis, used in the expansion, given in \cref{eq:fmpj}, satisfies the orthogonality relation \citep{Snider2017}

\begin{equation} \label{eq:orthYpLj}
\int d \bm s e^{- s^2}  L_j^{p+1/2}(s^2) \Y^{p'}(\bm s) L_{j'}^{p'+1/2}(s^2) \Y^{p}(\bm s)  \cdot \mathsfbi{T}^{p} = \delta_{pp'} \delta_{jj'} \pi^{3/2} \sigma^p_j  \mathsfbi{T}^{p},
\end{equation}
\\
with $\mathsfbi{T}^p$ an arbitrary $p$-th order tensor. We note that the dot product, appearing in \cref{eq:fmpj,eq:orthYpLj}, between two $p$-th order tensors, $\mathsfbi{T}^p$ and $\R^p$, i.e. $\mathsfbi{T}^p \cdot \R^p $, yields a scalar.

Projecting the spherical harmonic basis and using \cref{eq:fmpj}, one deduces that the expansion coefficient $\M_a^{pj} =\M_a^{pj}(\r) $, defined as the spherical harmonic \textit{particle} moments of species $a$, are given by

\begin{align} \label{eq:Mpjdef}
    \M_a^{pj} = \frac{1}{n_a }\int d \bm v f_a  \Y^{p}(\bm s_a) L_j^{p+1/2}(s_a^2),
\end{align}
\\
with $n_a $ is the particle density background of species $a$, with $n_a = N_a$.

Using the spherical harmonics particle moment expansion of $f_b$, one can obtain the spherical particle moment expansion of the Rosenbluth potentials, $H(f_b)$ and $G(f_b)$ \citep{Jorge2019},

\begin{subequations}\label{eq:RosenbluthPots}
\begin{align} 
H(f_b) & = \frac{4 n_b}{\vTb} \sum_{p=0}^\infty \sum_{j=0}^\infty\sum_{m=0}^j \frac{L_{jm}^p }{ \sigma_p^j} \frac{ \Y^{p}(\hat{\bm s}_b) \cdot \M_b^{pj}}{2p+1} \left[ \frac{I_{-}^{2p + 2m +2}}{\sb^{p+1}} + \sb^p I_{+}^{2m +1} \right], \\
G(f_b) & = 2 n_b \vTb \sum_{p=0}^\infty \sum_{j=0}^\infty\sum_{l=0}^j  \frac{L_{jl}^p}{\sigma_j^p}\frac{ \M_b^{pj} \cdot \Y^p(\hat{\bm \sb}) }{(2p+1)} \nonumber \\
& \times  \left[ \frac{1}{(2p +3)} \left(  \frac{I_-^{2p+2l +4}}{\sb^{p+1}} + \sb^{p+2} I_+^{2l+1}\right)      -\frac{1}{(2p-1)} \left(  \frac{I_-^{2p+2l +2}}{\sb^{p-1}} + \sb^{p} I_+^{2l+3}\right)  \right], 
\end{align}
\end{subequations}
\\
with $\hat{\bm s_a} = \bm s_a / s_a$ and the upper and lower incomplete gamma functions defined by $I_-^k = \int_0^x d s s^{(k-1)/2} e^{-s} / \sqrt{\pi}$ and $ I_+^k = \int_x^\infty d s s^{(k-1)/2} e^{-s} / \sqrt{\pi}$, respectively. In deriving \cref{eq:RosenbluthPots}, the associated Laguerre polynomials, $L_j^{p+1/2}(s_b^2)$, are expanded using \cref{eq:Laguerre}. 

The velocity derivatives of the Rosenbluth potentials appearing in the \textit{field} component of the Coulomb collision operator, \cref{eq:CCF}, can be analytically evaluated using \cref{eq:RosenbluthPots}, and are given by

\begin{subequations} \label{eq:derivationRosenbluth}
\begin{align}
\frac{\partial}{\partial s_b} H_b(f_b) &=  \frac{4 n_b}{\vTb} \sum_{p=0}^\infty \sum_{j=0}^\infty \sum_{m=0}^j \frac{L_{jm}^p }{ \sigma_p^j} \frac{ \Y^{p}(\hat{\bm s}_b) \cdot \M_b^{pj}}{2p+1} \left[ -(p+1)\frac{I_{-}^{2p + 2m +2}}{\sb^{p+2}} + p \sb^{p-1} I_{+}^{2m +1} \right],  \\
\frac{\partial^2}{\partial \sb^2} G_b(f_b ) &  = 2 n_b \vTb \sum_{p=0}^\infty \sum_{j=0}^\infty \sum_{l=0}^j  \frac{L_{jl}^p}{\sigma_j^p}\frac{ \M_b^{pj} \cdot \Y^p(\hat{\bm \sb}) }{(2p+1)}  \left[ \frac{(p+1)(p+2)}{(2p +3)} \left(  \frac{I_-^{2p+2l +4}}{\sb^{p+3}} + \sb^{p} I_+^{2l+1}\right)    \right.   \nonumber \\
 & \left.  -\frac{p(p-1)}{(2p-1)} \left(  \frac{I_-^{2p+2l +2}}{\sb^{p+1}} + \sb^{p-2} I_+^{2l+3}\right) \right],
\end{align}
\end{subequations}
\\
respectively. The spherical harmonic expansion of $f_a$, \cref{eq:fmpj}, can also be inserted into the \textit{test} component of the Coulomb operator, given in \cref{eq:CCT}. This finally yields the spherical harmonic moment expansion of the test and field components of the linearized Coulomb collision operator $C_{ab}$,

\begin{subequations}
\begin{align} \label{eq:cmomentexpansion}
C^T_{ab} & = \sum_{p=0}^\infty \sum_{j=0}^\infty \frac{f_{Ma}}{\sigma_j^p}  \M_a^{pj}\cdot \Y^p(\hat{\bm v })\nu_{ab}^{Tpj}, \\
C^F_{ab}  & = \sum_{p=0}^\infty \sum_{j=0}^\infty \frac{f_{Ma}}{\sigma_j^p} \M_b^{pj}\cdot \Y^p(\hat{\vi})\nu_{ab}^{Fpj}, \label{eq:CabFmoment}
\end{align}
\end{subequations}
\\
where the \textit{test} and \textit{field} speed functions, $ \nu_{ab}^{Tpj}$ and $\nu_{ab}^{Fpj}$, are defined in \cref{appendix:CoulombSpeedfunctions}. We remark that the $v$ dependence of $C_{ab}^T$ and $C_{ab}^F$ is entirely contained in the speed functions, while the angular $(\theta,\xi)$ dependence is isolated in the spherical harmonic tensor $\Y^{p}(\hat{\bm v})$.

We now focus on evaluating the gyro-average of the test component of the linearized Coulomb operator in \cref{eq:cmomentexpansion}. Since the gyro-average of $C_{ab}$, which appears in the linearized GK equation \cref{eq:LinGK}, is performed at constant $\R$, the spatial dependence of $\M^{pj}_a(\r)$ is first expressed as a function of $\R$. Since $\r = \R + \bm \rho_a$, the gyro-average can be carried out in Fourier space by observing that $\M_a^{pj}(\r) = \int  d \bm k \M_a^{pj} e^{i \bm k \cdot \bm \rho_a } e^{i \bm k \cdot \R} $ and by using the Jacobi-Anger identity to express the phase factor difference $e^{ i \bm k \cdot \bm \rho_a } = \sum_{n} i^n J_n(b_a\sqrt{x_a}) e^{in \theta}$ \citep{gradshteyn}. Thus, in Fourier space, the linearized test component of the Coulomb collision operator at $\R$ can be written as

\begin{align} \label{eq:CollFourier}
C_{ab}^T =  \sum_{p=0}^\infty  \sum_{p=j}^\infty \sum_{n=- \infty}^\infty  \frac{F_{Ma}}{\sigma_j^p}  \M_a^{pj} \cdot \Y^p(\hat{\bm v }) i^n J_n(b_a \sqrt{x_a}) e^{i n \theta}\nu_{ab}^{Tpj}.
\end{align}
\\
By expanding the spherical tensors, $\Y^{p}$ appearing in \cref{eq:CollFourier}, as shown in \cref{eq:Yp}, one obtains 
  
  \begin{align}\label{eq:gyaverCL}
\gyaver{C_{ab}^T}_{\R}  &= \sum_{p=0}^\infty \sum_{j=0}^\infty \sum_{m=-p}^p\sum_{n = -\infty}^{\infty} \frac{1}{\sigma^p_j} \sqrt{\frac{2 \pi^{3/2} p!}{2^p (p+1/2) !}}  F_{Ma}  \nu_{ab}^{Tpj} i^n  J_n(b_a\sqrt{x_a}) \nonumber  \\
& \times \gyaver{e^{i n \theta} Y_p^m(\xi,\theta)}_{\R} \M_a^{pj} \cdot \bm e^{pm}.
\end{align}
\\
 The scalar harmonic functions $Y_p^m(\xi,\theta)$ can then be expanded in terms of associated Legendre polynomials $P_p^{m}(\xi)$, as indicated by \cref{eq:Ypm2Ppm}. Thus, \cref{eq:gyaverCL} can be written as 

\begin{align} \label{eq:gyaverCBesselJm}
\gyaver{C_{ab}^T}_{\R}  &= \sum_{p=0}^\infty \sum_{j=0}^\infty \sum_{m=-p}^p  \frac{1}{\sigma^p_j} F_{Ma}  \nu_{ab}^{Tpj} i^m  J_m(\sqrt{x_a} b_a)  \sqrt{\frac{\pi^{1/2} p!}{2^p (p-1/2) !}} \nonumber \\ & \times\sqrt{\frac{(p-|m|)!}{ (p+|m|)!}}  P_p^{m}(\xi) \M_a^{pj} \cdot \bm e^{pm}.
\end{align}
\\
We remark that the same gyro-average procedure can be applied to the field component. In fact, the gyro-average of the field component is given by \cref{eq:gyaverCBesselJm}, having replaced $\nu_{ab}^{Tpj}(v)$ by $\nu_{ab}^{Fpj}(v)$ and $\M^{pj}_a$ by $\M^{pj}_b$, i.e.

\begin{align} \label{eq:gyaverCBesselJmField}
\gyaver{C_{ab}^F}_{\R}  &= \sum_{p=0}^\infty \sum_{j=0}^\infty \sum_{m=-p}^p  \frac{1}{\sigma^p_j} F_{Ma}  \nu_{ab}^{Fpj} i^m  J_m(\sqrt{x_a} b_a)  \sqrt{\frac{\pi^{1/2} p!}{2^p (p-1/2) !}} \nonumber \\ & \times\sqrt{\frac{(p-|m|)!}{ (p+|m|)!}}  P_p^{m}(\xi) \M_b^{pj} \cdot \bm e^{pm}.
\end{align}

\subsubsection{Gyrokinetic Hermite-Laguerre Expansion}

\Cref{eq:gyaverCBesselJm} and \cref{eq:gyaverCBesselJmField} are in a suitable form for the projection onto the Hermite-Laguerre basis, $\C_{ab}^{lk} = \C_{ab}^{lk}(\bm k,t)$, defined \cref{eq:Cpj}. We focus first on the expression for the test component in \cref{eq:gyaverCBesselJm}. This yields 

 \begin{align} \label{eq:CabTlkinterstep}
 \C_{ab}^{Tlk} & =  \C_{ab}^{Tlk}(\bm k,t) =  \sum_{p=0}^\infty \sum_{j=0}^\infty \sum_{m=-p}^p  \frac{1}{\sigma^p_j} i^m    \sqrt{\frac{\pi^{1/2} p!}{2^p (p-1/2) !}} \sqrt{\frac{(p-|m|)!}{ (p+|m|)!}}  \M_a^{pj} \cdot \bm e^{pm} \frac{ I_{abm}^{Tlkpj}}{\sqrt{2^l l!}},
 \end{align}
 \\
 where we introduce,
 
 \begin{align} \label{eq:IabmTlkpj}
  I_{abm}^{Tlkpj} = \int d \vi F_{Ma}  \nu_{ab}^{Tpj} J_m(\sqrt{x_a} b_a)  P_p^{m}(\xi) H_l(s_{\parallel a}) L_k(x_{ a}),
 \end{align}
 \\
 A similar expression can be obtained for the field component, $\C_{ab}^{Flk}$, having replaced $ \nu_{ab}^{Tpj}$ by  $\nu_{ab}^{Fpj}$ in \cref{eq:IabmTlkpj}, and $\M_a^{pj}$ by $\M_b^{pj}$ in \cref{eq:CabTlkinterstep}. In order to express the Hermite-Laguerre projection of the linearized GK Coulomb collision operator, given in \cref{eq:CabTlkinterstep}, in a closed analytical form in terms of gyro-moment, the spherical harmonic \textit{particle} moments, $\M_a^{pj}$, defined in \cref{eq:Mpjdef}, must be written as a function of the gyro-moments $N_a^{pj}$, and the velocity integral contained in \cref{eq:IabmTlkpj} must be evaluated.

 As a first step, we relate the spherical harmonics \textit{particle} moments, $\M_a^{pj}$ in \cref{eq:Mpjdef}, to the gyro-moments $N_a^{pj}$ defined in \cref{eq:Npjdef}. To proceed, we transform the velocity space integral in \cref{eq:Mpjdef} to a phase-space integral, i.e.
 
 \begin{align} \label{eq:Mapjphasespace}
 \M_a^{pj}(\r) = \frac{1}{n_a }\int d \bm v \int d \r' \delta (\r - \r') f_a (\r', \bm v) \Y^{p}(\bm s_a) L_j^{p+1/2}(s_a^2).
 \end{align}
 \\
The coordinate transformation $\r' = \r' (\R, \mu, \theta) = \R + \bm \rho_a(\mu, \theta)$ in \cref{eq:Mapjphasespace} allows us to write
 
  \begin{align} \label{eq:Mapjphasespace}
 \M_a^{pj}(\r) = \frac{1}{n_a } \int d \theta d \mu d v_\parallel d \bm R \frac{B}{m_a} \delta (\r - \R - \bm \rho_a ) f_a (\r', \bm v) \Y^{p}(\bm s_a) L_j^{p+1/2}(s_a^2),
 \end{align}
 \\
 where $f_a (\r', \bm v) = f_a(\r'(\R, \mu, \theta) , \bm  v(\mu, v_\parallel, \theta))$. We now aim to express the perturbed \textit{particle} distribution function, $f_a(\r',\bm  v)$, in terms of the perturbed \textit{gyrocenter} distribution function, $\g_a(\R,\mu, v_\parallel)$. The scalar invariance of the total particle and gyrocenter distribution functions yields
 
 \begin{align} \label{eq:fafctofga}
  f_a(\r', \bm v)  =\g_a(\R, \mu, v_\parallel) +  F_{Ma}(\R, \mu, v_\parallel) - f_{Ma}(\r', \bm v) .
 \end{align}
 \\
 Since the gyrocenter coordinates $(\R, \mu, v_\parallel, \theta)$ are obtained through a perturbative coordinate transformation from the particle phase-space $(\r, \bm v)$, within a perturbation approach known as Lie-transform perturbation method \citep{Cary1981,Brizard2009, Frei2020}), the functional form of any function $F$, defined on $(\R, \mu, v_\parallel, \theta)$, can be expressed in terms of the functional form of a function $f$, defined on $(\r, \bm v)$, by $f = \T F$, where the operator $\T$ is known as the pull-back operator \citep{Frei2020}. Taking the case of $f = f_{Ma}$ and $F = F_{Ma}$ such that $f_{Ma} = \T F_{Ma}$ in \cref{eq:fafctofga}, and at the leading order in $\epsilon \sim e \phi / T_e \ll 1$, one derives that the perturbed \textit{particle} and \textit{gyrocenter} distribution functions are related by 

\begin{align} \label{eq:fasofg}
    f_a(\r',\bm v) & = \g_a(\R,\mu,v_\parallel)   - \frac{q_a}{T_a} \left( \phi(\r') -\gyaver{\phi}_{\R}(\R)\right)  F_{Ma} .
\end{align}
\\
Since $\phi - \gyaver{\phi}_{\R} \sim (k_\perp \rho_s)^2  e \phi / T_e $, the last term in \cref{eq:fasofg} can be omitted in DK collision operators but, in general, it cannot be neglected in GK collisional theories, that consider $k_\perp \rho_s \sim 1$. The difference between the particle and gyrocenter Maxwellian distribution functions \citep{Madsen2013} leads to polarization effects in the collision operator \citep{Xu1991,Dimits1994}. On the other hand, we remark that polarization effects are neglected in the nonlinear full-F GK Coulomb collision operator developed in \citet{Jorge2019}, where it is assumed that $\epsilon_\nu  e \phi / T_e \ll \epsilon^2$ (with $\epsilon_\nu \sim \nu_{ii} / \Omega_i \sim \epsilon^2$), an ordering that we do not consider here (the same ordering is considered in the nonlinear gyrokinetic Dougherty collision operator derived in \citet{Frei2020}). \added{ We also note that polarization effects are also neglected in the linearized GK Coulomb operator developed in \citet{Li2011}}. Using  \cref{eq:fasofg} in the definition of the spherical harmonics particle moments given in \cref{eq:Mpjdef}, we can write

\begin{align} \label{eq:momentpolarized}
   \M_a^{pj}(\r) & = \m_a^{pj}(\r) + \sum_{i=1}^{2}\m^{pj}_{a \phi i}(\r),
\end{align}
\\
with

\begin{subequations}
\begin{align}
     \m_a^{pj}(\bm r) &=\frac{1}{N_a} \int  d \theta d \mu d v_\parallel   \frac{B}{m_a} \gyaver{\g_a \Y^p(\bm s_a) }_{\r}^\dagger  L_j^{p+1/2}(s_a^2), \label{eq:mpj} \\
      \m^{pj}_{ a \phi 1}(\r) & = - \frac{q_a}{N_a T_a}\int  d \theta d \mu d v_\parallel  \frac{B}{m_a} \gyaver{ \phi(\bm R + \bm \rho_a) \Y^p(\bm s_a) }^\dagger_{\bm r}  L_j^{p+1/2}(s_a^2)   F_{Ma},  \label{eq:mpjphi1} \\
       \m^{pj}_{ a \phi 2}(\r) & = \frac{q_a}{N_a T_a}\int  d \theta d \mu d v_\parallel  \frac{B}{m_a} \gyaver{ \gyaver{\phi}_{\R}(\R) \Y^p(\bm s_a) }^\dagger_{\bm r}  L_j^{p+1/2}(s_a^2)   F_{Ma},\label{eq:mpjphi2}
\end{align}
\end{subequations}
\\
where we introduce the adjoint gyro-average operator at the particle position $\r$, 

\begin{align} \label{eq:gyaverx}
\gyaver{\xi}_{\r}^\dagger  =  \int_0^{2 \pi} \frac{d \theta}{2 \pi} \int d \R \delta(\R + \bm \rho_a - \r)   \xi(\R, v_\parallel,\mu,\theta),
\end{align}
\\
defined over any gyrocenter phase-space function $\xi$. We remark that the operator in \cref{eq:gyaverx} is the adjoint of the gyro-average operator at constant $\R$, such that $\int d \bm R \gyaver{\chi}_{\R} \xi  = \int d \bm x \gyaver{\xi}^\dagger_{\r}\chi  $, and plays a central role in polarization and magnetization effects in the field equations of GK theories \citep{Frei2020}.

We now focus on evaluating the velocity integrals in $\m_a^{pj}$ defined \cref{eq:mpj}. As a first step, we compute $\gyaver{\g_a  \Y^p(\bm s_a) }_{\r}^\dagger$ in Fourier-space, such that

\begin{align} \label{eq:mapjFourier}
\m_a^{pj}(\bm r)  & = \frac{1}{N_a} \int d \bm k \int  d \theta  d \mu d v_\parallel  \frac{B}{m_a}   \g_a(\bm k, \mu, v_\parallel)    \int_0^{2 \pi} \frac{d \theta'}{2 \pi} e^{- \bm k \cdot \bm \rho_a'}\Y^p(\bm s_a')   L_j^{p+1/2}(s_a^2) e^{i \bm k \cdot \r} \nonumber \\
 & = \frac{1}{N_a} \sum_{n = - \infty}^\infty (-i)^n \int d \bm k \int  d \theta  d \mu d v_\parallel    \frac{B}{m_a}  \g_a(\bm k, \mu, v_\parallel) J_n(b_a \sqrt{x_a} )  \nonumber \\
&  \times   \int_0^{2 \pi} \frac{d \theta'}{2 \pi} e^{ i n \theta'}\Y^p(\bm s_a')   L_j^{p+1/2}(s_a^2) e^{i \bm k \cdot \r},
\end{align}
\\
where we use the Jacobi-Anger identity and where the $\bm \rho_a' $ and $\bm s_a'$ notation indicate that these quantities have to be evaluated at $\theta'$. The velocity integrals in \cref{eq:mapjFourier} can be performed by expanding the Bessel functions $J_n$ into associated Laguerre polynomials, thanks to \cref{eq:J02Laguerre}, and by using the definition of the gyro-moment, given in \cref{eq:Npjdef}. A basis transformation between associated Legendre and Laguerre polyomials to Hermite and Laguerre polynomials is necessary to express $\m_a^{pj}$ as a function of the gyro-moments $N_a^{pj}$. This basis transformation (and its inverse) can be expressed as \citep{Jorge2019},

\begin{subequations} \label{eq:basistransformation}
 \begin{align} \label{eq:ALL2HL}
 s_s^p P_p^{m}(\xi) L_j^{p+1/2}(s_s^2)&  = \sum_{l=0}^{p + m + 2j} \sum_{k=0}^{j + \lfloor (p +m)/2 \rfloor} T_{pjm}^{lk} H_{l}(s_{\parallel s}) L_k(x_s) x_s^{m/2},\\
 \label{eq:HL2ALL}
H_{l}(s_{\parallel s}) L_k(x_s) x_s^{m/2} & = \sum_{p=0}^{l + m +2k} \sum_{j=0}^{k + \lfloor (l+m)/2 \rfloor} (T^{-1})^{pjm}_{lk} s_s^p P_p^{m}(\xi) L_j^{p+1/2}(s_s^2),
 \end{align}
 \end{subequations}
 \\
 where the closed analytical expression of the coefficients $T_{pjm}^{lk}$ can be found in  \citet{Jorge2019}, and the coefficients $(T^{-1})^{pjm}_{lk}$ of the inverse basis transformation, \cref{eq:HL2ALL}, are given by 

 \begin{align} \label{eq:Tmclosed}
        (T^{-1})^{pjm}_{lk}
&         = \frac{\sqrt{\pi} 2^l l ! j!(p+1/2)}{(j+p+ 1/2)!} \frac{(p-m)!}{(p+m)!}  \times \sum_{k'=0}^{\min(j + \lfloor (p+m)/2 \rfloor,m +k)}   T_{pjm}^{lk'} d^{m}_{0kk'}    [ l \le p+ m +2j].
\end{align}
\\
with $[\cdot]$ the Iverson bracket, defined by $[A] = 1$ if $A$ is true, and $0$ otherwise. The basis transformation given by \cref{eq:ALL2HL} yields the gyro-moment expansion 

 \begin{align} 
 \m_a^{pj} =\m_a^{pj}(\bm k) =   \sqrt{\frac{ \pi^{1/2} p!}{2^p (p-1/2)!}}\sum_{m=-p}^{p }  (-i)^m  \sqrt{\frac{(p-|m|)!}{ (p+|m|)!}} \mathcal{M}_a^{pjm} \e^{pm},
 \end{align}
 \\
 where we introduce,
 
  \begin{align} \label{eq:curlyMspjm}
 \mathcal{M}_a^{pjm} = \sum_{g=0}^{p+|m|+2j} \sum_{h=0}^{j + \lfloor (p+|m|)/2\rfloor}\sum_{n=0}^\infty \sum_{s_1=0}^{n+|m|+h} d_{nhs_1}^{|m|} T_{pj|m|}^{gh}   \frac{ n!\sqrt{2^g g!}}{(n+|m|)!}  \left( \frac{b_a}{2}\right)^{ |m|} \kernel{n} (b_a)N_a^{gs_1}.
 \end{align}
 \\
 In \cref{eq:curlyMspjm}, the numerical coefficients $d_{nks_1}^{|m|}$, which arise from the product between Laguerre and associated Laguerre polynomials,
 
 \begin{align}\label{eq:LnmLkxm2Ls}
 L_n^{m}(x) L_k(x) x^{m} =\sum_{s=0}^{n+m +k} d^{m}_{nks} L_s(x),
 \end{align}
 \\
 are given by the closed formula \citep{Jorge2019},
 
  \begin{align} \label{eq:dmnks}
 d_{nks}^{m} =\sum_{n_1=0}^{n} \sum_{k_1=0}^k \sum_{s_1=0}^s L_{kk_1}^{-1/2} L_{nn_1}^{m-1/2} L_{ss_1}^{-1/2} (n_1 + k_1 + s_1 + m)!,
 \end{align}
 \\
with the coefficients $L_{jl}^{p}$ defined in \cref{eq:Ljlp}.

We now turn to the polarization term in \cref{eq:momentpolarized}, i.e. $\m_{a \phi 1}^{pj}$  and $\m_{a \phi 2}^{pj}$ defined in \cref{eq:mpjphi1,eq:mpjphi2}, respectively. Following the same steps considered in the evaluation of $\m_s^{pj}$, we first compute the terms in $\gyaver{  \phi(\bm R + \bm \rho_a) \Y^p(\bm s_a) }_{\r}^\dagger$ and $\gyaver{ \gyaver{\phi}_{\R}(\R) \Y^p(\bm s_a) }_{\r}^\dagger$. Using the coordinate transformation $\r = \bm R + \bm \rho_a$, we obtain 
 
 \begin{align} \label{eq:adjointoper1}
 \gyaver{  \phi(\bm R + \bm \rho_a) \Y^p(\bm s_a) }_{\r}^\dagger = \phi(\r)  \gyaver{ \Y^p(\bm s_a) }_{\r}^\dagger,
 \end{align}
 \\
 while using the Jacobi-Anger identity 
  
 \begin{align}\label{eq:adjointoper2}
 \gyaver{  \gyaver{\phi}_{\R}(\R) \Y^p(\bm s_a) }_{\r}^\dagger = \int d \bm k \sum_{n = - \infty}^\infty  (-i)^n  \phi(\bm k) J_0(b_a \sqrt{x_a})  J_n( b_s \sqrt{x_s}) \gyaver{  e^{in \theta} \Y^p(\bm s_a) }_{\r}^\dagger e^{i \bm k \cdot \r}.
 \end{align}
 \\
\Cref{eq:adjointoper1} and \cref{eq:adjointoper2} allow us to perform analytically the velocity integrals contained in \cref{eq:mpjphi1,eq:mpjphi2}. We first use the orthogonality relation, given in \cref{eq:orthYpLj}, and expand $\Y^p$ in associated Legendre polynomials with \cref{eq:Yp}, while expressing the Bessel functions in associated Laguerre polynomials according to \cref{eq:J02Laguerre}. Finally, we use the basis transformation given in \cref{eq:ALL2HL}. This yields

\begin{subequations}
\label{eq:polarizaionterms}
\begin{align}  
\m_{a \phi 1}^{pj} &  =  - \frac{q_a \phi}{T_a} \delta_{p}^0 \delta_j^0 \e^{p0}, \label{eq:mphipj1}\\
    \m_{a \phi 2}^{pj}    & = \frac{q_a \phi}{T_a}   \sum_{m'=-p}^{p }(-i)^{m'} \sqrt{\frac{ \pi^{1/2} p!}{2^p (p-1/2)!}} \sqrt{\frac{(p-|m'|)!}{  (p+|m'|)!}} \mathcal{\Pi}^{pjm'}_a
    \e^{pm'}\label{eq:mphipj2},
\end{align}
\end{subequations}
\\
where we introduce the coefficients

\begin{align}
    \mathcal{\Pi}^{pjm'}_a = \sum_{h=0}^{ j + \lfloor(p+|m'| )/2 \rfloor} \sum_{s=0}^\infty  \sum_{s'=0}^{s+|m'| +h} T^{0h}_{pj|m'|}\left(\frac{b_a}{2}\right)^{|m'|}\frac{s! \kernel{s}(b_a)\kernel{s'}(b_a)}{(s +|m'|)!} d^{|m'|}_{shs'}.
\end{align}
 \\
The electrostatic potential  $\phi$, appearing in \cref{eq:polarizaionterms}, can be self-consistently expressed as a function of $N_a^{pj}$ using the gyrokinetic quasineutrality condition, \cref{eq:GKPoissonNpj}.

As a last step, we compute the velocity integral in $ I_{abm}^{Tlkpj}$ defined in \cref{eq:IabmTlkpj}. To proceed, we first expand the Bessel functions $J_m$ in terms of associated Laguerre polynomials, according to \cref{eq:J02Laguerre}, and use the inverse basis transformation given in \cref{eq:HL2ALL}, to obtain
 
  \begin{align} \label{eq:IabmTlkpjclosed}
  I_{abm}^{Tlkpj} =  \sum_{n=0}^\infty  \sum_{f =0}^{n+k}  \sum_{g=0}^{l + m +2f} \sum_{h=0}^{f + \lfloor (l+m)/2 \rfloor} (T^{-1})^{ghm}_{lf}  \overline{d}^{m}_{nkf} \\
  \times \int d \vi F_{Ma}   \sa^g L_h^{g+1/2}(\sa^2) \nu_{ab}^{Tpj}  P_p^{m}(\xi)  P_g^{m}(\xi) ,
 \end{align}
 \\
 where the product of associated Laguerre and Laguerre polynomials is expressed as a single series of Laguerre polynomial thanks to \citep{Jorge2019}, 

\begin{equation} \label{eq:LrnLj2Lf}
L_n^{m}(x_a)  L_k(x_a)  = \sum_{f =0}^{n+k} \overline{d}^{m}_{nkf} L_f(x_a),
\end{equation}
\\
where 

\begin{equation} \label{eq:bardmnkf}
\overline{d}^{m}_{nkf} = \sum_{n_1 = 0}^n \sum_{k_1 =0}^k   \sum_{f_1=0}^f L_{kk
_1}^{-1/2}  L_{nn_1}^{m-1/2}  L_{f f_1}^{-1/2} (n_1 + k_1 + f_1)!.
\end{equation}
\\
Finally, using the orthogonality relations \citep{gradshteyn}

\begin{align}
\int_{-1}^{1} d \xi P_p^{m}(\xi)  P_g^{m}(\xi) = \frac{2 \delta_g^p}{(2p +1)} \frac{(p + m)!}{(p - m)!},
\end{align}
\\
and $\e^{pm} \cdot\e^{pm'} = (-1)^m \delta_{-m}^{m'}$ \citep{Snider2017}, the Hermite-Laguerre expansion of the linearized GK Coulomb collision operator, $\C_{ab}^{lk} = \C_{ab}^{lk}(\bm k,t)$, is obtained
 
 \begin{align} \label{eq:GKClk}
     \C_{ab}^{lk} = \C^{Tlk}_{ab0} +\C^{Flk}_{ab0} +   \sum_{i=1}^{2} \left( \C^{Tlk}_{ab\phi i}  +  \C^{Flk}_{ab\phi i} \right)
 \end{align}
 \\
with the Hermite-Laguerre expansion of the \textit{test} and \textit{field} components, $\C^{Tlk}_{ab0}$ and $\C^{Flk}_{ab0}$, given by
 
 \begin{align} \label{eq:CabTlkCoulomb}
    \C_{ab0}^{Tlk} & =\sum_{p=0}^\infty \sum_{j=0}^\infty \sum_{n=0}^\infty   \sum_{f=0}^{k+n} \sum_{m=0}^{p}  \sum_{t=0}^{f + \lfloor (l +m)/2\rfloor} \  \frac{ a_m }{\sigma^p_j}  \frac{2^{p}  (p!)^2 }{(2 p)! }    \frac{  \overline{d}_{nkf}^{m}}{(2p+1)}\nonumber \\
    &  \times
 \frac{ \left( T ^{-1} \right)^{ptm}_{lf} }{\sqrt{2^l l!}} \frac{ n! \kernel{n}(b_a)   }{(n + m)! } \left( \frac{b_a}{2}\right)^{ m}    \mathcal{M}_a^{pjm}  \overline{\nu}_{ab}^{Tpjt} \left[  p \leq l + m + 2f\right] ,\\
     \C_{ab0}^{Flk} & =\sum_{p=0}^\infty \sum_{j=0}^\infty \sum_{n=0}^\infty   \sum_{f=0}^{k+n} \sum_{m=0}^{p} \sum_{t=0}^{f + \lfloor (l +m)/2\rfloor}  \frac{ a_m }{\sigma^p_j}  \frac{2^{p}  (p!)^2 }{(2 p)! }    \frac{  \overline{d}_{nkf}^{m}}{(2p+1)}\nonumber \\
    &  \times 
 \frac{ \left( T ^{-1} \right)^{ptm}_{lf} }{\sqrt{2^l l!}} \frac{ n! \kernel{n}(b_a)   }{(n + m)! } \left( \frac{b_a}{2}\right)^{ m}    \mathcal{M}_b^{pjm}  \overline{\nu}_{ab}^{Fpjt}  \left[  p \leq l + m + 2f\right], \label{eq:CabFlkCoulomb}
\end{align}
\\
respectively, and the associated polarization contributions,

\begin{subequations} \label{eq:Cpjphi}
\begin{align} 
\mathcal{C}^{Tlk}_{ab \phi 1 } & =- \frac{q_a}{T_a} \phi \sum_{n=0}^\infty \sum_{s=0}^{n+k} \sum_{h=0}^{s +\lfloor l/2 \rfloor}d_{nks}^0 \kernel{n}(b_a) \frac{\left( T^{-1} \right)^{0h0}_{ls}}{\sqrt{2^l l!}} \overline{\nu}_{ab}^{T00h}, \\
\mathcal{C}^{Flk}_{ab \phi 1 } & =- \frac{q_b}{T_b} \phi \sum_{n=0}^\infty \sum_{s=0}^{n+k} \sum_{h=0}^{s +\lfloor l/2 \rfloor}d_{nks}^0 \kernel{n}(b_a) \frac{\left( T^{-1} \right)^{0h0}_{ls}}{\sqrt{2^l l!}} \overline{\nu}_{ab}^{F00h}, \\
\mathcal{C}_{ab \phi 2 }^{T lk} & =  \frac{q_a}{T_a }  \phi  \sum_{p=0}^\infty  \sum_{j=0}^\infty \sum_{m=0}^{p}  \sum_{n=0}^\infty\sum_{f =0}^{n+k}  \sum_{t=0}^{f + \lfloor (l+m)/2 \rfloor}  \frac{a_m}{\sigma_p^j} \frac{2^{p}  (p!)^2 }{(2 p)! }    \frac{ 1}{( 2p+1)} \frac{\mathcal{\Pi}_a^{pjm}}{\sqrt{2^l l!}} \nonumber \\
& \times  \overline{d}^{m}_{nkf} \frac{n!\kernel{n}(b_a)}{(n+m)!}\left( \frac{b_a}{2}\right)^{m}  (T^{-1})^{ptm}_{lf}  \overline{\nu}_{ab}^{Tpjt}  \left[  p \leq l + m + 2f\right], \\
\mathcal{C}_{ab \phi 2}^{Flk} & =  \frac{q_b}{T_b }  \phi  \sum_{p=0}^\infty  \sum_{j=0}^\infty \sum_{m=0}^{p}  \sum_{n=0}^\infty\sum_{f =0}^{n+k} \sum_{t=0}^{f + \lfloor (l+m)/2 \rfloor} \frac{a_m}{\sigma_p^j} \frac{2^{p}  (p!)^2 }{(2 p)! }    \frac{ 1}{( 2p+1)} \frac{\mathcal{\Pi}_b^{pjm}}{\sqrt{2^l l!}} \nonumber \\
& \times  \overline{d}^{m}_{nkf} \frac{n!\kernel{n}(b_a)}{(n+m)!}\left( \frac{b_a}{2}\right)^{m}   (T^{-1})^{ptm}_{lf}  \overline{\nu}_{ab}^{Fpjt}   \left[  p \leq l + m + 2f\right] .
\end{align}
\end{subequations}
\\
where $a_m = 2$ for $m \geq 1$ while $a_0 =1$. In \cref{eq:CabTlkCoulomb,eq:CabFlkCoulomb,eq:Cpjphi}, we introduce the velocity-integrated speed functions 

\begin{align}
\overline{\nu}_{ab}^{Tpjt} & = \sum_{d =0}^t L_{td}^p\overline{\nu}_{*ab}^{Tpjd}, \\
\overline{\nu}_{ab}^{Fpjt} & = \sum_{d =0}^t L_{td}^p\overline{\nu}_{*ab}^{Fpjd},
\end{align}
\\
where $ \overline{\nu}_{*ab}^{Tpjd} = \int d \vi s_a^{p+2d} \nu_{ab}^{Tpj} F_{Ma}$ and $\overline{\nu}_{*ab}^{Fpjd} = \int d \vi s_a^{p+2d} \nu_{ab}^{Fpj} F_{Ma}$. Closed analytical expressions for $\overline{\nu}_{ab*}^{Tpjd}$ and $\overline{\nu}_{ab*}^{Fpjd}$ are reported in \cref{appendix:CoulombSpeedfunctions}. \added{We remark that the Hermite-Laguerre expansions of the test and field components, \cref{eq:CabTlkCoulomb,eq:CabFlkCoulomb} respectively, can be further written in terms of the gyro-moments using the definition of $\mathcal{M}_a^{pjm}$ and $\mathcal{M}_b^{pjm}$ given in \cref{eq:curlyMspjm}. The remaining quantities in \cref{eq:CabTlkCoulomb,eq:CabFlkCoulomb,eq:Cpjphi} are numerical coefficients.}

\Cref{eq:GKClk}, with \cref{eq:CabTlkCoulomb,eq:CabFlkCoulomb,eq:Cpjphi}, define the gyro-moment expansion of the linearized GK Coulomb collison operator. This operator is valid for arbitrary mass and temperature ratios. It retains FLR effects associated with the difference between the particle and gyrocenter positions, $\r$ and $\R$, at arbitrary perpendicular wavenumber. These enter explicitly in the linearized GK Coulomb collision operator through terms proportional to $b_s^m \kernel{n}(b_s)$ in the test and field components. Coupled with the quasineutrality equation, \cref{eq:GKPoissonNpj}, and the definitions of the different numerical factors appearing in \cref{eq:CabTlkCoulomb,eq:CabFlkCoulomb}, $\C_{ab}^{lk}$ can be numerically implemented, and be used as a collision operator model in the linearized gyro-moment hierarchy, \cref{eq:momenthierachyEquationNormalized}, to describe collisions between like and unlike species.

\subsubsection{Drift-Kinetic Limit of Coulomb Collision Operator}

The DK limit of the GK linearized Coulomb collision operator, given in \cref{eq:GKClk}, can be deduced by neglecting the FLR effects in the test and field components as well as the polarization terms. For instance, in the zero gyroradius limit, \cref{eq:fasofg} reduces to $f_a =  \g_a$. In addition, the gyro-moment expansion of the spherical harmonic particle moments, $\M_a^{pj}$, given in \cref{eq:momentpolarized}, reduces to

 \begin{align} \label{eq:mpjDK}
 \M_a^{pj} =   \sqrt{\frac{ \pi^{1/2} p!}{2^p (p-1/2)!}} \mathcal{M}_a^{pj} \e^{p0},
 \end{align}
\\
where 

\begin{align}
    \mathcal{M}_a^{pj} = \sum_{g=0}^{p+2j} \sum_{h=0}^{j + \lfloor p/2\rfloor}  T_{pj}^{gh}   \sqrt{2^g g!} N_a^{gh} 
\end{align}
 \\
Neglecting also FLR effects in \cref{eq:gyaverCBesselJm}, such that $\gyaver{C_{ab}(\r,\vi)}_{\R} \simeq \gyaver{C_{ab}(\R,\vi)}_{\R}$, and projecting the resulting expression onto the Hermite-Laguerre basis, we obtain the linearized DK Coulomb collision operator with the test and field components given by 
 \\
 \begin{align} \label{eq:DKClk}
\C_{ab}^{Tlk} =  & \sum_{j=0}^{\infty} \sum_{p=0}^{l+2k}   \sum_{t=0}^{k + \lfloor l/2\rfloor} \sum_{d=0}^t   \frac{ 2^{p}  (p!)^2 }{(2 p)! \sigma_j^p}      \frac{ L_{td}^p}{(2p+1)} 
 \frac{\left( T ^{-1} \right)^{pt}_{lk}}{\sqrt{2^l l!}} \mathcal{M}_a^{pj}    \overline{\nu}_{*ab}^{Tpjd}, \\
 \C_{ab}^{Flk} =  & \sum_{j=0}^{\infty} \sum_{p=0}^{l+2k}   \sum_{t=0}^{k + \lfloor l/2\rfloor} \sum_{d=0}^t   \frac{ 2^{p}  (p!)^2 }{(2 p)! \sigma_j^p}      \frac{ L_{td}^p}{(2p+1)} 
 \frac{\left( T ^{-1} \right)^{pt}_{lk}}{\sqrt{2^l l!}} \mathcal{M}_b^{pj}    \overline{\nu}_{*ab}^{Fpjd},
 \end{align}
 \\
 respectively. \Cref{eq:DKClk} can also be obtained by taking explicitly the zero gyroradius limit of \cref{eq:GKClk}. We remark that the DK linearized Coulomb collision operator obtained here is equivalent to the DK collision operator derived in \citet{Jorge2018}.

\subsection{Gyrokinetic Sugama Collision Operator}
\label{sec:GKSugama}

The Sugama collision operator model \citep{Sugama2009} is an advanced \added{model} collision operator that approximates the full linearized Coulomb collision operator such that it conserves particle, momentum, and energy. It also satisfies the H-theorem and is subject to the self-adjoint relations for the test and field components, i.e.

\begin{subequations}\label{eq:selfadjointness}
\begin{align} 
\int d \vi \frac{f_a}{f_{Ma}} C_{ab}^T \left( g_a \right)  & = \int d \vi \frac{g_a}{f_{Ma}} C_{ab}^T \left( f_a \right), \\
T_a \int d \vi \frac{f_a}{f_{Ma}} C_{ab}^F \left( f_b \right)  & = T_b \int d \vi \frac{f_b}{f_{Mb}} C_{ba}^F \left( f_a \right).
\end{align}
\end{subequations}
\\
The Sugama collision operator model extends the \citet{Abel2008} collision operator to the general case of collisions between different species, and reduces to it in the case of like-species, i.e. $a = b$. We remark that, because the difference with the Sugama and Coulomb operator is expected to increase at high collisionality, the Sugama operator was recently improved and extended to the regime of high-collisionality in \citet{Sugama2019}. In this work, we focus on the original Sugama operator, presented in \citet{Sugama2009}, but remark that the gyro-moment expansion can be applied also to the improved Sugama by following the same analytical procedure detailed below.

In the particle coordinates $(\r, v,\xi, \theta)$, the \textit{test} component of the Sugama collision operator, $ C_{ab}^T(f_a) = C_{ab}^T(\r,\bm v) $ is defined by 

\begin{align} \label{eq:CabTS}
C_{ab}^{T}(f_a) = C_{ab}^S(f_a)+  \sum_{n=1}^3  X^n_{ab}
\end{align}
\\
where

\begin{align} \label{eq:CabS}
C_{ab}^S(f_a ) = - \nu_{ab}^D(v)\mathcal{L}^2 f_a  +  \frac{1}{v^2} \partialv \left(  \nu_{ab}^{\parallel}(v) v^4 F_{aM} \partialv \left( \frac{f_a}{f_{aM}}\right)\right),
\end{align}
\\
and

\begin{align}
X_{ab}^1 &= 2 (\theta_{ab} -1 )   f_{Ma} \left[ \frac{m_a}{T_a} \frac{1}{n_a} \int d \vi' \vi'  C_{ab}^S(f_a) \cdot \vi +  \left( \sa^2 - \frac{3}{2}\right)   \frac{1}{n_a} \int  d\vi  \frac{2}{3} \sa^2 C_{ab}^S(f_a) \right], \\
X_{ab}^2 &= 2 (\theta_{ab} -1 )  \left[  \frac{m_a}{T_a} \u_a(f_a) \cdot C_{ab}^S( f_{Ma} \vi) + \frac{\delta T_a (f_a)}{T_a} C_{ab}^S(f_{Ma}s_a^2) \right], \label{eq:X2ab} \\
X_{ab}^3 &= - \frac{2 \cdot 4 \nu_{ab}}{3 \sqrt{\pi}}f_{Ma} \frac{\chi(\theta_{ab} -1 )^2}{ (1 + \chi^2)^{1/2}} \left[ \frac{m_a}{T_a} \vi \cdot \u_a(f_a)  + \frac{\delta T_a(f_a)}{T_a}\left( \sa^2 -\frac{3}{2}\right)  \frac{2 }{ (1 + \chi^2)} \right],
\end{align}
\\
with $\u_a(f_a) = \int d \vi \vi f_a / n_a$, $\delta T_a(f_a) = T_a \int d \vi f_a \left( 2s_a^2/3 - 1 \right) /n_a$, $\theta_{ab} = \sqrt{ (\tau + \chi^2)/(1 + \chi^2)}$, $\tau = T_a / T_b $, $\chi = \vTa / \vTb = \sqrt{\tau / \sigma}$ and $\sigma = m_a / m_b$. In \cref{eq:CabTS}, we introduce the pitch-angle scattering (also denoted as deflection) and energy diffusion frequencies defined by $\nu_{ab}^D(v ) =  \nu_{ab}  \left[ \erf(\sb) - \Phi(\sb)\right]/s_a^3$ and $\nu_{ab}^{\parallel}(v)  = 2 \nu_{ab} \Phi(\sb) / \sa^3$, respectively, with  $\Phi(x) =  \left[ \erf(x) - x \erf'(x)\right]/(2 x^2)$ the Chandrasekhar function. The closed analytical expressions for $C_{ab}^{S}(f_{Ma} \vi ) $ and $C_{ab}^{S}( \sa^2 f_{Ma}) $ appearing in \cref{eq:X2ab} are given by \citep{Sugama2009}

\begin{subequations}\label{eq:closedCabSFam}
\begin{align}
C_{ab}^{S}(f_{Ma} \vi ) &= -\left( 1 + \chi^2\right) f_{Ma} \nu_{ab}^{\parallel}(v) \sa^2 \vi,\\
C_{ab}^{S}( \sa^2 f_{Ma}) & =-  \nu_{ab} f_{Ma} \frac{2}{\chi^2 \sa} \left[ \erf(\sb) - \sb \left( 1 + \chi^2 \right) \erf'(\sb)\right].
\end{align}
\end{subequations}
\\
We remark that $C_{ab}^{S}$, given in \cref{eq:CabS}, is actually equivalent to the test component of the linearized Coulomb operator in \cref{eq:CCT}, for the case of like-species collisions. More precisely, $C_{ab}^{S}$ in \cref{eq:CabS} is obtained by using the analytical equilibrium Rosenbluth potentials, $H(f_{Mb})$ and $G(f_{Mb})$, in \cref{eq:CCT} and neglecting the term proportional to $(1 - T_a/T_b)$.

The \textit{field} component of the Sugama collision operator is constructed from the test component $C_{ab}^T$, in \cref{eq:CabTS}, such that the local conservation laws, \cref{eq:particleconservation,eq:momentumconservation,eq:energyconservation}, and the self-adjoint relations, \cref{eq:selfadjointness}, are satisfied. \citet{Sugama2009} shows that this is the case if the \textit{field} component, $C_{ab}^F(f_b) = C_{ab}^F(\r,\bm v)$, is given by 

\begin{align} \label{eq:CabFS}
C_{ab}^{F}(f_b) = - \frac{m_a \bm V_{ab}(f_b)}{T_a} \cdot C_{ab}^{T}\left( f_{Ma} \vi\right)   - W_{ab}(f_b) C_{ab}^{T}( f_{Ma} \sa^2),
\end{align}
\\
where $C_{ab}^{T}$ is test component of the Sugama collision operator, given in \cref{eq:CabTS}. From \cref{eq:CabTS}, one has

\begin{subequations} \label{eq:CabTSclosed}
\begin{align}
C_{ab}^{T}( f_{Ma} \vi)& = - \theta_{ab} \left( 1 + \chi^2\right) \frac{4 \nu_{ab}}{3 \sqrt{\pi}}  f_{Ma} \vi  \left[ \frac{3 \sqrt{\pi}}{2}  \frac{\Phi(\sb)}{\sa} + \frac{\chi (\theta_{ab}-1)}{(1 + \chi^2)^{3/2}}\right], \\
C_{ab}^{T}(f_{Ma} \sa^2) &= -  \frac{4 \nu_{ab}}{3 \sqrt{\pi}}  \theta_{ab} f_{Ma} \left[  \frac{3 \sqrt{\pi}}{2} \frac{1}{\chi^2 \sa} \left\{  \erf(\sb) - \sb \erf'(\sb) (1 + \chi^2) \right\} \right. \nonumber \\
& \left. + \frac{2 \chi (\theta_{ab} -1 )}{(1 + \chi^2)^{3/2}} \left( \sa^2 - \frac{3}{2}\right)\right].
\end{align}
\end{subequations}
\\
The quantities, $\bm V_{ab}[f_b]$ and $W_{ab}[f_b] $, appearing in \cref{eq:CabFS}, are defined by 

\begin{align}
\bm V_{ab}[f_b]& = \frac{m_b}{\gamma_{ab}} \int d \vi \frac{f_b}{f_{Mb}} C_{ba}^{T}\left( f_{Mb} \vi \right), \\ 
W_{ab}[f_b] & = \frac{T_b}{\eta_{ab}} \int d \vi \frac{f_b}{f_{Mb}} C_{ba}^{T}(f_{Mb} s_b^2),
\end{align}
\\
where $\gamma_{ab}$ and $\eta_{ab}$ are given by 

\begin{align}
\gamma_{ab} & = -  \frac{4 \nu_{ab}}{3 \sqrt{\pi}}  \frac{n_a m_a \chi}{(1 + \chi^2)^{3/2}} \left( \frac{T_a}{T_b} + \chi^2\right), \\
\eta_{ab} & = -  \frac{4 \nu_{ab}}{3 \sqrt{\pi}}  \frac{3 \chi n_a T_a}{(1 + \chi^2)^{5/2}} \left( \frac{T_a}{T_b} + \chi^2\right).
\end{align}
\\
We remark that $C_{ba}^{T}( f_{Mb} \vi)$ and $C_{ba}^{T}(f_{Mb} s_b^2)$ correspond to $C_{ab}^T(f_{Ma} \vi)$ and $C_{ab}^{T}(f_{Ma} \sa^2)$, given in \cref{eq:CabTSclosed}, with $a$ replaced by $b$.

In order to express the linearized GK Sugama collision operator in the gyrocenter phase-space coordinates $(\R, \mu, v_\parallel, \theta)$, \citet{Sugama2009} considers the nonadiabatic part $h_a$ of the perturbed particle distribution function $f_a$, which is defined by 

\begin{align} \label{eq:fatoha}
 f_a(\r, \bm v)   =  h_a(\R,\mu,v_\parallel) - \frac{q_a }{T_a}\phi (\r)  f_{Ma}.
\end{align}
\\
Then, to evaluate the velocity derivatives contained \cref{eq:CabTS} on $h_a$ and to perform the gyro-average, the gyrokinetic transformation \citep{Xu1991, Abel2008}

\begin{align} \label{eq:gyCtransformation}
    \C^T_{ab} = \gyaver{ e^{ i \bm k \cdot \bm \rho_a} C^T_{ab}( h_a e^{- i \bm k \cdot \bm \rho_a})}_{\R}.
    \end{align}
\\
is used. This yields the test component of the GK Sugama operator $\C_{ab}^T = \C_{ab}^T(\bm k, \mu, v_\parallel)$, 

\begin{align} \label{eq:CabTGKS}
\C_{ab}^{T} & = C_{ab }^{S}(h_a) + R^\perp_{ab} + \sum_{n=1}^3 \overline{X}_{ab}^n, 
\end{align}
\\
with $ C_{ab }^{S}$ defined in \cref{eq:CabS} and 

\begin{align} \label{eq:Rabperp}
R_{ab}^\perp & = - \frac{k_\perp^2}{4 \Omega_a^2} h_a \left[ \nu_{ab}^D(v) \left( 2 v_\parallel^2 + v_\perp^2 \right) + \nu_{ab}^{\parallel}(v) v_\perp^2 \right], \\
\overline{X}_{ab}^1 & = \left( \theta_{ab}-1\right)  F_{Ma} \left[ J_0 \sparallel \mathcal{R}_{ab}^{\parallel 1} + J_1 \sqrt{x_a} \mathcal{R}_{ab}^{\parallel 2} + J_0 \left(\sa^2 - \frac{3}{2} \right) \mathcal{R}_{ab}^{\parallel 3} \right] , \\ 
\overline{X}_{ab}^2 & =\left( \theta_{ab}-1\right) \left[ \vTa C_{ab}^S(F_{Ma} m_a v_\parallel /T_a  )  \left(J_0  \overline{u}_{\parallel a}   + J_1 \frac{v_\perp}{v_\parallel} \overline{u}_{\perp a } \right)+ J_0 C_{ab}^S(F_{Ma} \sa^2)\overline{T}_a \right]  , \\
\overline{X}_{ab}^3 & =  - F_{Ma} \frac{8 \nu_{ab} (\theta_{ab} -1 )^2 \chi}{ 3 \sqrt{\pi}(1 + \chi^2)^{1/2}} \left[  J_0 \sparallel \overline{u}_{\parallel a} + J_1 \sqrt{x_a} \overline{u}_{\perp a} + \frac{J_0}{1 + \chi^2} \left(\sa^2 - \frac{3}{2} \right) \overline{T}_a  \right],
\end{align}
\\
with 

\begin{subequations} \label{eq:mathcalCs}
\begin{align}
\mathcal{R}_{ab}^{\parallel 1} & = \frac{1}{n_a}\int d \vi J_0 \frac{h_a}{F_{Ma}} C_{ab}^S(F_{Ma} m_a v_\parallel /T_a),\\
\mathcal{R}_{ab}^{\parallel 2} & =\frac{1}{n_a} \int d \vi J_1 \frac{h_a}{F_{Ma}} \frac{v_\perp}{v_\parallel} C_{ab}^S(F_{Ma} m_a v_\parallel /T_a),\\
\mathcal{R}_{ab}^{\parallel 3} & = \frac{1}{n_a}\int d \vi J_0 \frac{h_a}{F_{Ma}} \frac{2}{3} C_{ab}^S(F_{Ma} \sa^2),
\end{align}
\end{subequations}
\\
and 
\begin{subequations} \label{eq:mathcalMs}
\begin{align}
\overline{u}_{\parallel a} & = \frac{1}{n_a}\int d \vi J_0 h_a \sparallel, \\
\overline{u}_{\perp a} & = \frac{1}{n_a}\int d \vi J_1 \sqrt{x_a} h_a,\\
\overline{T}_{a} & = \frac{1}{n_a} \int d \vi J_0 h_a \frac{2}{3} \left( \sa^2  - \frac{3}{2} \right),
\end{align}
\end{subequations}
\\
being $J_0 =  J_0(b_a \sqrt{x_a})$ and $J_1 = J_1(b_a \sqrt{x_a})$. 

The GK \textit{field} component of the Sugama operator $\C_{ab}^T = \C_{ab}^T(\bm k, \mu, v_\parallel)$, expressed in terms of $h_b$, is also obtained by applying the transformation in \cref{eq:gyCtransformation} to \cref{eq:CabFS}. It yields, 
    
    \begin{align} \label{eq:CabFSGK}
\C_{ab}^{F} & = -  C_{ab}^{T}(F_{Ma}  v_{\parallel})  J_0 \overline{V}_{\parallel ab} -  C_{ab}^{T}( F_{Ma}  v_{\parallel}) J_1 \frac{v_\perp}{v_\parallel}\overline{V}_{ab \perp}  -  C_{ab}^{T}( F_{Ma} \sa^2)  J_0  \overline{W}_{ab},
\end{align}
\\
where 

\begin{subequations} \label{eq:GKBACKveloctiyintegrals}
\begin{align}
\overline{V}_{\parallel ab}  & =\frac{m_a m_b}{\gamma_{ab} T_a} \int d \vi J_0 \frac{h_b}{F_{Mb}} C_{ba}^{T} ( F_{Mb} v_\parallel), \\
\overline{V}_{\perp ab}  & = \frac{ m_a m_b}{T_a \gamma_{ab}} \int d \vi J_1 \frac{h_b}{F_{Mb}}\frac{v_\perp}{v_\parallel} C_{ba}^{T} \left( F_{bM} v_\parallel \right), \\
\overline{W}_{ab} & = \frac{T_b}{\eta_{ab}}\int d \vi J_0  \frac{h_b}{F_{Mb}} C_{ba}^{T} \left( F_{Mb} \sb^2\right).
\end{align}
\end{subequations}

\subsubsection{Gyrokinetic Hermite-Laguerre Expansion}

We now perform the Hermite-Laguerre expansion of the test and field components of the GK Sugama collision operator given in \cref{eq:CabTGKS,eq:CabFSGK}, respectively. Since the GK Sugama operator is formulated in terms of $h_a$, the Hermite-Laguerre expansion is written in terms of the nonadiabatic part of the gyro-moments $n_a^{pj}$, defined in \cref{eq:npjdef}, which can be expressed as a function of the gyro-moments $N_a^{pj}$, thanks to \cref{eq:defnspj}. 

The Hermite-Laguerre expansion of the test part of the GK Sugama collision operator is obtained by projecting \cref{eq:CabTGKS} onto the Hermite-Laguerre basis. The velocity integrals necessary for the projection are performed analytically in pitch-angle coordinates using the basis transformation from Hermite-Laguerre polynomials to Legendre and associated Laguerre polynomials given in \cref{eq:ALL2HL,eq:HL2ALL}. It is found that the gyro-moment expansion of the test component of the GK Sugama operator is given by

\begin{align} \label{eq:CabTHLS}
\C_{ab}^{Tlk} = \C_{ab}^{Slk}+ R_{ab}^{  \perp lk}  + \sum_{n=1}^3  \overline{X}^{nlk}_{ab},
\end{align}
\\
where 

\begin{align}  \label{eq:CabSlk}
\C_{ab}^{Slk} & =  \sum_{j=0}^{\infty} \sum_{p=0}^{l+2k} \sum_{h=0}^{k + \lfloor  l/2\rfloor} \sum_{g=0}^{p+2j} \sum_{r=0}^{j + \lfloor p/2\rfloor}   \sum_{d=0}^h \frac{2^{p} (p!)^2}{(2p)!} \frac{L_{hd}^p \left(T^{-1} \right)^{ph}_{lk} }{\sigma_p^j \sqrt{2^l l!}} \frac{T_{pj}^{gr}  \sqrt{2^g g!} }{(2p +1)} \overline{\nu}_{*ab}^{Spjd}  n_{a}^{gr},
\end{align}
\\
and

\begin{align}
R_{ab }^{ \perp lk}& = - \frac{b_a^2}{4}  \sum_{p=0}^\infty  \sum_{j=0}^\infty \sum_{g=0}^{l + 2 k} \sum_{h=0}^{k + \lfloor  l/2\rfloor} \sum_{r=0}^{ p + 2j} \sum_{s=0}^{j + \lfloor  p/2\rfloor}  \frac{\left( T^{-1} \right)^{gh}_{lk}}{\sqrt{2^l l!}}\frac{\left( T^{-1} \right)^{rs}_{pj}}{\sqrt{2^p p!}} \mathcal{A}_{ab\perp}^{gr}n_a^{pj}.  \label{eq:Rabperplk}
\end{align}
\\
The details of the calculations leading to the expressions of $\C_{ab}^{Slk}$ and $R_{ab}^{lk}$, given in \cref{eq:CabSlk} and \cref{eq:Rabperplk} respectively, are reported in \cref{appendix:HLexpansionofCabSandRperpab}. In \cref{eq:CabSlk}, the Sugama velocity-integrated speed function is introduced, $\overline{\nu}_{*ab}^{Spjd} =\overline{\nu}_{*ab}^{\parallel pjd} - p (p+1) \overline{\nu}_{*ab}^{D pjd}$ (expressions for $\overline{\nu}_{*ab}^{D pjd}$ and $\overline{\nu}_{*ab}^{ \parallel pjd}$ can be found in \cref{eq:nusabDpjd,eq:nusabparapjd}, respectively). In \cref{eq:Rabperplk}, we introduce the coefficients 

\begin{align}
 \mathcal{A}_{ab\perp}^{gr} = \sum_{s_1 =0}^s\sum_{h_1 =0}^h  L^{r}_{ss_1} L^{g}_{hh_1}  \mathcal{\alpha}_{ab\perp}^{grs_1 h_1} ,
\end{align}
\\
with

\begin{align}
\mathcal{\alpha}_{ab\perp}^{grs_1 h_1}& =  \frac{\delta_g^r}{(2g+1)}  \left( \overline{\nu}_{ab}^{\parallel g + 2 + s_1 + h_1}  + \overline{\nu}_{ab}^{D g  + 2 + s_1 + h_1 }\right) \nonumber \\
&   + \frac{d_{g}^r}{2} \left(\overline{\nu}_{ab}^{D(g+r)/2
+ 2 + s_1 + h_1 } -  \overline{\nu}_{ab}^{\parallel (g +r)/2 + 2 + s_1 + h_1} \right) , 
\end{align}
\\
where 

\begin{align} \label{eq:defdgr}
d_g^r =  \begin{cases}
\dfrac{2 (r + 1)(r+2)}{(2 r+1)(2r +3)(2r+5)}, & \mbox{for } g = r+2\\
\dfrac{2(2r^2 + 2r -1)}{(2r-1)(2r+1)(2r+3)}, & \hbox{for } g=r \\
\dfrac{2 r(r-1)}{(2r -3)(2r -1)(2r +1)}, & \hbox{for } g=r-2
\end{cases}
\end{align}
\\
and the speed integrated energy and deflection functions, $\overline{\nu}_{ab}^{\parallel k} = 4 \int_0^\infty d \sa \sa^{2k} \nu_{ab}^{\parallel}(v) e^{- \sa^2} / \sqrt{\pi}$ and $\overline{\nu}_{ab}^{Dk} =4 \int d \sa e^{- s^2} \sa^{2k} \nu_{ab}^D(v) / \sqrt{\pi}$, with closed analytical expressions, 

\begin{subequations} \label{eq:nuabparaandD}
\begin{align}
\overline{\nu}_{ab}^{\parallel k}& =\frac{4 \nu_{ab}}{\sqrt{\pi}\chi^2} \left(E_{ab}^{k-3} - \chi e_{ab}^{k-2} \right), \\
\overline{\nu}_{ab}^{Dk} &= \frac{4 \nu_{ab} }{\sqrt{\pi}}  \left[ E_{ab}^{k-2} - \frac{1}{2 \chi^2} E_{ab}^{k-3} + \frac{1}{2 \chi} e_{ab}^{k-2}\right],
\end{align}
\end{subequations}
\\
respectively. Finally, the $\overline{X}_{ab}^{nlk}$ terms in \cref{eq:CabTHLS} are defined by 

\begin{subequations} \label{eqs:SugamatestHL}
\begin{align}
\overline{X}^{1lk}_{ab} & =  \left( \theta_{ab} -1 \right) \left[  \mathcal{U}_{\parallel}^{lk}\mathcal{R}_{ab}^{\parallel 1} + \mathcal{U}_{\perp}^{lk} \mathcal{R}_{ab}^{\parallel 2} + \mathcal{T}_{\perp}^{lk}  \mathcal{R}_{ab}^{\parallel 3} \right] , \\ 
\overline{X}^{2lk}_{ab} & =  \left(\theta_{ab} -1 \right) \left( \mathcal{F}_{ab0\parallel}^{lk} \overline{u}_{\parallel a}   +\mathcal{F}_{ab1\parallel}^{lk}\overline{u}_{\perp a} + \mathcal{F}_{abE}^{lk} \overline{T}_a\right) , \label{eq:X2lkab} \\
\overline{X}^{3lk}_{ab} & = - \frac{8 \nu_{ab} (\theta_{ab}-1)^2 \chi}{3 \sqrt{\pi}(1 + \chi^2)^{1/2}}\left[ \mathcal{U}_{\parallel}^{lk} \overline{u}_{\parallel a} + \mathcal{U}_{\perp}^{lk} \overline{u}_{\perp a} + \frac{1}{1 + \chi^2} \mathcal{T}_{\perp }^{lk} \overline{T}_{a}\right] ,
\end{align}
\end{subequations}
\\
where we introduce

\begin{subequations} \label{eq:Rabparallel}
\begin{align} 
\mathcal{R}_{ab}^{\parallel 1} & = - \sum_{p=0}^\infty  \sum_{j=0}^\infty \sum_{n=0}^{\infty} \sum_{s=0}^{n+j} \sum_{h=0}^{s + \lfloor p/2 \rfloor} \sum_{h_1=0}^h   L_{hh_1}^{1}  d_{njs}^0  \frac{\left( T^{-1} \right)^{1h}_{ps}}{\sqrt{2^p p!}}\frac{2 (1 + \chi^2) }{3 }\overline{\nu }_{ab}^{\parallel 3 + h_1} \kernel{n}(b_a)n_a^{pj},  \\
\mathcal{R}_{ab}^{\parallel 2} & = - \sum_{p=0}^\infty  \sum_{j=0}^\infty \sum_{n=0}^{\infty} \sum_{s=0}^{n+j+1} \sum_{h=0}^{s + \lfloor p/2 \rfloor } \sum_{h_1=0}^h L_{hh_1}^0 d_{njs}^1 \frac{\left( T^{-1} \right)^{0h}_{ps}}{\sqrt{2^p p!}}  \frac{ (1 + \chi^2)}{ (n+1)}  \overline{\nu }_{ab}^{\parallel 2 + h_1} b_a \kernel{n}(b_a)  n_a^{pj},\\
\mathcal{R}_{ab}^{\parallel 3} & =  -  \frac{16 \nu_{ab}}{3 \chi^2 \sqrt{\pi}} \sum_{p=0}^\infty  \sum_{j=0}^\infty \sum_{n=0}^\infty \sum_{s=0}^{n+j} \sum_{h=0}^{s + \lfloor p/2 \rfloor} \sum_{h_1 =0}^h L_{hh_1}^0 d_{njs}^0 \frac{\left( T^{-1} \right)^{0h}_{ps}}{\sqrt{2^p p!}} \kernel{n}(b_a) \nonumber \\
& \times \left[ E_{ab}^{h_1} - \chi ( 1 + \chi^2) e_{ab}^{h_1+1}\right] n_a^{pj},
\end{align}
\end{subequations}
\\
and 

\begin{align}
\mathcal{U}_{\parallel}^{lk} &= \kernel{k}(b_a) \frac{\delta_{l}^1 }{\sqrt{2}},\\
\mathcal{U}_{\perp}^{lk} &= \frac{b_a}{2} \left( \kernel{k}(b_a) - \kernel{k-1}(b_a)\right)\delta_l^0, \\
\mathcal{T}_{\perp }^{lk}  & =\left( \frac{\delta_l^2}{\sqrt{2}} - \delta_l^0 \right) \kernel{k}+ \delta_{l}^0 \left[ (2k +1)\kernel{k} - k \kernel{k-1} - (k+1)\kernel{k+1}\right].
\end{align}
\\
In \cref{eq:X2lkab}, we define 

\begin{subequations}  \label{eq:curlyFab}
\begin{align} \label{eq:FabSlk}
\mathcal{F}_{ab 0 \parallel}^{lk} & = -  \frac{2 \left( 1 + \chi^2\right) }{3} \sum_{n=0}^\infty \sum_{s =0}^{n+k} \sum_{r =0}^{l + 2s} \sum_{t=0}^{s + \lfloor l/2\rfloor} \sum_{t_1=0}^t  d_{nks}^0 L^1_{tt_1} \frac{\left( T^{-1}\right)^{1t}_{ls}}{\sqrt{2^l l!}} \delta_{r}^1 \kernel{n}(b_a)\overline{ \nu}_{ab}^{\parallel 3 + t_1}, \\
\mathcal{F}_{ab 1 \parallel}^{lk} & =-  2 \left( 1 + \chi^2\right)  \sum_{n=0}^\infty \sum_{s =0}^{n+k+1}  \sum_{t=0}^{s + \lfloor l/2\rfloor} \sum_{t_1=0}^t d_{nks}^1 L_{tt_1}^{0} \frac{\left( T^{-1}\right)^{0t}_{ls}}{\sqrt{2^l l!}} \frac{b_a \kernel{n}(b_a)}{(n+1)} \overline{\nu}_{ab}^{\parallel 2  + t_1},\\
\mathcal{F}_{abE}^{lk} & = -\frac{8 \nu_{ab}}{\chi^2\sqrt{\pi}} \sum_{n=0}^\infty \sum_{s =0}^{n+k} \sum_{t=0}^{s + \lfloor l/2\rfloor}   \sum_{t_1=0}^t d_{nks}^0 \frac{\left( T^{-1}\right)^{0t}_{ls}}{\sqrt{2^l l!}} \kernel{n}(b_a) L_{tt_1}^0 \left[ E_{ab}^{t_1} - \chi ( 1 + \chi^2) e_{ab}^{t_1+1}\right],
\end{align}
\end{subequations}
\\
with 

\begin{subequations}
\begin{align}
\overline{u}_{\parallel a} & =  \sum_{j =0}^\infty \kernel{j}(b_a) \frac{n_a^{1j}}{\sqrt{2}}, \\
\overline{u}_{\perp a} & =    \frac{b_a}{2} \sum_{j=0}^\infty  \kernel{j}(b_a) \left( n_a^{0j} - n_a^{0j+1}\right), \\
\overline{T}_{a} & = \frac{2}{3} \sum_{j=0}^\infty \kernel{j}(b_a) \left[  \frac{n_a^{2n}}{\sqrt{2}} +  2 jn_a^{0j} - j n_a^{0j-1} - (j+1)n_a^{0j+1}  \right].
\end{align}
\end{subequations}
\\
In deriving \cref{eq:Rabparallel,eq:curlyFab}, the closed expressions of $C_{ab}^{S}(f_{Ma} \vi ) $ and $C_{ab}^{S}( \sa^2 f_{Ma})$, given in \cref{eq:closedCabSFam}, are used. 

We now consider the Hermite-Laguerre expansion of the \textit{field} component of the GK Sugama operator given in \cref{eq:CabFSGK}. Similarly to the test part of the collision operator, we multiply \cref{eq:CabFSGK} by the Hermite-Laguerre basis and integrate the result over the velocity space in pitch-angle coordinates using the basis transformation from Hermite-Laguerre polynomials to Legendre and associated Laguerre polynomials, expressed in \cref{eq:ALL2HL,eq:HL2ALL}. This yields

\begin{align} \label{eq:CabFSlk}
\C_{ab}^{Flk} =  - \overline{\mathcal{V}}_{\parallel ab}^{lk} \overline{V}_{\parallel ab} - \overline{\mathcal{V}}_{\perp ab}^{lk}\overline{V}_{\perp ab}   -  \overline{W}_{ab} \overline{\mathcal{W}}_{ab}^{lk},
\end{align}
\\
where we introduce 

\begin{subequations} \label{eqs:Sugamafieldlk}
\begin{align}
\overline{\mathcal{V}}_{\parallel ab}^{lk} & = - \frac{4 \theta_{ab}(1 + \chi^2)}{3 \sqrt{\pi}} \sum_{n=0}^{\infty} \sum_{s =0}^{n+k}\frac{d_{nks}^0}{\sqrt{2^l l!}} \kernel{n}(b_a) \mathcal{V}_{ab}^{ls}, \\
\overline{\mathcal{V}}_{\perp ab}^{lk} & =  \sum_{n=0}^{\infty}\sum_{s=0}^{n + k +1}\frac{ d_{nks}^1}{\sqrt{2^l l!}} \frac{ b_a \kernel{n}(b_a)}{ 2 (n+1)}   \mathcal{I}_{ab}^{ls},\\
\overline{\mathcal{W}}_{ ab}^{lk} & = \sum_{n=0}^\infty \sum_{s=0}^{n+k} \frac{d_{nks}^0}{\sqrt{2^l l!}} \kernel{n}(b_a) \mathcal{W}_{ab}^{ls},
\end{align}
\end{subequations}
\\
with

\begin{align} \label{eq:curlyVablk}
\mathcal{V}_{ab}^{lk} & = \frac{4 \nu_{ab}}{3 \sqrt{\pi}} \sum_{h =0}^{k + \lfloor l/2 \rfloor} \sum_{h_1 =0}^h L_{hh_1}^1 \left( T^{-1}\right)_{lk}^{1h}  \left[ \frac{1}{ \chi^2 }\left(E_{ab}^{h_1}- \chi e_{ab}^{h_1 +1} \right) + \frac{2 \chi (\theta_{ab}-1) (3/2 + h_1)!}{3 \sqrt{\pi}(1 + \chi^2)^{3/2}}\right] \nonumber \\ &  \times \left[ l > 0 \cup k > 0 \right], \\
\mathcal{I}_{ab}^{ls} &  =  - \frac{ 4 \nu_{ab}\theta_{ab} (1 + \chi^2)}{\chi^2\sqrt{\pi}} \sum_{h =0}^{s +\lfloor l/2 \rfloor} \left( T^{-1}\right)^{0h}_{ls} \sum_{h_1 =0}^h L_{hh_1}^0 \left( E_{ab}^{h_1 -1} - \chi e_{ab}^{h_1}\right)  \nonumber \\
&  - \frac{ \nu_{ab} \chi \theta_{ab}}{ 3 \sqrt{\pi}} \frac{(\theta_{ab}-1)}{(1 + \chi^2)^{1/2}} \delta_{l}^0 \delta_s^0 , \\
\mathcal{W}_{ab}^{lk} & = -  \frac{8\nu_{ab}\theta_{ab}}{\chi^2 \sqrt{\pi}}  \sum_{h=0}^{ k + \lfloor l/2\rfloor} \sum_{h_1=0}^h L_{hh_1}^0 \left( T^{-1}\right)_{lk}^{0h} \left(E_{ab}^{h_1}  - \chi (1 + \chi^2) e_{ab}^{h_1 +1}  \right) \nonumber \\
&   - \frac{8 \nu_{ab}\theta_{ab}}{3 \sqrt{\pi}} \frac{ \chi (\theta_{ab}-1)}{(1 + \chi^2)^{3/2}}  \left( 2 \delta_{l}^2 \delta_k^0 - \delta_l^0 \delta_k^1 \right).
\end{align}
\\
Finally, performing the velocity integrals in \cref{eq:GKBACKveloctiyintegrals} using the gyro-moment expansion for $h_b$ and the expressions of $C_{ba}^{T}( F_{Mb} v_\parallel)$ and $C_{ba}^{T}(F_{Mb} \sb^2)$, given in \cref{eq:CabTSclosed}, yields

\begin{align}
 \overline{V}_{\parallel ab}& =  -\sigma \tau  \theta_{ba}\frac{2 m_b}{ \gamma_{ab} } \left( 1 + 1/\chi\right)  \sum_{p=0}^\infty  \sum_{j=0}^\infty \sum_{n=0}^{\infty} \sum_{s =0}^{n+j} \frac{d_{njs}^0 }{\sqrt{2^p p!}} \kernel{n}(b_b) n_b^{pj} \mathcal{V}_{ba}^{ps}, \\
 \overline{V}_{\perp ab} & =  \frac{2m_b}{\chi\gamma_{ab}}  \sum_{p=0}^\infty  \sum_{j=0}^\infty \sum_{n=0}^{\infty} \sum_{s =0}^{n + j + 1}\frac{d^1_{njs}}{\sqrt{2^p p!}} \frac{b_b}{2(n+1)} \kernel{n} (b_b)n_b^{pj}\mathcal{I}_{ba}^{ps}, \\
 \overline{W}_{ab} & = \frac{T_b}{\eta_{ab}} \sum_{p=0}^\infty  \sum_{j=0}^\infty\sum_{n=0}^\infty \sum_{s =0}^{j + n} \frac{d_{njs}^0 }{\sqrt{2^p p!}} \kernel{n}(b_b) n_b^{pj} \mathcal{W}_{ba}^{ps}.
\end{align}
\\
The Hermite-Laguerre expansion of the test and the field components $\C_{ab}^{Tlk}$ and $\C_{ab}^{Flk}$, given in \cref{eq:CabTHLS,eq:CabFSlk} respectively, can also be expressed in terms of $N_a^{pj}$ thanks to \cref{eq:defnspj}. Similarly to the linearized GK Coulomb operator, collisional FLR effects enter through the terms proportional to $\kernel{n}(b_a)$.  

\subsubsection{Drift-kinetic Limit of the Sugama Collision Operator}

We now consider the DK limit of the GK Sugama collision operator. This limit can be deduced by neglecting FLR effects in \cref{eq:CabTHLS} and \cref{eq:CabFSlk}, which appear through the kernel functions $\kernel{n}(b_a)$. In addition, $f_a  = \g_a$ follows from \cref{eq:fasofg} in the DK limit, such that $n_a^{pj}   =  N_a^{pj}$. Hence, we obtain the DK limit of the Hermite-Laguerre expansion of the \textit{test} component,

\begin{align} \label{eq:CabTSlk}
\C_{ab}^{Tlk}  = \C_{ab}^{Slk} + \sum_{n=1}^{3} X_{ab}^{nlk},
\end{align}
\\
where $\C_{ab}^{Slk}$ is given by \cref{eq:CabSlk} with $n_a^{pj} =  N_a^{pj}$, and

\begin{subequations}\label{eqs:DKTestSugamaX}
\begin{align} 
X_{ab}^{1lk} & =   2 (\theta_{ab}-1)\left[-(1 + \chi^2)\sqrt{2} \delta_l^1 \delta_k^0 \sum_{p,j} \frac{I_{ab1}^{ \parallel pj}}{\sqrt{2^p p!}}N_a^{pj}+ \frac{2}{3} \left(  \frac{\delta_l^2 \delta_k^0}{\sqrt{2}}- \delta_{k}^1 \delta_l^0 \right)  \sum_{p,j} \frac{I_{ab2}^{pj}}{\sqrt{2^p p!}}  N_a^{pj} \right], \\
    X_{ab}^{2lk} & = 2  ( \theta_{ab}-1) \left[- \frac{\left( 1 + \chi^2\right)}{\sqrt{2^l l!}} \frac{ 2 u_{\parallel a}}{\vTa}  I_{ab1}^{ \parallel lk } + \frac{\delta T_a}{T_a} \frac{I_{ab2}^{lk}}{\sqrt{2^l l!}} \right], \\
X_{ab}^{3lk} & =  - ( \theta_{ab}-1)^2 \frac{ 2 \cdot 4 \nu_{ab}}{3 \sqrt{\pi}} \left[ \frac{\sqrt{2}\chi}{(1 + \chi^2)^{1/2}} \frac{u_{\parallel a}}{\vTa} \delta_l^1 \delta_k^0 + \frac{\delta T}{T_a}   \frac{2 \chi}{ (1 + \chi^2)^{3/2}}\left( \frac{\delta_l^2 \delta_k^0}{\sqrt{2}} - \delta_l^0 \delta_k^1\right) \right],
\end{align}
\end{subequations}
\\
where $ u_{\parallel a}  / \vTa=N_a^{10}/ \sqrt{2} $ and $\delta T_a / T_a =(\sqrt{2} N_a^{20} - 2N_a^{01})/3 $. In  \cref{eqs:DKTestSugamaX}, we define 

\begin{align}
    I_{ab1}^{ \parallel lk } & = \frac{1}{3} \sum_{h=0}^{ k + \lfloor l/2 \rfloor } \sum_{h_1=0}^h L_{hh_1}^1 \left( T^{-1}\right)_{lk}^{1h} \overline{\nu}_{ab}^{\parallel h_1 + 3} \left[ l>0 \cup k>0\right], \\
     I_{ab2}^{lk} & = - \frac{8 \nu_{ab}}{\chi^2 \sqrt{\pi} } \sum_{h=0}^{k + \lfloor l/2\rfloor}\sum_{h_1=0}^h L^{0}_{hh_1} \left( T^{-1}\right)_{lk}^{0h}  \left[E_{ab}^{h_1}- \chi  \left( 1 + \chi^2 \right)e_{ab}^{h_1+1}  \right].
\end{align}
\\
Similar considerations lead to the DK limit of the Hermite-Laguerre gyro-moment expansion of the \textit{field} component, given in \cref{eq:CabFSlk}, that is 

\begin{align} \label{eq:CabSFV2}
\C_{ab}^{Flk} & = - \frac{4\theta_{ab} \theta_{ba}}{\chi \sigma}  \frac{m_a}{ \gamma_{ab}}\frac{(1  + \chi^2)}{\sqrt{2^l l!}} \sum_{p=0}^\infty  \sum_{j=0}^\infty\frac{(1 + 1/\chi^{2})}{\sqrt{2^p p!}}  \mathcal{V}_{ab}^{lk}  \mathcal{V}_{ba}^{pj}  N_b^{pj}  \nonumber \\
& - \frac{1}{\sqrt{2^l l!}} \frac{2}{\tau} \frac{T_a}{\eta_{ab}} \sum_{p=0}^\infty  \sum_{j=0}^\infty \frac{N_b^{pj}}{\sqrt{2^p p!}} \mathcal{W}_{ab}^{lk} \mathcal{W}_{ba}^{pj}.
\end{align}

\section{Numerical implementations}
\label{sec:numericalimplementations}

\begin{figure}
 \centering
\includegraphics[scale = 0.5]{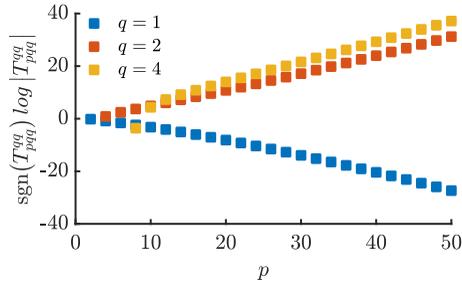}
\caption{Magnitude (with sign) of the coefficients of the basis transformation $T_{pqq}^{qq}$ in \cref{eq:ALL2HL} as a function of $p$ when $q = 1$, $2$, and $4$.}
\label{fig:Tpqqqq}
 \end{figure}
 
Within the adopted Hermite-Laguerre gyro-moment method, the numerical evaluation of the collision operators leads to the calculation of large sums of algebraic coefficients such as the ones appearing, for example, in \cref{eq:CabTlkCoulomb}. In the numerical implementation of the gyrokinetic pseudo-spectral method proposed by \citep{Mandell2018} where a GK Dougherty collision operator is considered, a closed analytical formula is used for the lowest order coefficients from the products between Laguerre polynomials (such as $d_{hs_1}^{0}$ and $\overline{d}_{0kf}^{0}$ defined in \cref{eq:dmnks,eq:bardmnkf} respectively). On the other hand, a Fourier dealiased scheme is employed to numerically compute the integral representation of the higher order coefficients. This method can be generalized to obtain $\overline{d}_{nkf}^{m}$ and $d_{n'hs_1}^{m}$, but it is unpractical for the basis transformation coefficients $T^{lk}_{pjm}$ that appear in the operators considered here. This is because of the rapid increase in magnitude of these coefficients and the enhancement of the oscillatory nature of the basis functions (such as the Hermite polynomials) with the order of the polynomials basis. In fact, due to the presence of positive and negative numerical coefficients of different magnitude that eventually cancel each other, the Hermite-Laguerre formulation of the present collision models is highly sensitive to round off errors. This becomes particularly pronounced as the order of the polynomials basis increases. As an example, \cref{fig:Tpqqqq} displays the magnitude (with sign) of a few of  the basis transformation coefficients, in particular $T_{pqq}^{qq}$ defined in \cref{eq:ALL2HL}. An increase of $40$ orders is observed by going from $p=1$ to polynomials of order $p =50$. We note that similar issues are observed in the numerical implementation of spectral methods based on generalized Laguerre polynomials in the energy variable $v^2$ \citep{Belli2011}. \added{In addition, it should be mentioned that convergence tends to be relatively slow when using Laguerre polynomials in $v^2$ because they have very similar shapes for the bulk of the distribution \citep{landreman2013}}. As a matter of fact, we find that the use of double and/or quadruple precision is not sufficient to carry out the evaluation of the coefficients required for the Coulomb and Sugama operators. 

To avoid spurious unphysical effects associated to precision loss in the linearized collision operators, we evaluate the closed analytical expressions of the numerical coefficients and perform the sums numerically using a multiple precision arithmetic package. To avoid redundancy, the species-independent numerical coefficients that depend on the mass and temperature ratios of the colliding species ($\overline{\nu}_{*ab}^{Tpjt} $, $\overline{\nu}_{*ab}^{Fpjt} $ and $\overline{\nu}_{*ab}^{Spjd}$ defined in \cref{appendix:CoulombSpeedfunctions,appendix:HLexpansionofCabSandRperpab}) are computed once and stored. Then, given the normalized perpendicular wavenumber $b_a$, the mass and temperature ratios, the GK linearized collision operators can be obtained by performing linear combinations  of the previously computed numerical coefficients.   
 
 \begin{figure}
 \centering
\includegraphics[scale = 0.45]{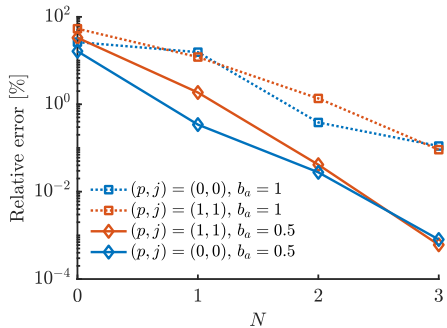}
\caption{Relative error of the diagonal elements, (blue lines) $(p,j) =(0,0)$ and (orange lines) $(p,j) = (1,1)$, of the GK Coulomb [see \cref{fig:Fig_GKCoulomb}], when (solid) $b_a = 0.5 $ and (dotted) $b_a = 1$, as a function of $N$. The relative error is measured with respect to the $N=4$.}
\label{fig:Fig_Nscan_v2}
 \end{figure}
 
Because of FLR effects, infinite sums arise in the GK Coulomb and Sugama operators from the expansion of the Bessel functions, $J_m$, in terms of associated Laguerre [see \cref{eq:J02Laguerre}]. These sums appear in the Hermite-Laguerre projections, such as the ones in \cref{eq:CabTlkCoulomb} and \cref{eq:Rabparallel}. For their numerical  implementations, we truncate these sums at a finite $N > 0$. While a large number of terms in the FLR terms might \textit{a priori} be needed for a good convergence, the nature of the kernel function, $\kernel{n}$ defined in \cref{eq:kernel}, ensures that only a small number of terms contribute in the sums at a given perpendicular wavenumber. This is because the kernel functions $\kernel{n}$ decreases rapidly with $n$, as $\kernel{n} \sim 1/ n!$ for large $n$ for all $b_a$, and $\kernel{n} \sim b_a^{2n}/ n!$ when $b_a \ll 1$ \citep{Frei2020}. Additionally, since the maximum of $\kernel{n}$ occurs at $b_a = 2 \sqrt{n} $, as a rule of thumb, one can choose $N$ such that $N >  b_a^2/4 $. \Cref{fig:Fig_Nscan_v2} displays the relative error of the diagonals elements of the GK Coulomb operator corresponding to the gyro-moments (blue lines) $(p,j) =(0,0)$ and (orange lines) $(p,j) = (1,1)$, for typical values of the normalized perpendicular wavenumber in the ion gyroscale when (solid lines) $b_a = 0.5 $ and (dotted lines) $b_a = 1$, as a function of $N$. The relative error is computed with respect to the $N =4$ case. As observed, the error decreases linearly with $N$, and it is of the order of $ \lesssim 0.1 \%$ or smaller at $N =3$, when $b_a = 1$. A similar behaviour is found with the GK Sugama operator.

\begin{figure}
 \centering
\includegraphics[scale = 0.55]{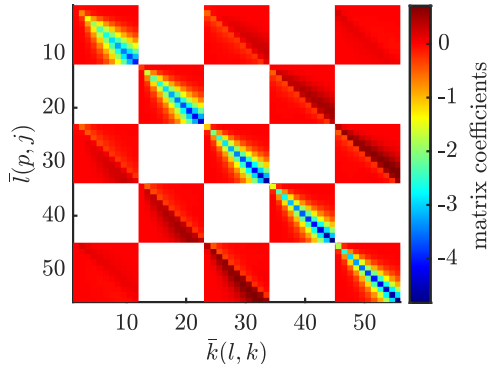}
\caption{Block matrix obtained from the GK Coulomb collision operator for like-species when $b_a = 0.5$. We show, on the vertical axis, the one-dimensional index $\bar{l}(p,j) = (J_a +1) p + 1 + j$ and, on the horizontal axis, the one-dimensional index $\bar{k}(l,k) = (J_a +1) l + 1 + k$.   }
\label{fig:Fig_GKCoulomb}
 \end{figure}

Finally, we discuss the coupling between gyro-moments introduced by the GK Coulomb collision operator. By mapping the gyro-moments into a one dimensional array, the GK Coulomb operator can be represented in matrix form, relating the projection of the collision operator with the gyro-moments. Given $(P_a,J_a)< \infty$, we can construct the collisional matrix by defining the one dimensional gyro-moment array as $[N_{a}^{00}, N_{a}^{01},N_{a}^{02}, \dots, N_{a}^{0J} , N_a^{10}, N_a^{11}, N_a^{12}, \dots , N_a^{P_a0}, \dots, N_a^{P_aJ_a}]^T$ and the one dimensional index $\bar{l}(p,j) = (J_a +1) p + 1 + j$. \Cref{fig:Fig_GKCoulomb} displays the block matrix obtained from the GK Coulomb collision operator for like-species when $b_a = 0.5$ for $(p,j) \le (P_a,J_a)$. It is visible that, the $p$th Hermite blocks is coupled with the $ l = p + 2n$ block (with $n$ a positive integer). We remark that the  block structure of the matrix representing the Coulomb collision operator results from the vanishing basis transformation coefficients $T^{lk}_{pjm}$. Notice also the negative elements in the diagonal with a magnitude that increases with $\bar{l}$ at fixed $p$, yielding damping of higher order gyro-moments. A similar block matrix is obtained for the GK Sugama.

\label{sec:numericalimplementations}
\section{Numerical Tests}
\label{sec:numericaltests}

This section presents the numerical tests and benchmarks of the implementation of the GK Coulomb and Sugama collision operators. First, we numerically investigate the linear growth rate of the ion-temperature gradient (ITG) instability in \Cref{sec:ITG}. For this test, we consider both the linearized GK Coulomb and Sugama collision operators and various perpendicular wavenumbers and levels of collisionality. As a second test, in \Cref{sec:ZF}, we address the collisional damping of zonal flows (ZF). We benchmark our results against the DK linearized Coulomb and GK Sugama collision operators \citep{Crandall2020} \added{currently available in the public release of} the gyrokinetic continuum code GENE \citep{Jenko2000}, finding good agreement.

\subsection{Ion-Temperature Gradient Instability}
\label{sec:ITG}

Belonging to the class of ion-gyroscale gradient-driven drift-wave instability, the ion temperature gradient (ITG) mode is widely recognised as the main candidate to explain the level of anomalous ion heat transport observed experimentally in the tokamak core \citep{Garbet2004}. As a matter of fact, the collisionless theory of ITG mode in core conditions is well established and documented \citep[see, e.g,][]{Romanelli1989,Hahm1989,Romanelli1990,Chen1991}. On the other hand, the role of the ITG mode is less clear in the edge and in the SOL region due to the limitation of Braginskii-like fluid models that are most often applied in these regions \citep{Hallatschek2000,Mosetto2015}. While a complete collisional theory of ITG modes using the gyro-moment approach will be subject of a future publication, we consider here the \added{slab} growth rate of this mode, as a function of the collisionality and the perpendicular wavenumber, when the GK Coulomb and Sugama collision operators are used. We also perform a comparison with GENE. 

We focus on a shearless slab configuration and measure the relative strength of the ion temperature and density gradients by the dimensionless parameter $\eta = L_N / L_{T_i}$, where $L_N = \left| \grad_\perp \ln N \right|^{-1}$ and $L_{T_i} = \left| \grad_\perp \ln T_i \right|^{-1}$. We assume that the electrons are adiabatic, and consider only like-species collisions, $C_i = C_{ii}$, neglecting the ion-electron collisions since they occur on a time scale larger by a factor $O(\sqrt{m_i / m_e})$ than the ion-ion collisions. We neglect magnetic gradient drifts and particle trapping, considering therefore the slab ITG branch. We define the normalized ion-ion collisionality by the parameter $\nu  = L_N \nu_{ii} / c_s$, with $c_s^2 = T_e/ m_i$ the ion sound speed, and we impose $T_e = T_i$. We normalize the perpendicular wavenumber, $k_\perp$, to $\rho_s = c_s /\Omega_i$ and the linear growth rate $\gamma$ to $c_s / L_N$. The parallel wavenumber is fixed at $k_\parallel = 0.1 / L_N$. We solve the gyro-moment hierarchy, in \cref{eq:momenthierachyEquationNormalized}, with $(P, J ) = (20,8)$ for the ion species by applying a closure by truncation such that $N_{i}^{pj} =0$ for $(p,j) > (P,J)$. \added{We have carefully checked that the considered number of ion gyro-moments provide a sufficient resolution of the velocity space for the numerical tests carried out in this section. We note the slab ITG mode relies mainly on parallel dynamics and thus emphasizes the faster convergence of the Hermite expansion in $v_\parallel$ relative to the Laguerre expansion.}

 \begin{figure}
\centering
\includegraphics[scale = 0.45]{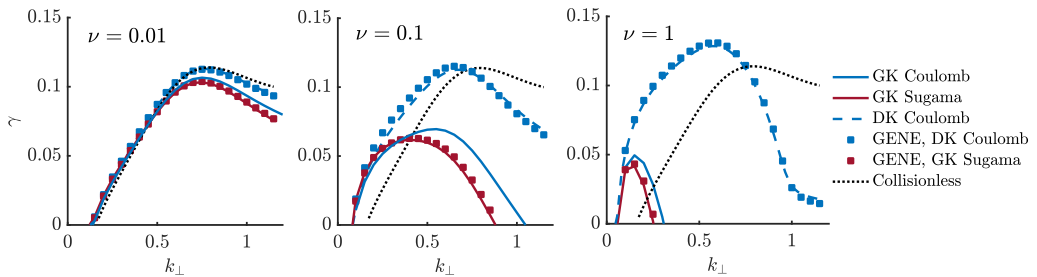}
\caption{Comparisons between the GK Coulomb (solid blue lines) and GK Sugama (solid red lines) operator models, given in \cref{eq:DKClk} and \cref{eq:CabTHLS,eq:CabFSlk}, and GENE (markers), as a function of the perpendicular wavenumber, $k_\perp$ for different collisionalities. The collisionless solution (dotted lines) is shown for comparison. A good agreement is found with the GENE operators. Here, $\eta = 3$. }
     \label{fig:Fig_GENEVSMOLI_kperpScan}
\end{figure}

The numerical results are shown in \cref{fig:Fig_GENEVSMOLI_kperpScan}. We consider the full linearized GK Coulomb collision operator given in \cref{eq:GKClk}, and the GK Sugama collision operator, with test and field components defined in \cref{eq:CabTHLS,eq:CabFSlk}, respectively. The temperature gradient strength is fixed at $\eta =3$, and a scan in the perpendicular wavenumber $k_\perp$ is performed at different collisionalities. The GK Coulomb is represented by the solid blue lines, and the GK Sugama by the red solid lines, while their DK limits are shown for comparison by the dashed lines. Markers represent GENE results with the DK Coulomb and GK Sugama. First of all, we note that a good agreement is found between our results and GENE. \added{We have also successfully compared the DK Sugama operator model with GENE (comparison not shown)}. This shows the ability of the gyro-moment approach to accurately describe FLR collisional damping in \added{advanced GK collision operators}. From a more physical perspective, we observe that the GK Sugama yields an underestimate of the ITG growth rate (of the order of $ 15 \%$ or less) compared to GK Coulomb for wavelengths smaller than the one corresponding to the peak growth rate. We also note, by comparing the DK operators, that FLR collisional effects provide a strong damping and that the DK Sugama follows closely the DK Coulomb\added{, similar to results obtained for TEMs \citep{Pan2020,pan2021}. In principle, comparison with the GK Coulomb operator in GENE \citep{Pan2020} is also possible when it becomes publicly available.}

\begin{figure}
    \centering
    \includegraphics[scale = 0.5]{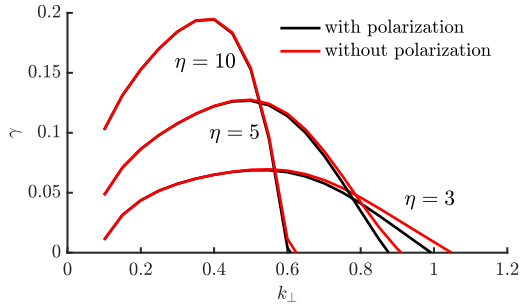}
    \caption{Effects of polarization given in \cref{eq:Cpjphi} in the GK Coulomb collision operator on the ITG growth rate for different temperature gradients as the function of the perpendicular wavenumber $k_\perp$ when $\nu = 0.1$.}
    \label{fig:ITG_FigGKPol}
\end{figure}

To conclude our analysis of the ITG mode, we illustrate the effect of the collisional polarization terms in the GK Coulomb collision operator, expressed by $\sum_{i=1}^{2} (\C^{Tlk}_{ab\phi i}  +  \C^{Flk}_{ab\phi i} )$ appearing in \cref{eq:Cpjphi}. We remark that $\C^{Tlk}_{aa\phi 1}  +\C^{Flk}_{aa\phi 1}  =0$ for like-species, such that only the $\C^{Tlk}_{aa\phi 2}$ and $\C^{Flk}_{aa\phi 2}$ terms in the polarization contribute in the GK linearized Coulomb operator. We perform a scan of the growth rate at a fixed collisionality ($\nu = 0.1$), as a function of the perpendicular wavenumber $k_\perp$ for different temperature gradient strengths $\eta$. The results are shown in \cref{fig:ITG_FigGKPol} where the growth rates with the GK Coulomb that include and exclude the polarization terms are shown by the solid black and red lines, respectively. It is observed that the collisional polarization terms have a negligible effect on the ITG peak growth rate. On the other hand, the difference increases up to $15\%$ for short wavelengths.

\subsection{Collisional  Damping of Zonal Flow}
\label{sec:ZF}

\begin{figure}
\centering
\includegraphics[scale = 0.5]{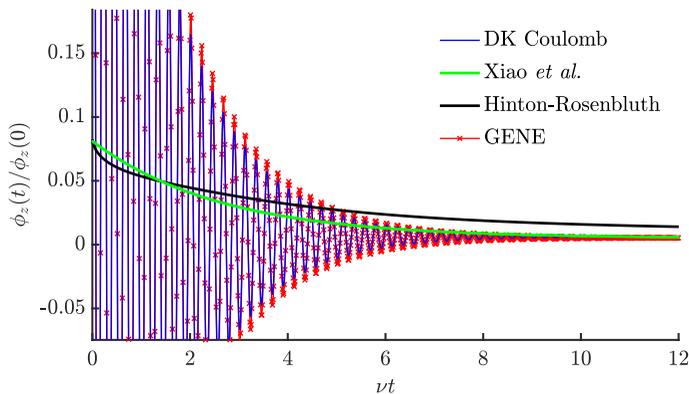}
\caption{Collisional ZF damping using (blue solid line) the DK Coulomb collision operator given in \cref{eq:DKClk} and (red solid with markers) the GENE results. The Hinton-Rosenbluth collisional damping, \cref{eq:HR}, is shown by the black line, and the collisional damping predicted by \citet{Xiao2007}, given in  \cref{eq:XiaoEq94}, is plotted by the solid green. A good agreement is found between GENE results and gyro-moment approach. Here, the parameters are $k_r = 0.05$, $\epsilon = 0.1$, $q = 1.4$ and $\nu_{i*} = 3.1$. }
\label{fig:ZF_res_coll_damping_Coulomb}
\end{figure}

\begin{figure}
\centering
\includegraphics[scale = 0.55]{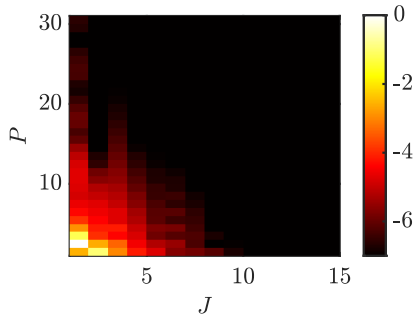}
\caption{Modulus of the normalized Hermite-Laguerre spectrum of the zonal flow residual at $\nu t =10 $ taken from \cref{fig:ZF_res_coll_damping_Coulomb} on a logarithmic scale.}
\label{fig:HL_spectrum_dkcoulomb}
\end{figure}

Zonal flows (ZF) can reduce the particle and heat exhaust from the core in magnetised plasma confinement devices \citep{hasegawa1979,hammett1993}. These self-generated, primarily poloidal and axisymmetric, flows help to shear apart structures that propagate radially, caused by the growth of an instability driven by, e.g., a steep pressure gradient. In fact, ZF regulate and, ultimately, can suppress the turbulent transports driven by, e.g., ITG and trapped electron modes \citep{candy2003anomalous,ernst2004}.
 
 The collisionless damping of self-generated ZF driven by ITG turbulence to a non vanishing residual in the large radial wavelength limit has been addressed and calculated by \citet{Rosenbluth1998a}, in a seminal work that considers a large aspect ratio and circular flux surface geometry. Later, including the effects of ion collisions modelled by a pitch angle scattering operator, \citet{Hinton1999} showed that the ZF component of the electrostatic potential, $\phi_z(t)$, decays to a smaller values than the one predicted by the collisionless theory on long time scales. Hinton-Rosenbluth's analytical prediction,

 \begin{align}  \label{eq:HR}
\frac{\phi_z(t)}{\phi_z(0)}  =  \varpi \exp\left( \varpi^2\alpha^2 t\right) \left[ 1 - \text{erf} \left( \varpi \alpha t^{1/2}\right)\right],
\end{align}
\\
has been the subject of numerous tests and benchmarks of gyrokinetic codes \citep[see, e.g.,][]{idomura2008,merlo2016linear}. In \cref{eq:HR}, we introduce $\varpi = 1/ (1 + 1.61 q^2 / \sqrt{\epsilon}) $ and $\alpha= 3 \pi q^2 \bar{\nu}^{1/2}/ (\epsilon \Lambda^{3/2})$, where $\Lambda \simeq \ln [16 (\epsilon / \bar{\nu} t)^{1/2}]$ and $\bar{\nu} = 0.61 \nu_{ii}$. In addition, $\epsilon$ is the local inverse aspect ratio and $q$ is the safety factor. As an extension of the work by \citet{Hinton1999} to arbitrary collisionality, \citet{Xiao2007} retains the velocity dependence of the deflection frequency, $\nu_{ii}^D(v)$, appearing in the pitch angle scattering operator, as well as a momentum restoring field component, in the limit of ZF with perpendicular wavelength larger than the ion poloidal Larmor radius. In fact, \citet{Xiao2007} shows that the time dependence of the collisional decay of the ZF residual is predicted by  

\begin{align}\label{eq:XiaoEq94}
\frac{\phi_z(t)}{\phi_z(0)} = \frac{\beta }{1 +\beta} \left[ 1 +  \frac{1 - \Theta}{\Theta + \beta} e^{- (1 + \beta) \Gamma \nu_{ii} t/ [\Theta + \beta]}\right], 
\end{align}
\\
with $\beta = \epsilon^2 /q^2$, $\Theta = 1.635 \epsilon^{3/2} + \epsilon^2 /2 + 0.360 \epsilon^{5/2}$ and $\Gamma = 0.4(1.46 \sqrt{\epsilon} + 1.32 \epsilon)$. Since the pitch angle scattering is the dominant collisional damping process in large aspect ratio tokamak \citep{Hinton1999}, energy diffusion - neglected in \cref{eq:XiaoEq94} - is expected to play a subdominant role. \Cref{eq:XiaoEq94} provides an accurate analytical estimate to benchmark the ZF collisional damping using the DK Coulomb and DK Sugama collision operators.

As a test of the implementation of the collisional gyro-moment hierarchy, given in \cref{eq:momenthierachyEquationNormalized}, we consider a shearless, toroidal circular concentric flux surface geometry, with an initial zonal density perturbation radially varying with a wavevector $\bm k = k_r \grad r $. Density and temperature equilibrium gradients are neglected. A finite number of ion gyro-moments, $(P,J)= (30,15)$, are evolved and electrons are assumed adiabatic. We consider a level of ion-ion neoclassical collisionality that is in the Pfirsch-Schluter regime, i.e. $\nu_{i*} \simeq 3.1$. In this parameter regime, collisions smear out fine structures in $v_\parallel$, which develop at low collsionality because of passing ions and finite orbit width effects and might require a larger number of gyro-moments \citep{idomura2008}.

 \Cref{fig:ZF_res_coll_damping_Coulomb} shows the ZF collisional damping predicted by the gyro-moment approach (blue solid line) using the DK Coulomb collision operator and its comparison with GENE results (red line with cross markers). We also plot Hinton and Rosenbluth's, \cref{eq:HR}, and Xiao's, \cref{eq:XiaoEq94}, predictions. An excellent match is observed between the gyro-moment approach and GENE simulation. In addition, the collisional ZF damping agrees with the analytical prediction, given in \cref{eq:XiaoEq94}. Ultimately, the Hermite-Laguerre spectrum at $t = 10 / \nu_{ii}$, in \cref{fig:HL_spectrum_dkcoulomb}, reveals that the velocity space dependence of the ion distribution function is well resolved. We remark that the decay of the spectrum is slower along the Hermite direction compared to the Laguerre direction. This is primarily because of the the ballistic mode response of the passing ion population \citep{idomura2008}.

\begin{figure}
\centering
\includegraphics[scale = 0.5]{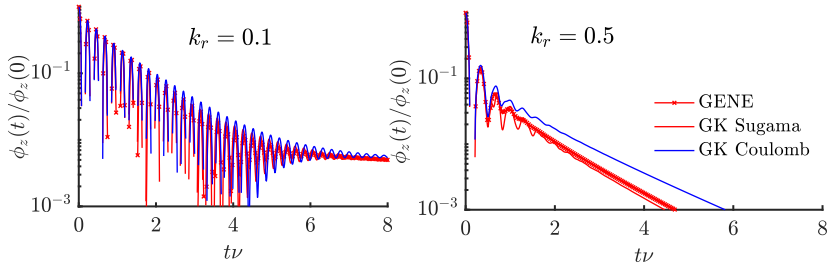}
\caption{Collisional ZF damping using (blue solid line) the GK Coulomb collision operator, given in \cref{eq:GKClk},  (red solid) the GK Sugama, given in \cref{eq:CabTHLS}, and  (red solid with markers) GENE when (left) $k_r = 0.1$ and (right) $k_r = 0.5$. Others parameters are the same in as \cref{fig:ZF_res_coll_damping_Coulomb}. }
\label{fig:ZF_damping_GKCoulomb_kx.0.1}
\end{figure}

We now focus on the collisional ZF damping when the linearized GK Coulomb and GK Sugama collision operators are considered. The results are shown in \cref{fig:ZF_damping_GKCoulomb_kx.0.1}, for $k_r = 0.1$ (left) and $k_r = 0.5$ (right). The ZF damping by the GK Sugama operator is compared with GENE. Other parameters are the same as in \cref{fig:ZF_res_coll_damping_Coulomb}. The GK Coulomb collision operator yields a smaller collisional damping of the ZF residual  than the one predicted by the GK Sugama. \added{The same qualitative result was obtained in \citep{Pan2020,pan2021}, but in the low collisionality banana regime.} We remark that the GK Sugama agrees well with GENE simulations. The observed difference between GK Coulomb and Sugama operators in the ZF residual collisional damping suggests that higher ion particle and heat fluxes can be expected in nonlinear simulations that use the GK Sugama operator rather than the GK Coulomb, particularly in the $\nu_{i*} \gtrsim 1$ Pfirsch-Schlüter regime. \added{However, the opposite trend in TEM turbulence was observed in \citet{pan2021}.}

Furthermore, we note that larger oscillations in the gyro-moment approach are observed compared to GENE when $k_r$ increases. This is related to the large number of gyro-moments required to accurately resolve the resonant passing particle dynamics and finite orbit width effects during the geodesic mode oscillation. The low resolution does not influence the collisional damping of the ZF residual.

\section{Convergence Test}
\label{sec:convergence}

\begin{figure}
\centering
\includegraphics[scale=0.5]{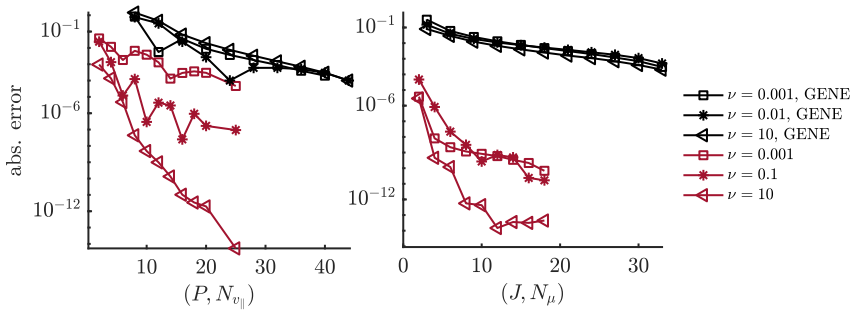}
\caption{Absolute error of the ITG linear growth rate as a function of the number of gyro-moments, $(P,J)$ (red), and the number of velocity-space grid points, $(N_{v_\parallel}, N_{\mu})$ used in GENE (black), for increasing collisionality regimes. \added{We display, on the same horizontal axis, the gyro-moment and GENE errors as a function of the discretization along the parallel direction given by $P$ for the gyro-moment method and $N_{v_\parallel}$ for GENE (left panel), and as a function of the discretization along the perpendicular direction given by $J$ for the gyro-moment method and $N_{\mu}$ for GENE (right panel), respectively.} Collisions are modelled by the DK Coulomb operator. Here, $k_\perp = 0.5$, $k_\parallel = 0.1$ and $\eta = 3$.}
\label{fig:GENEVSMOLI_conv}
\end{figure}


The numerical tests in \cref{sec:numericaltests} illustrate the ability of the gyro-moment approach to model collisional effects using advanced linearized GK collision operators. Additionally, the direct comparisons with the GK continuum GENE code show that the Hermite-Laguerre approach is able to resolve the velocity dependence of the distribution functions in the presence of collisions, as shown in \cref{fig:HL_spectrum_dkcoulomb}, with a finite number of gyro-moments even when a simple closure by truncation is adopted. 

As we now show, the number of gyro-moments necessary to describe the distribution function decreases with the collisionality, as collisions provide a natural way to smear out any sharp velocity space structures and damp the higher order gyro-moments in the Hermite-Laguerre spectrum. Thus, in the high collisional regime, only the lowest order gyro-moments are necessary, while, as the collisionality decreases, a larger number of gyro-moments is necessary to resolve the essential kinetic effects in order to retrieve the collisionless solution [see, e.g., dotted black line in \cref{fig:Fig_GENEVSMOLI_kperpScan}]. We remark that the presence of kinetic effects related to, e.g., finite magnetic gradient drifts, passing ions, particle trapping and finite orbit width can highly affect the velocity space structures, and ultimately deteriorates the convergence of the Hermite-Laguerre spectrum, when collisions are subdominant. These fine scale structures require a fine velocity space resolution. 

Contrary to velocity space grid methods, the Hermite-Laguerre gyro-moment approach features a spectral convergence rate. This convergence rate is faster than any algebraic schemes that characterize the grid methods \citep{boyd2001}\added{, though some basis functions converge faster than others, depending on the velocity space structure present in the distribution function \citep{landreman2013}}. To illustrate the convergence properties of the gyro-moment method, we compute the absolute error on the ITG growth rate as a function of the ion Hermite-Laguerre resolution $(P,J)$. Similarly, by using GENE, we consider convergence study with the number of velocity-space grid points, defined by $N_{v_\parallel}$ and  $N_{\mu}$ \added{(with $N_{v_\parallel}$ the number of grid points in the parallel velocity space direction $v_\parallel$, and $N_\mu$ the number of grid points in the perpendicular velocity space direction $\mu$)} for different collisionality using the DK Coulomb collision operator. We first note that the difference between the most resolved growth rate evaluated by GENE and by the gyro-moment method is less than $3 \%$ in all cases. The results are shown in \cref{fig:GENEVSMOLI_conv} ($k_\perp = 0.5$, $k_\parallel = 0.1$ and $\eta =3$). The convergence with respect to the parallel and perpendicular velocity direction refinement are plotted on the left and on the right panel, respectively.

As observed in \cref{fig:GENEVSMOLI_conv}, the gyro-moment approach features a convergence rate in both velocity-space directions that is sensitive to the collisionality $\nu$ (as discussed above), while the velocity-space grid method, used in GENE, shows a convergence rate that is much less sensitive to $\nu$. In fact, while in both methods, fewer number of gyro-moments and grid points in velocity-space is required in the high collisional regime as the perturbed distribution function does not feature sharp velocity-space structures, the effect of collision is much stronger on the number of gyro-moments required than on grid points. While the increase of the convergence rate of the gyro-moment method with $\nu$ is clearly visible in \cref{fig:GENEVSMOLI_conv}, we notice that the GENE error also decreases with the collisionality $\nu$, but is much less visible. We remark that, in the case of slab ITG mode, the perturbed distribution function does not feature sharp velocity-space structures which appear, e.g., in the ballistic mode response from passing particles and in the trapping of particles in velocity-space \citep{idomura2008}. However, preliminary studies show that the same conclusion on the convergence rate of the gyro-moment approach can be made even if stronger kinetic drives are present. Thus, the convergence study we carried out suggest that the gyro-moment approach offers a particularly efficient numerical scheme to simulate the plasma boundary, where the collisionality is larger (with respect to the strength of velocity space structure drive) than in the core, compared to a finite difference. \added{A careful analysis of the description of fine velocity-space structure using the gyro-moment approach will be subject of a future publication.}

\section{Conclusion}
\label{sec:conclusion}
In this work, we report on the derivation and numerical implementation of a linearized version of \added{a} full-F nonlinear gyrokinetic Coulomb collision operator \citep{Jorge2019} based on an Hermite-Laguerre moment decomposition of the perturbed gyrocenter distribution function. Thanks to a spherical harmonic moment expansion technique, the gyro-average of the linearized Coulomb collision operator is analytically evaluated, projected onto the Hermite-Laguerre basis, and ultimately expressed in terms of gyro-moments. The obtained linearized gyrokinetic Coulomb collision operator is valid at arbitrary perpendicular wavenumber, and at arbitrary mass and temperature ratios of the colliding species. Within the same formalism, we have projected the linearized gyrokinetic Sugama collision operator \citep{Sugama2009} onto the same basis. Additionally, neglecting the finite Larmor radius effects, the DK Coulomb and DK Sugama collision operators are deduced. One of the main advantages of the Hermite-Laguerre decomposition coupled with a basis transformation is that velocity integrals in the linearized GK collision operator models can be evaluated analytically, while a numerical approach is needed in continuum gyrokinetic codes \citep{Crandall2020,Pan2020}. It follows that the collision operators are expressed as linear combinations of gyro-moments, where the coefficients are only functions of perpendicular wavenumber, mass and temperature ratios. These coefficients can be evaluated numerically thanks to closed analytical formulas.

The implementation of the collision operator is described and numerical tests focused on the ITG growth rates and collisional ZF damping are reported. In particular, we show that the collisional damping of the ZF residual, predicted by the GK Sugama operator, is larger than the one obtained using the GK Coulomb operator. This points out that, since ZF residual plays an important role in the self-regulation of turbulent transport, the choice of collision operator might affect the predicitions of the nonlinear level of particle and heat fluxes. Additionally, the gyro-moment approach is tested and benchmarked, for the first time, against the continuum gyrokinetic code GENE \citep{Jenko2000}, reporting an excellent agreements between the DK Coulomb and GK Sugama operator. The comparisons between spectral and velocity space grid approach to collisional gyrokinetic modelling illustrate the main advantage of the Hermite-Laguerre approach, i.e. its ability to reduce the number of required gyro-moments, as the plasma collisionality increases. This hints that the gyro-moment formulation provides an ideal framework for the development of a multi-fidelity scheme to simulate the turbulent plasma dynamics in the boundary region of tokamak devices where collisions play an important role and cannot be ignored \citep{Frei2020}.

Finally, we remark that the Hermite-Laguerre expansion presented here can be easily applied to other gyrokinetic collision operators, such as the improved GK Sugama operator \citep{Sugama2019}. Ultimately, it allows for comparisons between collision operator models over a wide range of physical phenomena.

\section{Acknowledgement}

The authors acknowledge helpful discussions with S. Brunner and P. Donnel. The authors thank the anonymous reviewers whose suggestions helped improve and clarify this work. The simulations presented herein were carried out in part at the Swiss National Supercomputing Center (CSCS) under the project ID s882 and in part on the CINECA Marconi supercomputer under the GBSedge project. This research was supported in part by the Swiss National Science Foundation, and has been carried out within the framework of the EUROfusion Consortium and has received funding from the Euratom research and training programme 2014 - 2018 and 2019 - 2020 under grant agreement No 633053. The views and opinions expressed herein do not necessarily reflect those of the European Commission. 

\section{Declaration of Interests}

The authors report no conflict of interest.

\appendix

\section{Coulomb Speed Functions}
\label{appendix:CoulombSpeedfunctions}

We report the analytical expressions of the speed functions, $\nu_{ab}^{Tpj}$ and $\nu_{ab}^{Fpj}$, appearing in \cref{eq:cmomentexpansion}. The speed function $\nu_{ab}^{Tpj}$ associated with the \textit{test} component of the Coulomb operator is given by 

\begin{align} \label{eq:nuabTpj}
\nu_{ab}^{Tpj} =  \sum_{l=0}^j L_{jl}^p \nu_{*ab}^{Tpjl}
\end{align}
\\
with
\begin{align} \label{eq:nusabTpjl}
\nu_{*ab}^{Tpjl} =  \chi\nu_{ab}\sa^{2l+p}\sum_{n=0}^2 \left[ \frac{\erf(s_b)}{ s_b} \frac{a^{pl}_{abn}}{\sa^{2n}} + \erf'(\sb) \frac{a_{abn}^{'pl}}{\sa^{2n}} \right],
\end{align}
\\
where we introduce the coefficients 

\begin{subequations} \label{eq:testcoeff}
\begin{align}
a_{ab0}^{pl} &=-\frac{4 \sigma  (\tau -1)}{\tau },\\
a^{pl}_{ab1} &= 4 \frac{\sigma}{\tau}l(\tau -2)-p (p  -2 \sigma   +\frac{4 \sigma}{\tau} +1),\\
a_{ab2}^{pl} & =\frac{\sigma}{\tau}  \left[4 l^2+4 l (p-1)+\frac{3}{2} p(p-1) \right], \\
a^{'pl}_{ab0} &=\frac{4 (\tau -1) (\sigma +\tau )}{\tau },\\
a^{'pl}_{ab1} &=-2 (p+2 l) \left[\frac{\sigma}{\tau}  (\tau -2)-1 \right],\\
a_{ab2}^{'pl} & =- \frac{\sigma}{\tau}  \left[4l^2+4 l (p-1)+\frac{3}{2}p (p-1) \right],
\end{align}
\end{subequations}
\\
with $\sigma = m_a / m_b$, $\tau = T_a / T_b $, and $\chi = \vTa / \vTb = \sqrt{\tau / \sigma}$. Using \cref{eq:nusabTpjl}, a closed analytical expression for 

\begin{align}
\overline{\nu}_{*ab}^{Tpjd} = \int d \vi s_a^{p+2d} \nu_{ab}^{Tpj} F_{Ma}
\end{align}
\\
can be derived. Performing the integral over the speed $v$ variable, one obtains 

\begin{align} \label{eq:nuTpjgd}
 \overline{\nu}_{*ab}^{Tpjd} = \sum_{j_1=0}^j L_{jj_1}^p \sum_{n=0}^2 \left[ a_{abn}^{pj_1} \alpha_{Eab}^{dj_1np} + a_{abn}^{'pj_1} \alpha_{eab}^{dj_1np}\right],
\end{align}
\\
where we introduce the quantities

\begin{align}
\alpha_{Eab}^{dj_1np} &= \frac{4 \nu_{ab}}{\sqrt{\pi}} E_{ab}^{d+j_1 +p-n},\\
\alpha_{eab}^{dj_1np} &= \frac{4 \chi \nu_{ab}}{\sqrt{\pi}} e_{ab}^{d+j_1 + p-n +1},
\end{align}
\\
with

\begin{subequations} \label{eq:integralerrorfunctions}
\begin{align}
e_{ab}^k &= \int_0^\infty d s_a s_a^{2k} \erf'(\sb) e^{-s_a^2} = \frac{(k-1/2)!}{(-1/2)!(1+\chi^2)^{k+1/2}}, \\ 
E_{ab}^k &= \int_0^\infty d s_a s_a^{2k+1} \erf(s_b)e^{-s_a^2} = \frac{\chi}{2} \sum_{j=0}^k \frac{k!}{j!}e_{ab}^j.
\end{align}
\end{subequations}

The speed function $\nu_{ab}^{Fpj}$ associated with the \textit{field} component of the Coulomb operator is given by 

\begin{align}
\nu_{ab}^{Fpj} = \sum_{l=0}^j L_{jl}^p \nu_{*ab}^{Fpjl},
\end{align}
\\
with
\begin{align}\label{eq:nuabsFpjl}
\nu_{*ab}^{Fpjl} & = \frac{\nu_{ab}}{\chi} \sum_{n=0}^1 \left[ \sb^{p+2(l + n)} b_{abn}^{'pl} \erf'(\sb) + I_+^{2l+1} \sb^{p + 2n} b_{+abn}^{pl} + b_{-abn}^{pl} \frac{I_-^{2(p + l + 1) }}{\sb^{p + 1 - 2n}}\right],
\end{align}
\\
where 

\begin{subequations} \label{eq:backreactioncoeff}
\begin{align}
b_{ab0}^{'pl} &= 4 \frac{\tau^2}{\sigma},\\
b_{ab1}^{'pl} &= - 8\frac{(p^2 + p -1)}{(2p-1)(2p+3)},\\
b_{+ab0}^{pl} & =  -8 \frac{p(p-1)(l+1)}{(2p+1)(2p-1)} - \frac{\tau 8 ( 1 + p - \sigma p)}{\sigma (2p+1)},\\
b_{+ab1}^{pl} & =\frac{8(p+1)(p+2)}{(2p+1)(2p+3)},\\
b_{-ab0}^{pl} & =  \frac{4(p+1)(p+2)(2p+2l+3)}{(2p+1)(2p+3)}+\frac{8 \chi^2(p - \sigma p - \sigma)}{(2p+1)}, \\
b_{-ab1}^{pl} & =  - 8 \frac{p(p-1)}{(2p+1)(2p-1)}.
\end{align}
\end{subequations}
\\
Using \cref{eq:nuabsFpjl},  a closed analytical expression for 

\begin{align}
\overline{\nu}_{*ab}^{Fpjd} = \int d \vi s_a^{p+2d} \nu_{ab}^{Fpj} F_{Ma},
\end{align}
\\
can be obtained by performing the integral over the speed variable $s_a$, yielding

\begin{align} \label{eq:nuFpjgd}
\overline{\nu}_{*ab}^{Fpjd} = \sum_{j_1=0}^{j} L_{jj_1}^p \sum_{n=0}^1 \left[ b_{abn}^{'pj_1} \beta_{eab}^{pdnj_1} + b_{+abn}^{pj_1} \beta_{+eab}^{pdnj_1} + b_{-abn}^{pj_1} \beta_{-Eeab}^{pdnj_1}\right],
\end{align}
\\
where we introduce 

\begin{align}
\beta_{eab}^{pdnj_1} &= \frac{4 \nu_{ab}}{\sqrt{\pi}}\chi^{p+2j_1 +2n-1 }e_{ab}^{p +1 + d+j_1+ n}, \\
 \beta_{+eab}^{pdnj_1} & = \frac{4 \nu_{ab}}{ \sqrt{\pi}}\chi^{p+2n -1}\sum_{j_2=0}^{j_1} \frac{\chi^{2j_2} j_1!}{j_2!} \frac{e_{ab}^{p + n + 1 +d + j_2}}{2},\\
 \beta_{-Eeab}^{pdnj_1} & = \frac{4 \nu_{ab}}{ \sqrt{\pi} }\chi^{2n - p-2}\left[ \frac{(p+j_1+1/2)!}{(1/2)!} \frac{E_{ab}^{d+n}}{2} \right. \nonumber \\
&\left. - \sum_{j_2=0}^{p+j_1} \frac{\chi^{2j_2 +1}(p+j_1+1/2)! }{(j_2 + 1/2)!} \frac{e_{ab}^{d+n+j_2+1}}{2}\right].
\end{align}

\section{Hermite-Laguerre Expansion of $C_{ab}^{S}$ and $R^\perp_{ab}$}
\label{appendix:HLexpansionofCabSandRperpab}

We report the details of the Hermite-Laguerre expansion of $C_{ab}^{S}$, $R^\perp_{ab}$, $\C_{ab}^{Slk}$ and $R^{\perp lk}_{ab}$ given in \cref{eq:CabSlk,eq:Rabperplk}, that appear in the test component of the linearized gyrokinetic Sugama operator, \cref{eq:CabTHLS}. The derivation below can be used as an example to derive the Hermite-Laguerre expansions of the remaining terms in the GK Sugama collision operator given in \cref{eqs:SugamatestHL} and \cref{eqs:Sugamafieldlk}.

In order to obtain the gyro-moment expansion of $C_{ab}^S$ given in \cref{eq:CabS}, we first expand $h_a$ as 

\begin{align} \label{eq:haexpansion}
    h_a = \sum_{p=0}^\infty \sum_{j=0}^\infty  \frac{F_{Ma}c _p}{\sigma_j^p} \mathcal{H}_a^{pj} s_a^p  P_p(\xi) L_j^{p+1/2}(s_a^2),
\end{align}
\\
with $c_p = \sqrt{ \pi^{1/2} p! /[2^{p} (p-1/2)!]}   $, and $\mathcal{H}_a^{pj}$ defined by

\begin{align} \label{eq:Habpj}
\mathcal{H}_a^{pj}  = c_p  \int d \vi h_a s_a^p  P_p(\xi) L_j^{p+1/2}(s_s^2).
\end{align}
\\
Using the basis transformation, \cref{eq:basistransformation} with $m = 0$, the coefficients $H_a^{pj}$ are related to the nonadiabatic moments $n_a^{lk}$ by

 \begin{align} \label{eq:Hapjtonlk}
\mathcal{H}_a^{pj} =   c_p \sum_{g=0}^{p+2j} \sum_{h=0}^{j + \lfloor p/2\rfloor}  T_{pj}^{gh}   \sqrt{2^g g!} n_a^{gh}.
\end{align}
\\
We remark that $c_p s_a^p  P_p(\xi)  = \gyaver{\Y^p}_{\R}  \cdot \bm \e^{p0}$ (or equivalently $\gyaver{\Y^p}_{\r}  \cdot \bm \e^{p0}$, as the spatial dependence of $\Y^p$ on $\bm s_a = \bm v / v_{Ts} (\R)$ is neglected), and is motivated by the fact that the velocity space derivatives, contained in \cref{eq:CabTS}, can be easily evaluated using the pitch angle and energy variables $(\xi, s_a)$. In fact, inserting \cref{eq:haexpansion} into \cref{eq:CabTS} and expanding the associated Laguerre polynomials in monomial basis thanks to \cref{eq:Laguerre} yield

\begin{align} \label{eq:momentCabS}
C_{ab}^{S}(h_a)  & = \sum_{p,j} \frac{F_{aM} c_p}{\sigma^p_j}  \mathcal{H}_{a}^{pj}  P_p(\xi)  \nu_{ab}^{Spj}(v ),
\end{align}
\\
where the speed dependent frequency, $\nu_{ab}^{Spj}(v) = \nu_{ab}^{\parallel pj}(v) - p(p+1) \nu_{ab}^{Dpj}(v)$, with

\begin{align} \label{eq:nuabparallelpj}
 \nu_{ab}^{\parallel pj}(v) &= \nu_{ab} \sum_{j_1 =0}^j L_{jj_1}^p   s_a^{p+2j_1}\sum_{n=0}^{1} \left[ \erf(\sb) \sa^{ - ( 3+ 2n)} \alpha_{abe}^{pj_1n}+ \erf'(\sb) \alpha_{abE}^{pj_1n} \sa^{ -2(n+1)}\right],
 \end{align}
\\
with

\begin{subequations}
\begin{align}
\alpha_{abE}^{pj_10} & = - \frac{2 \sigma}{\tau} (p + 2j_1), \\
\alpha_{abE}^{pj_11} & =  \frac{\sigma}{\tau}(p + 2j_1)( 2 j_1 + p -2), \\
\alpha_{abe}^{pj_10} & = \frac{2}{\chi} (p + 2j_1) \left(1 + \chi^2 \right), \\
\alpha_{abe}^{pj_11} & =  \frac{1}{\chi}(p + 2j_1)\left( 2 -2 j_1 - p  \right),
\end{align}
\end{subequations}
\\
and

\begin{align}
 \nu_{ab}^{D pj}(v)  &=  \nu_{ab}\sum_{j_1 =0}^j L_{jj_1}^p \sa^{p+2j_1} \nu_{ab}^D(v) \label{eq:nuabDpj},
 \end{align}
 \\
 Multiplying \cref{eq:momentCabS} by the Hermite-Laguerre basis, using the basis transformation in \cref{eq:HL2ALL} with $m =0$, and performing the velocity integrals in the pitch angle coordinates yields 
 
 \begin{align}  \label{eq:CabSlkHapj}
 C_{ab}^{Slk} =  \sum_{j=0}^{\infty} \sum_{p=0}^{l+2k} \sum_{h=0}^{k + \lfloor  l/2\rfloor} \sum_{d=0}^h \frac{c_p}{\sigma_p^j} \frac{L_{hd}^p \left(T^{-1} \right)^{ph}_{lk} }{ \sqrt{2^l l!}} \frac{1}{(2p +1)} \mathcal{H}_{a}^{pj} \overline{\nu}_{*ab}^{Spjd},
\end{align}
\\
where the speed integrated frequency $\overline{\nu}_{*ab}^{Spjd}$ is defined by

\begin{align}
\overline{\nu}_{*ab}^{Spjd} &= \int d \vi F_{aM} \sa^{p + 2d}\nu_{ab}^{Spj}(v)  \nonumber \\
& = \overline{\nu}_{*ab}^{\parallel pjd} - p (p+1) \overline{\nu}_{*ab}^{D pjd},
\end{align}
\\
with $\overline{\nu}_{*ab}^{Dpjd}  =  \int d \vi F_{aM} s_a^{p+2d} \nu_{ab}^{Dpj}(v)$ and $\overline{\nu}_{*ab}^{\parallel pjd} = \int d \vi F_{aM} s_a^{p+2d} \nu_{ab}^{\parallel pj}(v)$. We derive the closed analytical expressions of $\overline{\nu}_{*ab}^{Dpjd} $ and $\overline{\nu}_{*ab}^{\parallel pjd} $ using \cref{eq:integralerrorfunctions}, yielding

\begin{align} \label{eq:nusabDpjd}
\overline{\nu}_{*ab}^{Dpjd} & =\frac{4\nu_{ab}}{\sqrt{\pi}}\sum_{j_1=0}^j L_{jj_1}^p \left[ E_{ab}^{p+d+j_1 -1} - \frac{1}{2 \chi^2} E_{ab}^{p+d + j_1 -2} + \frac{1}{2 \chi } e_{ab}^{p+d +j_1 -1} \right],
\end{align}
\\
and

\begin{align}\label{eq:nusabparapjd}
\overline{\nu}_{*ab}^{\parallel pjd} = \frac{4 \nu_{ab}}{\sqrt{\pi}} \sum_{j_1 =0}^j L_{jj_1}^p \sum_{n=0}^1\left[ \alpha_{abE}^{pj_1n}E_{ab}^{p+d +j_1 -n -1} + \alpha_{abe}^{pj_1n} e_{ab}^{p+d+j_1 -n }\right],
\end{align}
\\
respectively. Finally, expressing the coefficients $\mathcal{H}_a^{pj}$ appearing in \cref{eq:CabSlkHapj} in terms of $n_a^{lk}$ thanks to \cref{eq:Hapjtonlk}, the gyro-moment expansion of $C_{ab}^{S}$, i.e. $\C_{ab}^{Slk}$ given in \cref{eq:CabSlk}, is obtained.

We now aim to derive \cref{eq:Rabperplk} defined by 

\begin{align} \label{eq:Rablkdef}
R_{ab}^{\perp lk} =   - \frac{k_\perp^2}{4 \Omega_a^2} \int d \vi \frac{H_{l}(\sparallel) L_k(x_a)}{\sqrt{2^p p!}} h_a \left[ \nu_{ab}^D(v) \left(2 v_\parallel^2 + v_\perp^2 \right)  +  \nu_{ab}^{\parallel}(v) v_\perp^2 \right].
\end{align}
\\
Because of the speed $(v)$ dependence of the deflection and energy diffusion frequencies, $\nu_{ab}^{D}(v) $ and $\nu_{ab}^{\parallel}(v) $ respectively, the velocity integral in \cref{eq:Rablkdef} can be computed in pitch angle coordinates. To this aim, we use the Hermite-Laguerre expansion of the non-adiabatic part of the distribution function, $h_a$ [see \cref{eq:npjdef}], which can be written in terms of Legendre and associated Laguerre polynomials thanks to the basis transformation in \cref{eq:HL2ALL} with $m =0$, such that

\begin{align} \label{eq:hapitchangle}
h_a= \sum_{p=0}^\infty  \sum_{j=0}^\infty \sum_{r=0}^{ p + 2j} \sum_{s=0}^{j + \lfloor  p/2\rfloor} F_{Ma} n_a^{pj}\frac{\left( T^{-1} \right)^{rs}_{pj}}{\sqrt{2^p p!}} s_a^r P_r(\xi) L_s^{r+1/2}(\sa^2).
\end{align}
\\
Using again the same basis transformation in \cref{eq:Rablkdef} with \cref{eq:hapitchangle} yields

\begin{align}
R_{ab }^{\perp lk} &  = - \frac{k_\perp^2}{4 \Omega_a} \sum_{p=0}^\infty  \sum_{j=0}^\infty  \sum_{g=0}^{ l + 2k} \sum_{h=0}^{k + \lfloor  l/2\rfloor} \sum_{r=0}^{ p + 2j} \sum_{s=0}^{j + \lfloor  p/2\rfloor} \sum_{s_1 =0}^s\sum_{h_1 =0}^h L^{r}_{ss_1} L^{g}_{hh_1} \frac{\left( T^{-1} \right)^{gh}_{lk}}{\sqrt{2^l l!}}\frac{\left( T^{-1} \right)^{rs}_{pj}}{\sqrt{2^p p!}} \gyN^{pj}  \nonumber \\
& \times \frac{2}{\sqrt{\pi}} \int d\xi \int d \sa e^{- \sa^2} \sa^{r+g+4 + 2 s_1 + 2h_1} P_r(\xi) P_g(\xi)  \nonumber \\
& \times  \left[ \nu_{ab}^D(v) \left( \xi^2 +1 \right)  +  \nu_{ab}^{\parallel}(v) (1 - \xi^2) \right].
\end{align}
\\
Finally, using the Legendre orthogonality relation and the identity given by respectively,

\begin{align}
\int d \xi P_{g}(\xi) P_r(\xi) = \frac{2 \delta_{g}^r}{2g+1},\\
\int d \xi P_g(\xi ) P_r(\xi) \xi^2 = d_{g}^r,
\end{align}
\\
with $d_{g}^r$ defined in \cref{eq:defdgr}, and introducing the speed integrated frequencies,  $\overline{\nu}_{ab}^{\parallel k} = 4 \int_0^\infty d \sa \sa^{2k} \nu_{ab}^{\parallel}(v) e^{- \sa^2} / \sqrt{\pi}$ and $\overline{\nu}_{ab}^{Dk} =4 \int d \sa e^{- s^2} \sa^{2k} \nu_{ab}^D(v) / \sqrt{\pi}$ with closed formulas found in \cref{eq:nuabparaandD}, the Hermite-Laguerre expansion of $R_{ab}^\perp$, i.e. $R_{ab}^{\perp lk}$ is derived and is given in \cref{eq:Rabperplk}.

 \bibliographystyle{jpp}
 \bibliography{biblio}

\begin{thebibliography}{74}
\expandafter\ifx\csname natexlab\endcsname\relax\def\natexlab#1{#1}\fi
\def\au#1{#1} \def\ed#1{#1} \def\yr#1{#1}\def\at#1{#1}\def\jt#1{\textit{#1}}
  \def\bt#1{#1}\def\bvol#1{\textbf{#1}} \def\vol#1{#1} \def\pg#1{#1}
  \def\publ#1{#1}\def\arxiv#1{#1}\def\org#1{#1}\def\st#1{\textit{#1}}

\bibitem[Abel {\em et~al.\/}(2008)Abel, Barnes, Cowley, Dorland \&
  Schekochihin]{Abel2008}
{\sc \au{Abel, I.~G.}, \au{Barnes, M.}, \au{Cowley, S.~C.}, \au{Dorland, W.} \&
  \au{Schekochihin, A.~A.}} \yr{2008}  \at{Linearized model {F}okker-{P}lanck
  collision operators for gyrokinetic simulations. i. theory}.  \jt{Physics of
  Plasmas}  \bvol{15}~(12),  \pg{122509}.

\bibitem[Barnes {\em et~al.\/}(2009)Barnes, Abel, Dorland, Ernst, Hammett,
  Ricci, Rogers, Schekochihin \& Tatsuno]{Barnes2009}
{\sc \au{Barnes, M.}, \au{Abel, I.~G.}, \au{Dorland, W.}, \au{Ernst, D.~R.},
  \au{Hammett, G.~W.}, \au{Ricci, P.}, \au{Rogers, B.~N.}, \au{Schekochihin,
  A.~A.} \& \au{Tatsuno, T.}} \yr{2009}  \at{Linearized model
  {F}okker--{P}lanck collision operators for gyrokinetic simulations. ii.
  numerical implementation and tests}.  \jt{Physics of Plasmas}  \bvol{16}~(7),
   \pg{072107}.

\bibitem[Belli \& Candy(2011)]{Belli2011}
{\sc \au{Belli, E.~A.} \& \au{Candy, J.}} \yr{2011}  \at{Full linearized
  {F}okker--{P}lanck collisions in neoclassical transport simulations}.
  \jt{Plasma physics and controlled fusion}  \bvol{54}~(1),  \pg{015015}.

\bibitem[Belli \& Candy(2017)]{Belli2017}
{\sc \au{Belli, E.~A.} \& \au{Candy, J.}} \yr{2017}  \at{Implications of
  advanced collision operators for gyrokinetic simulation}.  \jt{Plasma Physics
  and Controlled Fusion}  \bvol{59}~(4),  \pg{045005}.

\bibitem[Boyd(2001)]{boyd2001}
{\sc \au{Boyd, J.~P.}} \yr{2001} {\em Chebyshev and Fourier spectral
  methods\/}.  \publ{Courier Corporation}.

\bibitem[Brizard \& Mishchenko(2009)]{Brizard2009}
{\sc \au{Brizard, A.~J.} \& \au{Mishchenko, A.}} \yr{2009}  \at{{Guiding-center
  recursive {V}lasov and Lie-transform methods in plasma physics}}.
  \jt{Journal of Plasma Physics}  \bvol{75}~(3),  \pg{675}.

\bibitem[Candy \& Waltz(2003)]{candy2003anomalous}
{\sc \au{Candy, J.} \& \au{Waltz, R.~E.}} \yr{2003}  \at{Anomalous transport
  scaling in the diii-d tokamak matched by supercomputer simulation}.
  \jt{Physical review letters}  \bvol{91}~(4),  \pg{045001}.

\bibitem[Cary(1981)]{Cary1981}
{\sc \au{Cary, J.~R.}} \yr{1981}  \at{{Lie transform perturbation theory for
  Hamiltonian systems}}.  \jt{Physics Reports}  \bvol{79}~(2),  \pg{129}.

\bibitem[Chang {\em et~al.\/}(2017)Chang, Ku, Tynan, Hager, Churchill,
  Cziegler, Greenwald, Hubbard \& Hughes]{Chang2017}
{\sc \au{Chang, C.~S.}, \au{Ku, S}, \au{Tynan, G.~R.}, \au{Hager, R.},
  \au{Churchill, R.~M.}, \au{Cziegler, I.}, \au{Greenwald, M.}, \au{Hubbard,
  AE} \& \au{Hughes, J.~W.}} \yr{2017}  \at{Fast low-to-high confinement mode
  bifurcation dynamics in a tokamak edge plasma gyrokinetic simulation}.
  \jt{Physical Review Letters}  \bvol{118}~(17),  \pg{175001}.

\bibitem[Chen {\em et~al.\/}(1991)Chen, Briguglio \& Romanelli]{Chen1991}
{\sc \au{Chen, L.}, \au{Briguglio, S.} \& \au{Romanelli, F.}} \yr{1991}
  \at{The long-wavelength limit of the ion-temperature gradient mode in tokamak
  plasmas}.  \jt{Physics of Fluids B: Plasma Physics}  \bvol{3}~(3),
  \pg{611--614}.

\bibitem[Crandall {\em et~al.\/}(2020)Crandall, Jarema, Doerk, Pan, Merlo,
  G{\"o}rler, Navarro, Told, Maurer \& Jenko]{Crandall2020}
{\sc \au{Crandall, P.}, \au{Jarema, D.}, \au{Doerk, H.}, \au{Pan, Q.},
  \au{Merlo, G.}, \au{G{\"o}rler, T.}, \au{Navarro, A.~Ba{\~n}{\'o}n},
  \au{Told, D.}, \au{Maurer, M.} \& \au{Jenko, F.}} \yr{2020}
  \at{Multi-species collisions for delta-f gyrokinetic simulations:
  Implementation and verification with gene}.  \jt{Computer Physics
  Communications}  \bvol{255},  \pg{107360}.

\bibitem[Dimits \& Cohen(1994)]{Dimits1994}
{\sc \au{Dimits, A.~M.} \& \au{Cohen, B.~I.}} \yr{1994}  \at{Collision
  operators for partially linearized particle simulation codes}.  \jt{Phys.
  Rev. E}  \bvol{49},  \pg{709}.

\bibitem[Dimits {\em et~al.\/}(1996)Dimits, Williams, Byers \&
  Cohen]{Dimits1996}
{\sc \au{Dimits, A.~M.}, \au{Williams, T.~J.}, \au{Byers, J.~A.} \& \au{Cohen,
  B.~I.}} \yr{1996}  \at{Scalings of ion-temperature-gradient-driven anomalous
  transport in tokamaks}.  \jt{Physical review letters}  \bvol{77}~(1),
  \pg{71}.

\bibitem[Donnel {\em et~al.\/}(2019)Donnel, Garbet, Sarazin, Grandgirard,
  Asahi, Bouzat, Caschera, Dif-Pradalier, Ehrlacher, Ghendrih {\em
  et~al.\/}]{Donnel2019}
{\sc \au{Donnel, P.}, \au{Garbet, X.}, \au{Sarazin, Y.}, \au{Grandgirard, V.},
  \au{Asahi, Y.}, \au{Bouzat, N.}, \au{Caschera, E}, \au{Dif-Pradalier, G.},
  \au{Ehrlacher, C.}, \au{Ghendrih, P.} \& \au{others}} \yr{2019}  \at{A
  multi-species collisional operator for full-f global gyrokinetics codes:
  Numerical aspects and verification with the {G}ysela code}.  \jt{Computer
  Physics Communications}  \bvol{234},  \pg{1}.

\bibitem[Dorf {\em et~al.\/}(2014)Dorf, Cohen, Dorr, Hittinger \&
  Rognlien]{Dorf2014}
{\sc \au{Dorf, M.~A.}, \au{Cohen, R.~H.}, \au{Dorr, M}, \au{Hittinger, J.} \&
  \au{Rognlien, T.~D.}} \yr{2014}  \at{Progress with the {COGENT} edge kinetic
  code: Implementing the {F}okker-{P}lanck collision operator}.
  \jt{Contributions to Plasma Physics}  \bvol{54}~(4-6),  \pg{517}.

\bibitem[Dougherty(1964)]{Dougherty1964}
{\sc \au{Dougherty, J.~P.}} \yr{1964}  \at{Model {F}okker-{P}lanck equation for
  a plasma and its solution}.  \jt{The Physics of Fluids}  \bvol{7}~(11),
  \pg{1788}.

\bibitem[Dudson {\em et~al.\/}(2009)Dudson, Umansky, Xu, Snyder \&
  Wilson]{dudson2009}
{\sc \au{Dudson, B.~D.}, \au{Umansky, M.~V.}, \au{Xu, X.~Q.}, \au{Snyder,
  P.~B.} \& \au{Wilson, H.~R.}} \yr{2009}  \at{Bout++: A framework for parallel
  plasma fluid simulations}.  \jt{Computer Physics Communications}
  \bvol{180}~(9),  \pg{1467}.

\bibitem[Ernst {\em et~al.\/}(2004)Ernst, Bonoli, Catto, Dorland, Fiore,
  Granetz, Greenwald, Hubbard, Porkolab, Redi {\em et~al.\/}]{ernst2004}
{\sc \au{Ernst, D.~R.}, \au{Bonoli, P.~T.}, \au{Catto, P.~J.}, \au{Dorland,
  W.}, \au{Fiore, CL}, \au{Granetz, RS}, \au{Greenwald, M}, \au{Hubbard, AE},
  \au{Porkolab, M}, \au{Redi, MH} \& \au{others}} \yr{2004}  \at{Role of
  trapped electron mode turbulence in internal transport barrier control in the
  alcator c-mod tokamak}.  \jt{Physics of Plasmas}  \bvol{11}~(5),  \pg{2637}.

\bibitem[Francisquez {\em et~al.\/}(2020)Francisquez, Bernard, Mandell, Hammett
  \& Hakim]{Francisquez2020}
{\sc \au{Francisquez, M.}, \au{Bernard, T.~N.}, \au{Mandell, N.~R.},
  \au{Hammett, G.~W.} \& \au{Hakim, A.}} \yr{2020}  \at{Conservative
  discontinuous {G}alerkin scheme of a gyro-averaged dougherty collision
  operator}.  \jt{Nuclear Fusion}  \bvol{60}~(9),  \pg{096021}.

\bibitem[Frei {\em et~al.\/}(2020)Frei, Jorge \& Ricci]{Frei2020}
{\sc \au{Frei, B.~J.}, \au{Jorge, R.} \& \au{Ricci, P.}} \yr{2020}  \at{A
  gyrokinetic model for the plasma periphery of tokamak devices}.  \jt{Journal
  of Plasma Physics}  \bvol{86}~(2),  \pg{905860205}.

\bibitem[Garbet {\em et~al.\/}(2004)Garbet, Mantica, Angioni, Asp, Baranov,
  Bourdelle, Budny, Crisanti, Cordey, Garzotti {\em et~al.\/}]{Garbet2004}
{\sc \au{Garbet, X.}, \au{Mantica, P.}, \au{Angioni, C}, \au{Asp, E},
  \au{Baranov, Y}, \au{Bourdelle, C}, \au{Budny, R}, \au{Crisanti, F},
  \au{Cordey, G}, \au{Garzotti, L} \& \au{others}} \yr{2004}  \at{Physics of
  transport in tokamaks}.  \jt{Plasma Physics and Controlled Fusion}
  \bvol{46}~(12B),  \pg{B557}.

\bibitem[Giacomin {\em et~al.\/}(2020)Giacomin, Stenger \& Ricci]{Giacomin2020}
{\sc \au{Giacomin, M.}, \au{Stenger, L.~N.} \& \au{Ricci, P.}} \yr{2020}
  \at{Turbulence and flows in the plasma boundary of snowflake magnetic
  configurations}.  \jt{Nuclear Fusion}  \bvol{60}~(2),  \pg{024001}.

\bibitem[G{\"o}rler {\em et~al.\/}(2011)G{\"o}rler, Lapillonne, Brunner,
  Dannert, Jenko, Aghdam, Marcus, McMillan, Merz, Sauter {\em
  et~al.\/}]{Gorler2011}
{\sc \au{G{\"o}rler, Tobias}, \au{Lapillonne, Xavier}, \au{Brunner, Stephan},
  \au{Dannert, Tilman}, \au{Jenko, Frank}, \au{Aghdam, Sohrab~Khosh},
  \au{Marcus, Patrick}, \au{McMillan, Ben~F}, \au{Merz, Florian}, \au{Sauter,
  Olivier} \& \au{others}} \yr{2011}  \at{Flux-and gradient-driven global
  gyrokinetic simulation of tokamak turbulence}.  \jt{Physics of Plasmas}
  \bvol{18}~(5),  \pg{056103}.

\bibitem[Gradshteyn \& Ryzhik(2014)]{gradshteyn}
{\sc \au{Gradshteyn, I.~S.} \& \au{Ryzhik, I.~M.}} \yr{2014} {\em Table of
  integrals, series, and products\/}.  \publ{Academic press}.

\bibitem[Hager {\em et~al.\/}(2016)Hager, Yoon, Ku, D'Azevedo, Worley \&
  Chang]{Hager2016}
{\sc \au{Hager, R.}, \au{Yoon, E.~S.}, \au{Ku, S.}, \au{D'Azevedo, Eduardo~F.},
  \au{Worley, Patrick~H.} \& \au{Chang, C.-S.}} \yr{2016}  \at{A fully
  non-linear multi-species {F}okker-{P}lanck-landau collision operator for
  simulation of fusion plasma}.  \jt{Journal of Computational Physics}
  \bvol{315},  \pg{644}.

\bibitem[Hahm \& Tang(1989)]{Hahm1989}
{\sc \au{Hahm, T.~S.} \& \au{Tang, W.~M.}} \yr{1989}  \at{Properties of ion
  temperature gradient drift instabilities in h-mode plasmas}.  \jt{Physics of
  Fluids B: Plasma Physics}  \bvol{1}~(6),  \pg{1185--1192}.

\bibitem[Hallatschek \& Zeiler(2000)]{Hallatschek2000}
{\sc \au{Hallatschek, K.} \& \au{Zeiler, A.}} \yr{2000}  \at{Nonlocal
  simulation of the transition from ballooning to ion temperature gradient mode
  turbulence in the tokamak edge}.  \jt{Physics of Plasmas}  \bvol{7}~(6),
  \pg{2554}.

\bibitem[Halpern {\em et~al.\/}(2016)Halpern, Ricci, Jolliet, Loizu, Morales,
  Mosetto, Musil, Riva, Tran \& Wersal]{Halpern2016}
{\sc \au{Halpern, F.}, \au{Ricci, P.}, \au{Jolliet, S.}, \au{Loizu, J.},
  \au{Morales, J.}, \au{Mosetto, A.}, \au{Musil, F.}, \au{Riva, F.}, \au{Tran,
  T.-H.} \& \au{Wersal, C.}} \yr{2016}  \at{The gbs code for tokamak scrape-off
  layer simulations}.  \jt{Journal of Computational Physics}  \bvol{315},
  \pg{388}.

\bibitem[Hammett {\em et~al.\/}(1993)Hammett, Beer, Dorland, Cowley \&
  Smith]{hammett1993}
{\sc \au{Hammett, G.~W.}, \au{Beer, M.~A.}, \au{Dorland, W.}, \au{Cowley,
  S.~C.} \& \au{Smith, S.~A.}} \yr{1993}  \at{Developments in the gyrofluid
  approach to tokamak turbulence simulations}.  \jt{Plasma physics and
  controlled fusion}  \bvol{35}~(8),  \pg{973}.

\bibitem[Hasegawa {\em et~al.\/}(1979)Hasegawa, Maclennan \&
  Kodama]{hasegawa1979}
{\sc \au{Hasegawa, A.}, \au{Maclennan, C.~G} \& \au{Kodama, Y.}} \yr{1979}
  \at{Nonlinear behavior and turbulence spectra of drift waves and rossby
  waves}.  \jt{The Physics of Fluids}  \bvol{22}~(11),  \pg{2122}.

\bibitem[Hazeltine \& Meiss(2003)]{Hazeltine2003}
{\sc \au{Hazeltine, R.~D.} \& \au{Meiss, J.~D.}} \yr{2003} {\em Plasma
  confinement\/}.  \publ{Courier Corporation}.

\bibitem[Helander \& Sigmar(2002)]{Helander2002}
{\sc \au{Helander, P.} \& \au{Sigmar, D.~J.}} \yr{2002} Collisional transport
  in magnetized plasmas cambridge university press.

\bibitem[Held {\em et~al.\/}(2015)Held, Kruger, Ji, Belli. \& Lyons]{held2015}
{\sc \au{Held, E.~D.}, \au{Kruger, S.~E.}, \au{Ji, J.-Y.}, \au{Belli., E.~A.}
  \& \au{Lyons, B.~C.}} \yr{2015}  \at{Verification of continuum drift kinetic
  equation solvers in nimrod}.  \jt{Physics of Plasmas}  \bvol{22}~(3),
  \pg{032511}.

\bibitem[Hinton \& Rosenbluth(1999)]{Hinton1999}
{\sc \au{Hinton, F.~L.} \& \au{Rosenbluth, M.~N.}} \yr{1999}  \at{{Dynamics of
  axisymmetric (E × B) and poloidal flows in tokamaks}}.  \jt{Plasma Physics
  and Controlled Fusion}  \bvol{41}~(3A).

\bibitem[Hirshman \& Sigmar(1976)]{Hirshman1976}
{\sc \au{Hirshman, S.~P.} \& \au{Sigmar, D.~J.}} \yr{1976}  \at{Approximate
  {F}okker-{P}lanck collision operator for transport theory applications}.
  \jt{Physics of Fluids}  \bvol{19}~(10),  \pg{1532}.

\bibitem[Holtkamp(2009)]{Holtkamp2009}
{\sc \au{Holtkamp, N.}} \yr{2009}  \at{The status of the iter design}.
  \jt{Fusion Engineering and Design}  \bvol{84}~(2-6),  \pg{98}.

\bibitem[Idomura {\em et~al.\/}(2008)Idomura, Ida, Kano, Aiba \&
  Tokuda]{idomura2008}
{\sc \au{Idomura, Y.}, \au{Ida, M.}, \au{Kano, T.}, \au{Aiba, N.} \&
  \au{Tokuda, S.}} \yr{2008}  \at{Conservative global gyrokinetic toroidal
  full-f five-dimensional vlasov simulation}.  \jt{Computer Physics
  Communications}  \bvol{179}~(6),  \pg{391}.

\bibitem[Idomura {\em et~al.\/}(2003)Idomura, Tokuda \& Kishimoto]{Idomura2003}
{\sc \au{Idomura, Y.}, \au{Tokuda, S.} \& \au{Kishimoto, Y.}} \yr{2003}
  \at{Global gyrokinetic simulation of ion temperature gradient driven
  turbulence in plasmas using a canonical maxwellian distribution}.
  \jt{Nuclear Fusion}  \bvol{43}~(4),  \pg{234}.

\bibitem[Jenko {\em et~al.\/}(2000)Jenko, Dorland, Kotschenreuther \&
  Rogers]{Jenko2000}
{\sc \au{Jenko, F.}, \au{Dorland, W.}, \au{Kotschenreuther, M.} \& \au{Rogers,
  B.~N.}} \yr{2000}  \at{Electron temperature gradient driven turbulence}.
  \jt{Physics of plasmas}  \bvol{7}~(5),  \pg{1904}.

\bibitem[Ji \& Held(2006)]{Ji2006}
{\sc \au{Ji, J.-Y.} \& \au{Held, E.~D.}} \yr{2006}  \at{Exact linearized
  {C}oulomb collision operator in the moment expansion}.  \jt{Physics of
  plasmas}  \bvol{13}~(10),  \pg{102103}.

\bibitem[Jorge {\em et~al.\/}(2019)Jorge, Frei \& Ricci]{Jorge2019}
{\sc \au{Jorge, R.}, \au{Frei, B.~J.} \& \au{Ricci, P.}} \yr{2019}
  \at{Nonlinear gyrokinetic {C}oulomb collision operator}.  \jt{Journal of
  Plasma Physics}  \bvol{85}~(6),  \pg{905850604}.

\bibitem[Jorge {\em et~al.\/}(2017)Jorge, Ricci \& Loureiro]{Jorge2017}
{\sc \au{Jorge, R.}, \au{Ricci, P.} \& \au{Loureiro, N.~F.}} \yr{2017}  \at{A
  drift-kinetic analytical model for scrape-off layer plasma dynamics at
  arbitrary collisionality}.  \jt{Journal of Plasma Physics}  \bvol{83}~(6).

\bibitem[Jorge {\em et~al.\/}(2018)Jorge, Ricci \& Loureiro]{Jorge2018}
{\sc \au{Jorge, R.}, \au{Ricci, P.} \& \au{Loureiro, N.~F.}} \yr{2018}
  \at{Theory of the drift-wave instability at arbitrary collisionality}.
  \jt{Physical review letters}  \bvol{121}~(16),  \pg{165001}.

\bibitem[Kolesnikov {\em et~al.\/}(2010)Kolesnikov, Wang, Hinton, Rewoldt \&
  Tang]{Kolesnikov2010}
{\sc \au{Kolesnikov, R.~A.}, \au{Wang, W.~X.}, \au{Hinton, F.~L.}, \au{Rewoldt,
  G.} \& \au{Tang, W.~M.}} \yr{2010}  \at{Drift-kinetic simulation of
  neoclassical transport with impurities in tokamaks}.  \jt{Physics of Plasmas}
   \bvol{17}~(2),  \pg{022506}.

\bibitem[Landreman \& Ernst(2012)]{landreman2012}
{\sc \au{Landreman, M.} \& \au{Ernst, D.~R.}} \yr{2012}  \at{Local and global
  fokker--planck neoclassical calculations showing flow and bootstrap current
  modification in a pedestal}.  \jt{Plasma Physics and controlled fusion}
  \bvol{54}~(11),  \pg{115006}.

\bibitem[Landreman \& Ernst(2013)]{landreman2013}
{\sc \au{Landreman, M.} \& \au{Ernst, D.~R.}} \yr{2013}  \at{New velocity-space
  discretization for continuum kinetic calculations and fokker--planck
  collisions}.  \jt{Journal of Computational Physics}  \bvol{243},  \pg{130}.

\bibitem[Li \& Ernst(2011)]{Li2011}
{\sc \au{Li, B.} \& \au{Ernst, D.~R.}} \yr{2011}  \at{Gyrokinetic
  {F}okker-{P}lanck collision operator}.  \jt{Physical Review Letters}
  \bvol{106}~(19).

\bibitem[Madsen(2013)]{Madsen2013}
{\sc \au{Madsen, J.}} \yr{2013}  \at{Full-f gyrofluid model}.  \jt{Physics of
  Plasmas}  \bvol{20}~(7),  \pg{072301}.

\bibitem[Mandell {\em et~al.\/}(2018)Mandell, Dorland \&
  Landreman]{Mandell2018}
{\sc \au{Mandell, N.~R.}, \au{Dorland, W.} \& \au{Landreman, M.}} \yr{2018}
  \at{Laguerre--{H}ermite pseudo-spectral velocity formulation of
  gyrokinetics}.  \jt{Journal of Plasma Physics}  \bvol{84}~(1).

\bibitem[Merlo {\em et~al.\/}(2016)Merlo, Sauter, Brunner, Burckel, Camenen,
  Casson, Dorland, Fable, G{\"o}rler, Jenko {\em et~al.\/}]{merlo2016linear}
{\sc \au{Merlo, G.}, \au{Sauter, O.}, \au{Brunner, S.}, \au{Burckel, A.},
  \au{Camenen, Y.}, \au{Casson, F.~J.}, \au{Dorland, W.}, \au{Fable, E.},
  \au{G{\"o}rler, T.}, \au{Jenko, F.} \& \au{others}} \yr{2016}  \at{Linear
  multispecies gyrokinetic flux tube benchmarks in shaped tokamak plasmas}.
  \jt{Physics of plasmas}  \bvol{23}~(3),  \pg{032104}.

\bibitem[Mosetto {\em et~al.\/}(2015)Mosetto, Halpern, Jolliet, Loizu \&
  Ricci]{Mosetto2015}
{\sc \au{Mosetto, A.}, \au{Halpern, F.~D.}, \au{Jolliet, S.}, \au{Loizu, J.} \&
  \au{Ricci, P.}} \yr{2015}  \at{Finite ion temperature effects on scrape-off
  layer turbulence}.  \jt{Physics of Plasmas}  \bvol{22}~(1),  \pg{012308}.

\bibitem[Nakata {\em et~al.\/}(2015)Nakata, Nunami, Watanabe \&
  Sugama]{nakata2015improved}
{\sc \au{Nakata, M.}, \au{Nunami, M.}, \au{Watanabe, T.-H.} \& \au{Sugama, H.}}
  \yr{2015}  \at{Improved collision operator for plasma kinetic simulations
  with multi-species ions and electrons}.  \jt{Computer Physics Communications}
   \bvol{197},  \pg{61--72}.

\bibitem[Nunami {\em et~al.\/}(2015)Nunami, Nakata, Watanabe \&
  Sugama]{nunami2015development}
{\sc \au{Nunami, M.}, \au{Nakata, M.}, \au{Watanabe, T.-H.} \& \au{Sugama, H.}}
  \yr{2015}  \at{Development of linearized collision operator for multiple ion
  species in gyrokinetic flux-tube simulations}.  \jt{Plasma and Fusion
  Research}  \bvol{10},  \pg{1403058--}.

\bibitem[Pan \& Ernst(2019)]{Pan2019}
{\sc \au{Pan, Qingjiang} \& \au{Ernst, Darin~R}} \yr{2019}  \at{Gyrokinetic
  landau collision operator in conservative form}.  \jt{Physical Review E}
  \bvol{99}~(2),  \pg{023201}.

\bibitem[Pan {\em et~al.\/}(2020)Pan, Ernst \& Crandall]{Pan2020}
{\sc \au{Pan, Q.}, \au{Ernst, D.~R.} \& \au{Crandall, P.}} \yr{2020}  \at{First
  implementation of gyrokinetic exact linearized {L}andau collision operator
  and comparison with models}.  \jt{Physics of Plasmas}  \bvol{27}~(4),
  \pg{042307}.

\bibitem[Pan {\em et~al.\/}(2021)Pan, Ernst \& Hatch.]{pan2021}
{\sc \au{Pan, Q.}, \au{Ernst, D.~R.} \& \au{Hatch., D.~R}} \yr{2021}
  \at{Importance of gyrokinetic exact fokker-planck collisions in fusion plasma
  turbulence}.  \jt{Physical Review E}  \bvol{103}~(5),  \pg{L051202}.

\bibitem[Paruta {\em et~al.\/}(2018)Paruta, Ricci, Riva, Wersal, Beadle \&
  Frei]{Paruta2018}
{\sc \au{Paruta, P.}, \au{Ricci, P.}, \au{Riva, F.}, \au{Wersal, C.},
  \au{Beadle, C.} \& \au{Frei, B.}} \yr{2018}  \at{Simulation of plasma
  turbulence in the periphery of diverted tokamak by using the gbs code}.
  \jt{Physics of Plasmas}  \bvol{25}~(11),  \pg{112301}.

\bibitem[Romanelli(1989)]{Romanelli1989}
{\sc \au{Romanelli, F.}} \yr{1989}  \at{Ion temperature-gradient-driven modes
  and anomalous ion transport in tokamaks}.  \jt{Physics of Fluids B: Plasma
  Physics}  \bvol{1}~(5),  \pg{1018--1025}.

\bibitem[Romanelli \& Briguglio(1990)]{Romanelli1990}
{\sc \au{Romanelli, F.} \& \au{Briguglio, S.}} \yr{1990}  \at{Toroidal
  semicollisional microinstabilities and anomalous electron and ion transport}.
   \jt{Physics of Fluids B: Plasma Physics}  \bvol{2}~(4),  \pg{754--763}.

\bibitem[Rosenbluth {\em et~al.\/}(1972)Rosenbluth, Hazeltine \&
  Hinton]{rosenbluth1972}
{\sc \au{Rosenbluth, M.~N.}, \au{Hazeltine, R.~D.} \& \au{Hinton, F.~L.}}
  \yr{1972}  \at{Plasma transport in toroidal confinement systems}.  \jt{The
  Physics of Fluids}  \bvol{15}~(1),  \pg{116--140}.

\bibitem[Rosenbluth \& Hinton(1998)]{Rosenbluth1998a}
{\sc \au{Rosenbluth, M.~N.} \& \au{Hinton, F.~L.}} \yr{1998}  \at{{Poloidal
  flow driven by ion-temperature-gradient turbulence in tokamaks}}.
  \jt{Physical Review Letters}  \bvol{80}~(4),  \pg{724--727}.

\bibitem[Rosenbluth {\em et~al.\/}(1957)Rosenbluth, MacDonald \&
  Judd]{Rosenbluth1957}
{\sc \au{Rosenbluth, M.~N.}, \au{MacDonald, W.~M.} \& \au{Judd, D.~L.}}
  \yr{1957}  \at{Fokker-{P}lanck equation for an inverse-square force}.
  \jt{Physical Review}  \bvol{107}~(1),  \pg{1}.

\bibitem[Shimada {\em et~al.\/}(2007)Shimada, Campbell, Mukhovatov, Fujiwara,
  Kirneva, Lackner, Nagami, Pustovitov, Uckan, Wesley {\em
  et~al.\/}]{Shimada2007}
{\sc \au{Shimada, M.}, \au{Campbell, D.~J.}, \au{Mukhovatov, V.}, \au{Fujiwara,
  M.}, \au{Kirneva, N.}, \au{Lackner, K.}, \au{Nagami, M.}, \au{Pustovitov,
  V.~D.}, \au{Uckan, N.}, \au{Wesley, J.} \& \au{others}} \yr{2007}
  \at{Progress in the iter physics basis-chapter 1: overview and summary}.
  \jt{Nuclear Fusion}  \bvol{47},  \pg{S1}.

\bibitem[Snider(2017)]{Snider2017}
{\sc \au{Snider, R.~F.}} \yr{2017} {\em Irreducible Cartesian Tensors\/}. {\em
  De Gruyter Studies in Mathematical Physics\/} 1.  \publ{De Gruyter}.

\bibitem[Stegmeir {\em et~al.\/}(2019)Stegmeir, Ross, Francisquez, Zholobenko,
  Coster, Maj, Manz, Jenko, Rogers {\em et~al.\/}]{Stegmeir2019}
{\sc \au{Stegmeir, A.}, \au{Ross, A .and~Body, T}, \au{Francisquez, M.},
  \au{Zholobenko, W.}, \au{Coster, D.}, \au{Maj, O.}, \au{Manz, P.}, \au{Jenko,
  F.}, \au{Rogers, B.~N.} \& \au{others}} \yr{2019}  \at{Global turbulence
  simulations of the tokamak edge region with grillix}.  \jt{Physics of
  Plasmas}  \bvol{26}~(5),  \pg{052517}.

\bibitem[Sugama {\em et~al.\/}(2019)Sugama, Matsuoka, Satake, Nunami \&
  Watanabe]{Sugama2019}
{\sc \au{Sugama, H.}, \au{Matsuoka, S.}, \au{Satake, S.}, \au{Nunami, M.} \&
  \au{Watanabe, T.-H.}} \yr{2019}  \at{Improved linearized model collision
  operator for the highly collisional regime}.  \jt{Physics of Plasmas}
  \bvol{26}~(10),  \pg{102108}.

\bibitem[Sugama {\em et~al.\/}(2009)Sugama, Watanabe \& Nunami]{Sugama2009}
{\sc \au{Sugama, H.}, \au{Watanabe, T.-H.} \& \au{Nunami, M.}} \yr{2009}
  \at{Linearized model collision operators for multiple ion species plasmas and
  gyrokinetic entropy balance equations}.  \jt{Physics of Plasmas}
  \bvol{16}~(11),  \pg{112503}.

\bibitem[Tamain {\em et~al.\/}(2016)Tamain, Bufferand, Ciraolo, Colin, Galassi,
  Ghendrih, Schwander \& Serre]{tamain2016}
{\sc \au{Tamain, P.}, \au{Bufferand, H.}, \au{Ciraolo, G.}, \au{Colin, C.},
  \au{Galassi, D.}, \au{Ghendrih, P.}, \au{Schwander, F.} \& \au{Serre, E.}}
  \yr{2016}  \at{The tokam3x code for edge turbulence fluid simulations of
  tokamak plasmas in versatile magnetic geometries}.  \jt{Journal of
  Computational Physics}  \bvol{321},  \pg{606}.

\bibitem[Villard {\em et~al.\/}(2013)Villard, Angelino, Bottino, Brunner,
  Jolliet, McMillan, Tran \& Vernay]{Villard2013}
{\sc \au{Villard, L.}, \au{Angelino, P.}, \au{Bottino, A.}, \au{Brunner, S.},
  \au{Jolliet, S.}, \au{McMillan, B.~F.}, \au{Tran, T.-M.} \& \au{Vernay, T.}}
  \yr{2013}  \at{Global gyrokinetic ion temperature gradient turbulence
  simulations of {ITER}}.  \jt{Plasma Physics and Controlled Fusion}
  \bvol{55}~(7),  \pg{074017}.

\bibitem[Wong(1998)]{Wong1998}
{\sc \au{Wong, M.~W.}} \yr{1998} {\em The Weyl Transform\/}.  \publ{Springer}.

\bibitem[Xiao {\em et~al.\/}(2007)Xiao, Catto \& Molvig]{Xiao2007}
{\sc \au{Xiao, Y.}, \au{Catto, P.~J.} \& \au{Molvig, K.}} \yr{2007}
  \at{{Collisional damping for ion temperature gradient mode driven zonal
  flow}}.  \jt{Physics of Plasmas}  \bvol{14}~(3).

\bibitem[Xu \& Rosenbluth(1991)]{Xu1991}
{\sc \au{Xu, X.~Q.} \& \au{Rosenbluth, M.~N.}} \yr{1991}  \at{Numerical
  simulation of ion-temperature-gradient-driven modes}.  \jt{Physics of Fluids
  B: Plasma Physics}  \bvol{3}~(3),  \pg{627}.

\bibitem[Zeiler {\em et~al.\/}(1997)Zeiler, Drake \& Rogers]{Zeiler1997}
{\sc \au{Zeiler, a.}, \au{Drake, J.~F.} \& \au{Rogers, B.}} \yr{1997}
  \at{{Nonlinear reduced Braginskii equations with ion thermal dynamics in
  toroidal plasma}}.  \jt{Physics of Plasmas}  \bvol{4}~(6),  \pg{2134}.

\bibitem[Zhu {\em et~al.\/}(2018)Zhu, Francisquez \& Rogers]{zhu2018}
{\sc \au{Zhu, B.}, \au{Francisquez, M.} \& \au{Rogers, B.~N.}} \yr{2018}
  \at{{GDB}: A global 3{D} two-fluid model of plasma turbulence and transport
  in the tokamak edge}.  \jt{Computer Physics Communications}  \bvol{232},
  \pg{46}.

\end{thebibliography}

\end{document}